\documentclass{article}
\usepackage{amsmath, amssymb}
\usepackage{graphicx}
\usepackage{caption}
\usepackage{geometry}
\usepackage{indentfirst}
\usepackage{amsthm}
\usepackage{changepage}
\usepackage{hyperref}
\usepackage{cite}
\usepackage{ctex}
\usepackage{cases}
\usepackage{accents}
\newtheorem{theorem}{Theorem}
\newtheorem{definition}{Definition}
\newtheorem{example}{Example}

\title{\bf{Rigid body kinematics in an intuitive group-theoretic approach, as completely as possible}\\[10pt] \large{Part I: Rotational Phenomena}}
\author{Ziyuan Wang\\\textit{Department of physics, Tianjin University, Tianjin 300350, China}}
\renewcommand{\today}{\ifcase\month\or
January\or February\or March\or April\or May\or June\or
July\or August\or September\or October\or November\or December\fi
~\number\day, \number\year}
\begin{document}
\maketitle

\begin{center}

{\Large ABSTRACT}
\end{center}

This paper focuses on rotational phenomena of rigid body kinematics. It discusses them in a group-theoretic approach as completely as possible, using methods and notations as intuitive as possible. With a review of current literature, this article also covers some original parts that remain largely unexplored.

\newpage
\tableofcontents
\newpage
\maketitle

\section{Introduction}\label{sec:intro}
Rigid body mechanics is a well-established discipline with a history of several hundred years, and there are some classic textbooks on it, like Crouch (1981)\cite{crouch1981}, Hahn (2002)\cite{hahn2002}, Heard (2008)\cite{heard2008}, and Huang (2012)\cite{huang2012}; and some on classical mechanics that involves this part, including Goldstein (2001)\cite{goldstein}, Taylor (2005)\cite{taylor}, Liu (2019)\cite{liu2024}, Gao (2024)\cite{gao2024}, and L\"{u} (2025)\cite{lu2025}. However, those textbooks are not complete for the discipline, in terms of the diversity of representations and illustrations of physical meanings, and some of them use physical languages opaque to starters, like always viewing rigid-body rotations under passive viewpoint. Furthermore, this discipline has a very tight relation with group theory, and it can be discussed in a group-theoretic approach systematically, which is an analytical paradigm and can be applied in many other disciplines like special relativity. L\"{u} (2025) is a good example adding group-theoretic approach to the frame, but one can go further. This is the reason why the author choose to write an article on rigid body mechanics.

Specifically, this article focuses on rigid body kinematics. It primarily serves as a review of it in a group-theoretic approach, while incorporating so many original insights. The author aim to give a complete, rigorous but intuitive discussion about this part of knowledge. While the article tries to elucidate rigid body kinematics systematically as long as it can, it also uses physical and mathematical languages as intuitive as possible. In the meanwhile, it clearly points out many subtleties that may cause faults, which is friendly to starters. This article use active viewpoint mainly.

The author divide the article into two parts. This paper serves as the first part of this article, focusing on rotational phenomena; Part II focuses on general rigid body motions and manifolds of them. This paper omits proofs to some basic mathematical theorems, e.g. that similar matrices share the same eigenvalues. Readers can easily find them in mathematical textbooks. 

There are 7 sections in the body of this paper, from Section~\ref{sec:overv} to~\ref{sec:descsu2}. At first, this article introduces some basic knowledge on rigid body transformations in Section~\ref{sec:overv}, aiming to let readers familiar with them. The following sections are all about rotations. Section~\ref{sec:o3so3} makes clear the necessary and sufficient condition for a matrix to express a rotation. For further discussion, Section~\ref{sec:view} originally summarizes the complete relationship among 4 equivalent scenarios of rotations. Section~\ref{sec:repro} introduces 3 equivalent expressions of rotations, and derives the relationship among them. In Section~\ref{sec:propso3}, we investigate local properties of SO(3) rotations, including the Lie algebras and angular velocity matrices of SO(3) Group, and then demonstrate the Euler-angles expressions for continuous rotations. For further understandings, we discuss composition of rotations in Section~\ref{sec:compo}. Section~\ref{sec:overv} to~\ref{sec:compo} constitute a complete system of description of rotations. In the last, Section~\ref{sec:descsu2} tries another representation of elements in SO(3) Group, which leads to SU(2) Group, originally serving as an equivalent description of rotations. Furthermore, it also has own physical meaning, as it brings out another kind of quantities named spinors. 
\section{Overview of rigid body transformations}\label{sec:overv}
\subsection{Meaning of rigid body transformations}\label{sec:defro}
A rigid body transformation is a physically realizable isometric transformation, encompassing both passive and active transformations. The active viewpoint keeps the coordinate frame fixed, with the change in the vector itself giving rise to the transformation of its components. In contrast, in a passive viewpoint, any vector itself remains unchanged while the coordinate frame is deformed, leading to a change in the vector's components as a consequence of the altered perspective, i.e. the vector is expressed in another representation. In physics, passive transformations are more commonly employed; but this article mainly focuses on the intrinsic nature and results of rigid-body motion\textemdash active transformations, with only a few under passive viewpoint involved due to actual scenarios. Accordingly, unless otherwise specified, transformations in this paper are to be understood as active ones.

In general, the set of rigid body transformation is composed of translations, rotations, and their composite operations, as will be proved in Part II.
\subsection{Necessary condition for a rigid body transformation}\label{sec:nececonrbt}
A rigid body transformation constitutes an isometry, namely, a transformation that preserves distances. This is equivalent to preserving both \textbf{distances} from a fixed point and \textbf{angles} with vertex at that point. Furthermore, since the motion of matter occurs locally, any physical transformation can be realized through a continuous transformation. Considering each tiny step in the continuous transformation, the orientation of the coordinate axes can only change by a small amount, so they cannot be flipped, and thus the coordinate frame is still right-handed if the original one is. As a result, a rigid body transformation is needed to be \textbf{right-handedness-preserving}.

In fact, as revealed in Section~\ref{sec:euler}, these three conditions also constitute sufficient conditions for a rotation to qualify. This still holds true for a general rigid-body motion.
\subsection{Translations}
Choose any two points in the rigid body: if the relative position vector between them is always a constant after a realizable isometric transformation, then the transformation is called a translation. In a translation, if the change in coordinates of a point in the rigid body is fixed, those of any others are known\textemdash exactly the same.
\subsubsection{Description}
In 3D Euclidean space, the change of a point is
\begin{equation}\label{eqt:tsexp}
\hat{T}:\ \boldsymbol{r}\to\boldsymbol{r}+\boldsymbol{r_0}
\end{equation}
where $\boldsymbol{r_0}$ is the deformation.
A translation can be described by the coordinate change of any point in the rigid body, i.e. the expression above.
\subsubsection{Commutative law}
In Euclidean space, position coordinates are addable. Consider a series of points $A_i$ ($i$ from 0 to n), i.e. the coordinates of $A_n$ in the $A_0$ frame is the sum of the position vector of $A_j$ to $A_{j-1}$ where $j$ is from 1 to $n$. Thus, consider two translations that $\hat{T}_1:\ \boldsymbol{r}\to\boldsymbol{r}+\boldsymbol{r_1}$ and that $\hat{T}_2:\ \boldsymbol{r}\to\boldsymbol{r}+\boldsymbol{r_2}$
\begin{equation}
\hat{T}_2\hat{T}_1=\bigl(\boldsymbol{r}\to\boldsymbol{r}+\boldsymbol{r_1}+\boldsymbol{r_2}\bigr)=\bigl(\boldsymbol{r}\to\boldsymbol{r}+\boldsymbol{r_2}+\boldsymbol{r_1}\bigr)=\hat{T}_1\hat{T}_2
\end{equation}
so translations commute.
\subsection{Rotations}
A rotation is a physically realizable isometric transformation with a fixed point somewhere, and the origin is chosen to be that point in default throughout the article unless specified.
\subsubsection{Planar rotation matrix}\label{sec:planar}
It's easy to derive from plane geometry that under active viewpoint the coordinates transform like
\begin{equation}
\begin{pmatrix}
x' \\ y'
\end{pmatrix}
=
\begin{pmatrix}
\cos\phi & -\sin\phi \\
\sin\phi & \cos\phi
\end{pmatrix}
\begin{pmatrix}
x \\ y
\end{pmatrix}
\end{equation}
where $\phi$ is the rotation angle of the vector.

And under passive viewpoint they transform like
\begin{equation}
\begin{pmatrix}
x' \\ y'
\end{pmatrix}
=
\begin{pmatrix}
\cos\phi & \sin\phi \\
-\sin\phi & \cos\phi
\end{pmatrix}
\begin{pmatrix}
x \\ y
\end{pmatrix}
\end{equation}
where $\phi$ is the rotation angle of the coordinate frame.
\subsubsection{Spatial rotation matrix}\label{sec:spatial}
A rotation constitutes an isometry. More weakly, it is worth noting that two vectors pointing in the same direction remain aligned after a rotational transformation. Together with the isometric property of rotations, this implies that the rotation operator $\hat{R}$ satisfies
\begin{equation}
\begin{pmatrix}
x' \\ y' \\ z'
\end{pmatrix}
=\hat{R}\lambda
\begin{pmatrix}
x \\ y \\ z
\end{pmatrix}
=\lambda\hat{R}
\begin{pmatrix}
x \\ y \\ z
\end{pmatrix}
\end{equation}
so $\hat{R}$ is a linear operator. Consequently, any spatial rotation operator can be expressed in matrix form.
\subsubsection{Rodrigues' formula and commutative law of infinitesimal angles}\label{sec:commt}
Next, we turn to the vectorial character of infinitesimal rotations. For clarity, we restrict attention to the case of two successive rotations, each about a single axis (the general validity of this construction will be established through Euler's Theorem in Section~\ref{sec:eig}). In general, rotations are non-commutative and thus cannot be regarded as vectors. Nevertheless, as will become clear, in the infinitesimal limit the rotation angle can be regarded as a vector\textemdash more precisely, as a pseudo-vector.

Next in this subsubsection we follow the derivations in L\"{u} (2025)\cite{lu2025}. For simplicity yet without loss of generality, one defines the axis of rotation to be the $z$-axis, and the angle to be $\phi$. The coordinates transform like
\begin{equation}
\begin{pmatrix}
x' \\ y' \\ z'
\end{pmatrix}
=\begin{pmatrix}
\cos\phi & -\sin\phi & 0 \\
\sin\phi & \cos\phi & 0 \\
0&0&1
\end{pmatrix}
\begin{pmatrix}
x \\ y \\ z
\end{pmatrix}
=\cos\phi
\begin{pmatrix}
x \\ y \\ z
\end{pmatrix}
+(1-\cos\phi)
\begin{pmatrix}
0 \\ 0 \\ z
\end{pmatrix}
+\sin\phi
\begin{pmatrix}
-y \\ x \\ 0
\end{pmatrix}
\end{equation}

Written in a covariant form, the equation becomes the \textbf{Rodrigues' rotation formula}
\begin{equation}\label{eqt:rodrigues}
\boldsymbol{r'}=\boldsymbol{r}\cos\phi+(\boldsymbol{\hat{n}}\cdot\boldsymbol{r})\boldsymbol{\hat{n}}(1-\cos\phi)+\boldsymbol{\hat{n}}\times\boldsymbol{r}\sin\phi
\end{equation}
where $\boldsymbol{\hat{n}}$ is the unit vector of the axis of rotation.
Take the angle $\phi$ to be an infinitesimal $\delta\phi$, then
\begin{equation}
\boldsymbol{r'}=\boldsymbol{r}(1-\mathcal{O}(\delta\phi^2))+(\boldsymbol{\hat{n}}\cdot\boldsymbol{r})\boldsymbol{\hat{n}}\ \mathcal{O}(\delta\phi^2)+\boldsymbol{\hat{n}}\times\boldsymbol{r}\delta\phi
\end{equation}
and it is easy to see
\begin{equation} \label{eqt:infmrot}
\mathrm{d}\boldsymbol{r} = \boldsymbol{\hat{n}}\times\boldsymbol{r}\mathrm{d}\phi
\end{equation}

One can define the infinitesimal angle ``vector'' $\mathrm{d}\boldsymbol{\phi} = \boldsymbol{\hat{n}}\,\mathrm{d}\phi$ , and then the equation goes to $\mathrm{d}\boldsymbol{r} = \mathrm{d}\boldsymbol{\phi}\times\boldsymbol{r}$. Since $\boldsymbol{r}$ and $\mathrm{d}\boldsymbol{r}$ are vectors, $\mathrm{d}\boldsymbol{\phi}$ may be a pseudo-vector.
Consider two successive infinitesimal rotations: the first by
$\mathrm{d}\boldsymbol{\phi}_1$ and the second by $\mathrm{d}\boldsymbol{\phi}_2$.
Then, after these two rotations, the variation of the position vector is
\begin{equation}
\mathrm{d}\boldsymbol{r}
= \mathrm{d}\boldsymbol{\phi}_1 \times \boldsymbol{r}
+ \mathrm{d}\boldsymbol{\phi}_2 \times \bigl(\boldsymbol{r}
   + \mathrm{d}\boldsymbol{\phi}_1\times\boldsymbol{r}\bigr)
+ \mathcal{O}(||\mathrm{d}\boldsymbol{\phi}_i||^{2})
= \bigl(\mathrm{d}\boldsymbol{\phi}_1
+ \mathrm{d}\boldsymbol{\phi}_2\bigr)\times\boldsymbol{r}
+ \mathcal{O}(||\mathrm{d}\boldsymbol{\phi}_i||^{2})
\end{equation}

Taking into account that the overall effect is equivalent to a total
rotation angle $\mathrm{d}\boldsymbol{\phi}_1 + \mathrm{d}\boldsymbol{\phi}_2$,
we may conclude that an infinitesimal rotation angle is indeed a
pseudo-vector.

Differentiate both sides of the equation with respect to time and define the angular velocity
\begin{equation}
\boldsymbol{\omega}=\frac{\mathrm{d}\boldsymbol{\phi}}{\mathrm{d}t}=\dot{\phi}\boldsymbol{\hat{n}}
\end{equation}
which is also a pseudo-vector. Then equation (\ref{eqt:infmrot}) goes to
\begin{equation}
\boldsymbol{v}=\boldsymbol{\omega}\times\boldsymbol{r}
\end{equation}

The angular velocity can also be written in matrix form, as will be noted in Section~\ref{sec:angv}.
\subsection{Degrees of freedom}\label{sec:degfr}
A rigid body has three rotational degrees of freedom. In Euclidean space, with one point fixed, arbitrarily take three non-collinear points $A, B, C$ on the body, which define the vectors $\overrightarrow{AB}$ and $\overrightarrow{AC}$. The direction of $\overrightarrow{AB}$ can be specified by spherical coordinates $(\Theta, \Phi)$, providing two degrees of freedom. The direction of $\overrightarrow{AC}$ must maintain a fixed angle with $\overrightarrow{AB}$. Once $\overrightarrow{AB}$ is fixed, the locus of point $C$ lies on a circle in a determined plane, and its orientation can be described by a polar angle $\alpha$, giving one more degree of freedom.

We have seen and will see many phenomena in agreement with this result, repeatedly in Sections~\ref{sec:commt}, ~\ref{sec:sgaietr}, ~\ref{sec:indep}, ~\ref{sec:caylay}: Infinitesimal rotations can be decomposed into 3 independent components about the Cartesian axes; Euler angles total 3; the number of solutions is finite (though not only one) once the 3 diagonal elements of a rotation matrix is fixed; independent real numbers (Caylay-Klein parameters) to describe an arbitrary rotation total 3. This phenomenon will be analyzed systematically in Part II.

Then, choose a point (often chosen as the center of mass) in the rigid body, and the relative position of any other point in the body to it is determined by the three rotational degrees of freedom. On the other hand, the position of this point in the reference frame is determined by the three pointal degrees of freedom, as characterized directly by $\boldsymbol{r}_0$ in (\ref{eqt:tsexp}). With the position of a point, and any other's relative position to it, the configuration is determined. Thus, the number of degrees of freedom is 6.

Actually, after translating the rigid body by the transformation of the point mentioned, then conducting the rotation mentioned above, or conversely, one can get the final configuration. Thus, the pointal degrees of freedom is more commonly named translational degrees. In conclusion, rigid body transformations has 6 degrees of freedom, 3 translational and 3 rotational.
\subsection{Similar transformation}\label{sec:repiner00}
Consider an invertible passive transformation. The essence of it is the basis transformation. But for convenience, we characterize the transformation by a global coordinate transformation operator $\hat{C}$.
\begin{theorem}\label{eqt:similar}
If an active transformation in the $\mathrm{A}$ frame is represented by $\hat{F}$, while the representation transformation operator for coordinates from $\mathrm{A}$ to $\mathrm{B}$ is $\hat{C}$, then the active transformation in the $\mathrm{B}$ frame is represented by $\hat{C}\hat{F}\hat{C}^{-1}$.
\end{theorem}
\begin{proof}
The coordinates of the $A$ frame are denoted by lower 1-s, while those of $B$ are not; the coordinates after the rotation are denoted by primes, while those before are not.
\[
\begin{pmatrix}
x'\\y'\\z'
\end{pmatrix}
=\hat{C}
\begin{pmatrix}
x_1'\\y_1'\\z_1'
\end{pmatrix}
,\
\begin{pmatrix}
x\\y\\z
\end{pmatrix}
=\hat{C}
\begin{pmatrix}
x_1\\y_1\\z_1
\end{pmatrix}
\]
And because
\[
\begin{pmatrix}
x_1'\\y_1'\\z_1'
\end{pmatrix}
=\hat{F}
\begin{pmatrix}
x_1\\y_1\\z_1
\end{pmatrix}
\]
we can conclude that
\[
\begin{pmatrix}
x'\\y'\\z'
\end{pmatrix}
=\hat{C}\hat{F}\hat{C}^{-1}
\begin{pmatrix}
x\\y\\z
\end{pmatrix}
\]
i.e.
\begin{equation}\label{eqt:genesimil}
\hat{B}=\hat{C}\hat{F}\hat{C}^{-1}
\end{equation}
where $\hat{B}$ is the operator in the $B$ representation.
\end{proof}

This result is essential for representation transformation, and will be further applied in Section~\ref{sec:repiner1}.
\section{O(3) Group and SO(3) Group}\label{sec:o3so3}
\subsection{Equivalent condition for an orthogonal matrix}\label{sec:equiv}
\begin{definition}
A matrix $\mathbf{M}$ is orthogonal if $\mathbf{M}^{-1}=\mathbf{M}^\mathrm{T}$.
\end{definition}
\begin{theorem}
A length- and angle-preserving matrix $\Leftrightarrow$ An orthogonal matrix

\end{theorem}

\begin{proof}
\noindent Take any two vectors $\boldsymbol{r_1}$ and $\boldsymbol{r_2}$ in Euclidean space.

\noindent 1)``$\Leftarrow$"

\noindent If $\mathbf{M}$ is orthogonal, then
\[
(\mathbf{M}\boldsymbol{r_1})^\mathrm{T}(\mathbf{M}\boldsymbol{r_2})=\boldsymbol{r_1}^\mathrm{T}\mathbf{M}^\mathrm{T}\mathbf{M}\boldsymbol{r_2}=\boldsymbol{r_1}^\mathrm{T}\mathbf{M}^{-1}\mathbf{M}\boldsymbol{r_2}=\boldsymbol{r_1}^\mathrm{T}\boldsymbol{r_2}
\]

\noindent Let $\boldsymbol{r_1}=\boldsymbol{r_2}$, then
\[
||\mathbf{M}\boldsymbol{r_1}||=\sqrt{(\mathbf{M}\boldsymbol{r_1})^\mathrm{T}(\mathbf{M}\boldsymbol{r_1})}=\sqrt{\boldsymbol{r_1}^\mathrm{T}\boldsymbol{r_1}}=||\boldsymbol{r_1}||
\]

\noindent On the other hand,
\[
\cos\langle\mathbf{M}\boldsymbol{r_1},\mathbf{M}\boldsymbol{r_2}\rangle=\frac{(\mathbf{M}\boldsymbol{r_1})^\mathrm{T}(\mathbf{M}\boldsymbol{r_2})}{||\mathbf{M}\boldsymbol{r_1}||\ ||\mathbf{M}\boldsymbol{r_2}||}=\frac{\boldsymbol{r_1}^\mathrm{T}\boldsymbol{r_2}}{||\boldsymbol{r_1}||\ ||\boldsymbol{r_2}||}=\cos\langle\boldsymbol{r_1},\boldsymbol{r_2}\rangle
\]

\noindent 2)``$\Rightarrow$"

\noindent Given the condition, it is easy to know $\forall \boldsymbol{r_1},\boldsymbol{r_2}$, $(\mathbf{M}\boldsymbol{r_1})^\mathrm{T}(\mathbf{M}\boldsymbol{r_2})=\boldsymbol{r_1}^\mathrm{T}\boldsymbol{r_2}$, i.e.
\[
\forall \boldsymbol{r_1},\boldsymbol{r_2},\ \boldsymbol{r_1}^\mathrm{T}\mathbf{M}^\mathrm{T}\mathbf{M}\boldsymbol{r_2}=\boldsymbol{r_1}^\mathrm{T}\boldsymbol{r_2}
\]

\noindent Denote
\[
\mathbf{M}^\mathrm{T}\mathbf{M}=
\begin{pmatrix}
a_{11} & a_{12}&a_{13}\\
a_{21} & a_{22} & a_{23} \\
a_{31}&a_{32}&a_{33}
\end{pmatrix}
,\
\boldsymbol{r_1}=
\begin{pmatrix}
x_1 \\ y_1 \\ z_1
\end{pmatrix}
,\
\boldsymbol{r_2}=
\begin{pmatrix}
x_2 \\ y_2 \\ z_2
\end{pmatrix}
\]

\noindent then $\boldsymbol{r_1}^\mathrm{T}\boldsymbol{r_2}=x_1x_2+y_1y_2+z_1z_2$, and
\begin{align*}
\boldsymbol{r_1}^\mathrm{T}\mathbf{M}^\mathrm{T}\mathbf{M}\boldsymbol{r_2}
&=
\begin{pmatrix}
x_1 & y_1 & z_1
\end{pmatrix}
\begin{pmatrix}
a_{11} & a_{12} & a_{13} \\
a_{21} & a_{22} & a_{23} \\
a_{31} & a_{32} & a_{33}
\end{pmatrix}
\begin{pmatrix}
x_2 \\ y_2 \\ z_2
\end{pmatrix} \\
&= a_{11} x_1 x_2 + a_{22} y_1 y_2 + a_{33} z_1 z_2
 + a_{12} x_1 y_2 + a_{21} x_2 y_1 \\
&\quad + a_{13} x_1 z_2 + a_{31} x_2 z_1
 + a_{23} y_1 z_2 + a_{32} y_2 z_1
\end{align*}

\noindent Two polynomials are equal if and only if $\forall i,\ a_{ii}=1$ and $\forall i\ne j,\ a_{ij}=0$ ($i$, $j$=1, 2, 3). Then $\mathbf{M}^\mathrm{T}\mathbf{M}=\mathbf{I}$.
\end{proof}
\subsection{Definitions of groups and Lie groups}\label{sec:defglg}
\begin{definition}
A group, based on a definition of multiplication between elements, is a set that satisfies the following conditions:
\begin{enumerate}
\item \textbf{Closure:} The product of any two elements is also contained in the set.

\item \textbf{Associativity:} Changing the way elements are grouped in the multiplication does not change the result.

\item \textbf{Identity element:} There exists an element that leaves every element of the set unchanged when multiplied.

\item \textbf{Inverse element:} For each element, there exists another element whose product with it gives the identity element.
\end{enumerate}
\end{definition}
\begin{definition}
For a given group, take any two elements and form a sequence with them as the starting and ending points. If there exists a way to insert a number of elements between them, in a certain order, such that for any two adjacent group elements in the sequence, the transformation between them still belongs to the group, and that adjacent elements can become arbitrarily close as more elements are inserted; then this group is called a Lie group.
\end{definition}

Here, ``arbitrarily close" means that the transformation between the two elements approaches the identity element.
For matrices, this is equivalent to saying that all corresponding matrix entries of the two matrices become arbitrarily close.

In an intuitive but informal sense, a Lie group is a group in which the transformation between any two elements can be generated by accumulating infinitesimal transformations within the group, i.e. through a continuous transformation.
\subsection{Group-theoretic description of orthogonal matrices: O(3) Group}\label{sec:gthpropo}
\begin{theorem}
All 3D orthogonal matrices constitute a group.

\end{theorem}

\begin{proof}
\
\begin{enumerate}
\item \textbf{Closure:} If $\mathbf{A}$ and $\mathbf{B}$ are both orthogonal matrices, then
\[
(\mathbf{A}\mathbf{B})^\mathrm{T} (\mathbf{A}\mathbf{B})
= \mathbf{B}^\mathrm{T} \mathbf{A}^\mathrm{T} \mathbf{A} \mathbf{B}
= \mathbf{B}^{-1} \mathbf{A}^{-1} \mathbf{A} \mathbf{B}
= \mathbf{I}
\]
so $\mathbf{A}\mathbf{B}$ is also an orthogonal matrix.

\item \textbf{Associativity:} By the associativity of matrix multiplication,
\[
(\mathbf{A}\mathbf{B}) \mathbf{C} = \mathbf{A} (\mathbf{B}\mathbf{C})
\]

\item \textbf{Identity element:} Due to the properties of matrices, there exists a unique identity matrix $\mathbf{I}$.

\item \textbf{Inverse element:} For an orthogonal matrix $\mathbf{A}$, $\mathbf{A}^\mathrm{T}$ serves as its inverse, as $\mathbf{A}^\mathrm{T}=\mathbf{A}^{-1}$ and $\mathbf{A}^{-1}$ is defined such that
\[
\mathbf{A}^{-1} \mathbf{A} = \mathbf{A} \mathbf{A}^{-1} = \mathbf{I}
\]
\end{enumerate}
\end{proof}
The group is named O(3) Group.

\begin{theorem}
$\mathrm{O(3)}$ Group is not a Lie Group.
\end{theorem}
\begin{proof}
There exists one element of the group, $
\mathbf{P_1}=
\begin{pmatrix}
+1 & 0 & 0 \\
0 & +1 & 0 \\
0 & 0 & -1
\end{pmatrix}
$ cannot be obtained through continuous transformation, because it changes a right-handed frame to a left-handed one (See in Section~\ref{sec:nececonrbt}). 
\end{proof}
\subsection{Equivalent condition for a rotation matrix}\label{sec:nns}
\subsubsection{Condition for right-handedness' preservation}\label{sec:rhpre}
The determinant of an orthogonal matrix $\mathbf{M}$ is $\pm 1$:
\[
\mathbf{M}^\mathrm{T}\mathbf{M}=\mathbf{I}\Rightarrow(\mathrm{det}\ \mathbf{M})^2=1\Rightarrow\mathrm{det} \ \mathbf{M}=\pm 1
\]

If the coordinate transformation matrix is an orthogonal matrix $\mathbf{M}=
\begin{pmatrix}
m_{11} & m_{12} & m_{13} \\
m_{21} & m_{22} & m_{23} \\
m_{31} & m_{32} & m_{33}
\end{pmatrix}
$, then the basis transformation relationship is

\begin{equation}\label{eqt:basis}
\begin{pmatrix}
\boldsymbol{\hat{x}'} & \boldsymbol{\hat{y}'} & \boldsymbol{\hat{z}'}
\end{pmatrix}
=
\begin{pmatrix}
\boldsymbol{\hat{x}} & \boldsymbol{\hat{y}} & \boldsymbol{\hat{z}}
\end{pmatrix}
\mathbf{M}^{-1}
\end{equation}
where $\mathbf{M}^{-1}=
\begin{pmatrix}
m_{11} & m_{21} & m_{31} \\
m_{12} & m_{22} & m_{32} \\
m_{13} & m_{23} & m_{33}
\end{pmatrix}
$
, then
\[
\begin{cases}
\boldsymbol{\hat{x}'} = m_{11}\boldsymbol{\hat{x}} + m_{12}\boldsymbol{\hat{y}} + m_{13}\boldsymbol{\hat{z}} \\
\boldsymbol{\hat{y}'} = m_{21}\boldsymbol{\hat{x}} + m_{22}\boldsymbol{\hat{y}} + m_{23}\boldsymbol{\hat{z}} \\
\boldsymbol{\hat{z}'} = m_{31}\boldsymbol{\hat{x}} + m_{32}\boldsymbol{\hat{y}} + m_{33}\boldsymbol{\hat{z}}
\end{cases}
\]
Then
\begin{align}
\boldsymbol{\hat{x}'}\times\boldsymbol{\hat{y}'}&=(m_{12}m_{23}-m_{13}m_{22})\boldsymbol{\hat{x}}+(m_{13}m_{21}-m_{11}m_{23})\boldsymbol{\hat{y}}+(m_{11}m_{22}-m_{12}m_{21})\boldsymbol{\hat{z}} \nonumber\\
&=A_{31}\boldsymbol{\hat{x}}+A_{32}\boldsymbol{\hat{y}}+A_{33}\boldsymbol{\hat{z}} \nonumber\\
&=\frac{1}{\mathrm{det}\ \mathbf{M}}(m_{31}\boldsymbol{\hat{x}}+m_{32}\boldsymbol{\hat{y}}+m_{33}\boldsymbol{\hat{z}}) \nonumber\\
&=\frac{1}{\mathrm{det}\ \mathbf{M}}\boldsymbol{\hat{z}'}
\end{align}
where $A_{ij}$ is the cofactor of $m_{ij}$.

It can be known that the matrix is right-handedness-preserved if $\mathrm{det}\ \mathbf{M}=1$, demonstrating it \textbf{may} be a rotation matrix; while turning a right-handed frame into a left-handed one if $\mathrm{det}\ \mathbf{M}=-1$, meaning that it is a reflecting matrix, and thus it would mean a discrete transformation.
\subsubsection{Euler angles ($zxz$-convention) and Tait-Bryan angles ($xyz$-convention)}\label{sec:euler}
A general rotation can be obtained through chained rotations. For convenience, we establish a fixed coordinate frame, and a body-attached frame representing the configuration of the rigid body. Based on this, the motion of a point on the rigid body is represented by the transformation in the inertial representation, of a fixed vector according to the body-attached frame. Initially, the two frames coincide, and the vector of the coordinates is denoted as $
\begin{pmatrix}
x&y&z
\end{pmatrix}
^\mathrm{T}$.

1) The rigid body rotates counterclockwise about the $z$-axis by $\varphi \in [0,2\pi)$, yielding new coordinates $\begin{pmatrix}
x_1&y_1&z_1
\end{pmatrix}
^\mathrm{T}$.

2) The rigid body rotates counterclockwise about the $x_{1}$-axis by $\theta \in [0,\pi]$, yielding new coordinates $\begin{pmatrix}
x_2&y_2&z_2
\end{pmatrix}
^\mathrm{T}$.

3) The rigid body rotates counterclockwise about the $z_{2}$-axis by $\psi \in [0,2\pi)$, yielding new coordinates $\begin{pmatrix}
x'&y'&z'
\end{pmatrix}
^\mathrm{T}$.

\begin{figure}[htbp]
  \centering
  \includegraphics[width=0.3\linewidth]{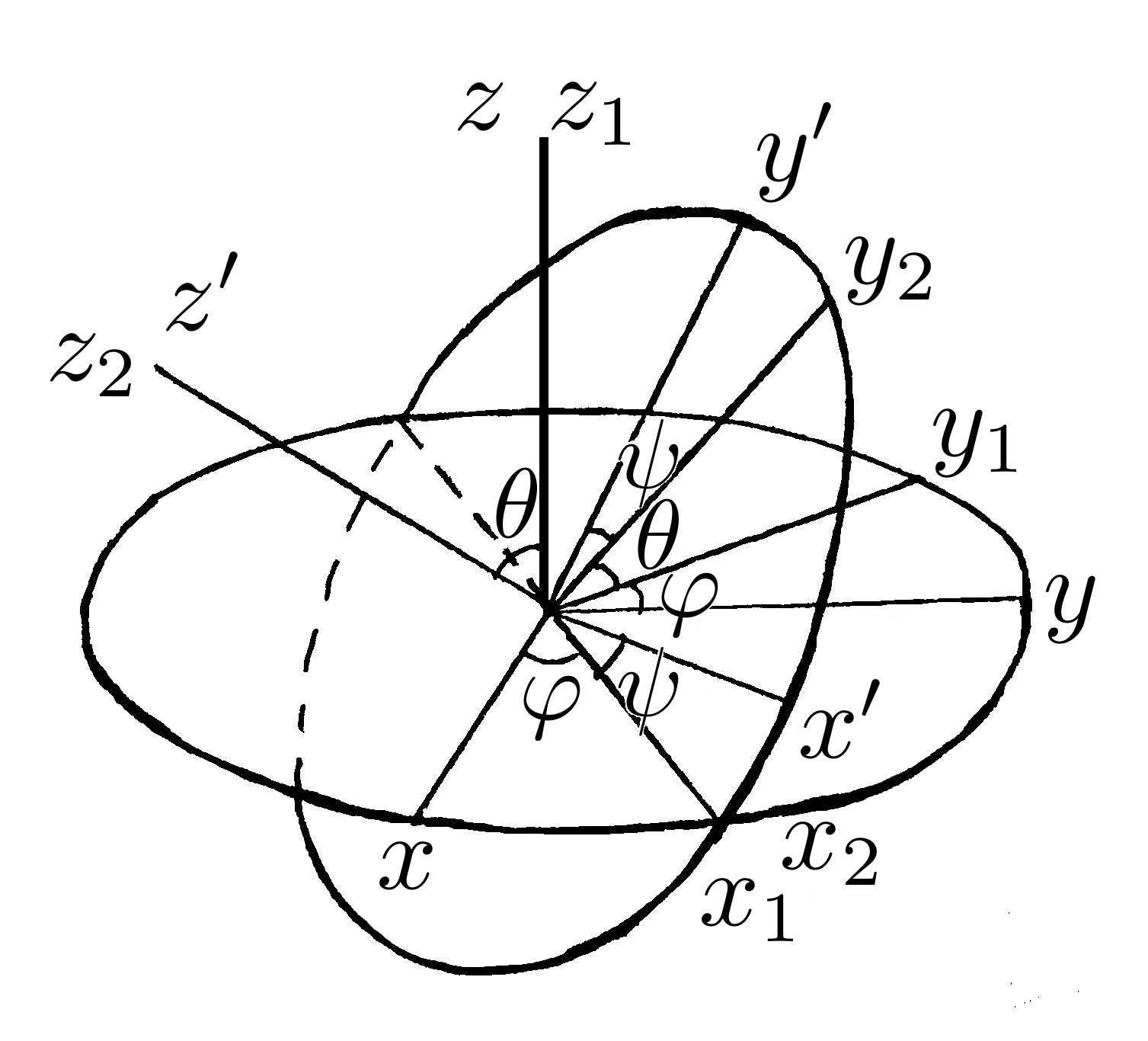}
  \caption{Euler angles.}
  \label{fig:eulerangles}
\end{figure}

Likewise, each rotation can also be described by 3 Tait-Bryan angles, as is defined by the angles that satisfy

1) The rigid body starts from the initial coordinates and rotates counterclockwise about the $z$-axis by $\varphi \in [0,2\pi)$, yielding new coordinates
$\begin{pmatrix}
x_1&y_1&z_1
\end{pmatrix}
^\mathrm{T}$.

2) The rigid body rotates counterclockwise about the $y_{1}$-axis by $\theta \in [-\frac{\pi}{2},\frac{\pi}{2}]$, yielding new coordinates
$\begin{pmatrix}
x_2&y_2&z_2
\end{pmatrix}
^\mathrm{T}$.

3) The rigid body rotates counterclockwise about the $x_{2}$-axis by $\psi \in [0,2\pi)$, yielding new coordinates $\begin{pmatrix}
x'&y'&z'
\end{pmatrix}
^\mathrm{T}$
where $\begin{pmatrix}
x'&y'&z'
\end{pmatrix}
^\mathrm{T}$
 can express the final coordinates of any point in the rigid body.
\begin{figure}[htbp]
  \centering
  \includegraphics[width=0.4\linewidth]{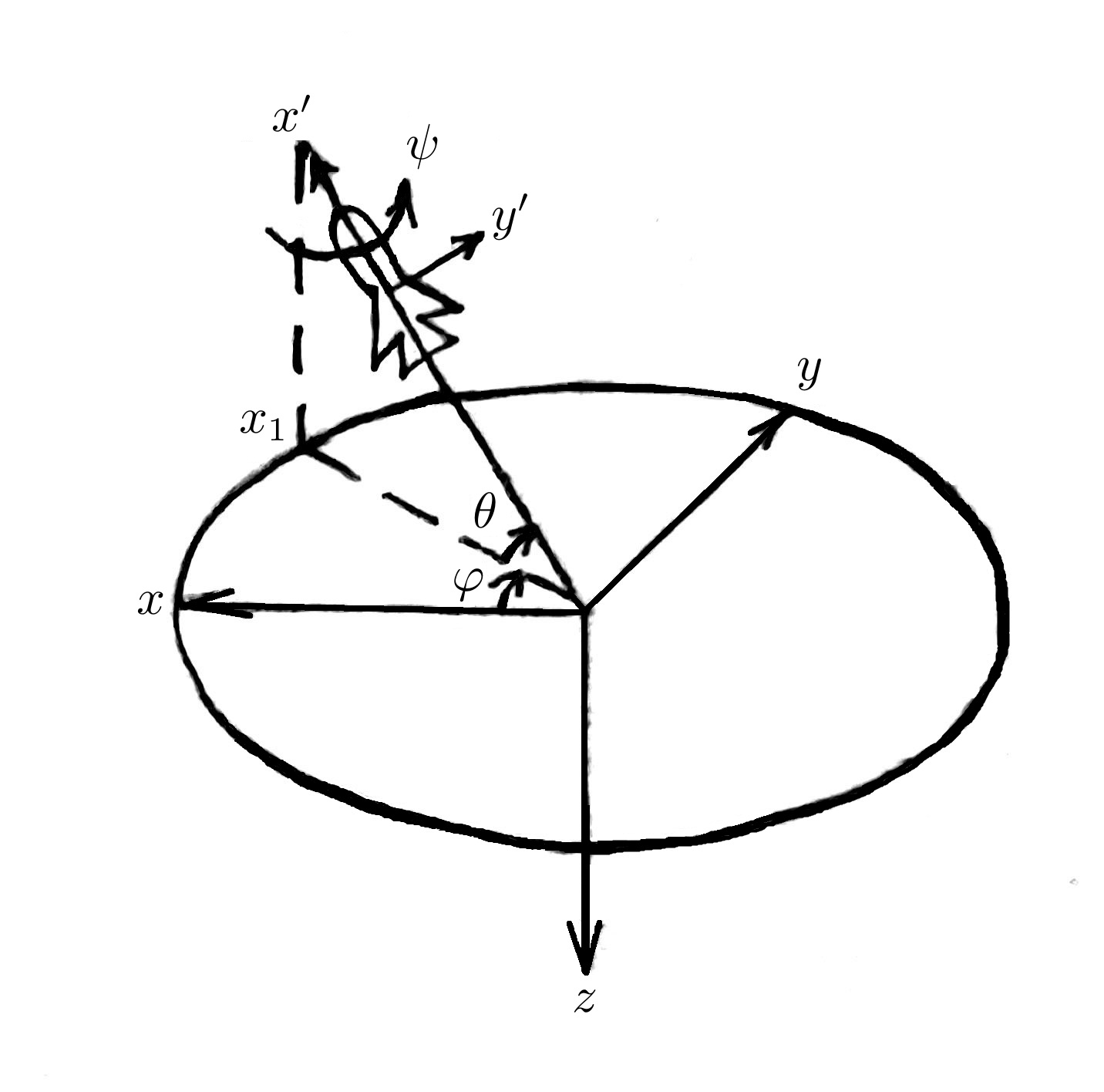}
  \caption{Tait-Bryan angles.}
  \label{fig:taitbryanangles}
\end{figure}
\subsubsection{SO(3) Group and its equivalence to rotations}\label{sec:sgaietr}
\begin{theorem}
All 3D orthogonal matrices with its determinant $+1$ constitute a group.
\end{theorem}
\begin{proof}
\
\item \textbf{Closure:} If orthogonal matrices $\mathbf{A},\ \mathbf{B}$ satisfy $\mathrm{det}\ \mathbf{A}=\mathrm{det}\ \mathbf{B}=1$, then $\mathrm{det}\ (\mathbf{AB})=\mathrm{det}\ \mathbf{A} \cdot \mathrm{det}\ \mathbf{B}=1$. Based on the closure of O(3), orthogonal matrices whose determinant equals to 1 is also closed.
\item \textbf{Other properties:} Following the ones of the O(3) Group.
\end{proof}
The group is named SO(3) Group, and we know the correctness of the theorem below.
\begin{theorem}
The inverse of a rotation matrix is also a rotation matrix.
\end{theorem}

Section~\ref{sec:rhpre} claims that every orthogonal matrix $\mathbf{M}$ with $\mathrm{\det}\ \mathbf{M} = 1$ may be a rotation matrix. We will show that it is indeed a rotation matrix below.

Following the thinking of Section~\ref{sec:euler} conversely, for any two length- and angle-preserving right-handed coordinate systems: if the $x'Oy'$ plane coincides with the $xOy$ plane, the transformation can be achieved by a planar rotation; if it do not, there always exists a line of intersection and an angle between the two planes. By comparison, the angle between the two planes is $\theta$, the angle between the $x'$-axis and the line of intersection is $\psi$, and the angle between the $x$-axis and the line of intersection is $\varphi$. Since both the $z$-axis and the $z'$-axis are defined by the right-hand rule, although this comparison only realizes the transformation of the $x$- and $y$-axes into $x'$- and $y'$-axes, the transformation of the $z$-axis into the $z'$-axis automatically follows. In addition, for a given configuration, every Euler angle is unique, unless $\theta=0$ or $\pi$.

Therefore, the transformation between any two length- and angle-preserving right-handed frames must be a physically realizable isometric transformation, i.e. a rotation; each rotation can be described by 3 Euler angles. Thus, as stated before, not only are all rotation matrices orthogonal, but also, every orthogonal transformation preserving the right-handedness is necessarily a rotation. That is to say, \textbf{SO(3) orthogonal matrices correspond one-to-one with rotations}.

Because every rotation matrix can be realized by integral of infinitesimal rotations due to the Euler-angles approach,
\begin{theorem}
$\mathrm{SO(3)}$ Group is a Lie Group.
\end{theorem}
\subsection{Reflection matrix}\label{sec:refm}
Reflection matrices are not closed as the determinant of the product of any two of them equals to +1, so they do not constitute a group.
\begin{definition}
One defines the parity transformation, written in matrix form
\begin{equation}
\mathbf{P}=
\begin{pmatrix}
-1 & 0 & 0 \\
0 & -1 & 0 \\
0 & 0 & -1
\end{pmatrix}
\end{equation}
\end{definition}

It is easy to examine that the determinant of the parity transformation is orthogonal and equals to $-1$, so this transformation is a special kind of reflection.

\begin{theorem}
The inverse of the parity transformation matrix is itself. $($Easy to prove$)$
\end{theorem}

\begin{theorem}
The parity transformation is equivalent to performing a rotational transformation before or after a simple reflection $\mathrm{(}$a reflection with a reflective surface$\mathrm{)}$.
\end{theorem}
\begin{proof}
Let the ``simple reflection" $\mathbf{P_1}$ be the $z$-axis reflection, i.e.
\begin{equation}\label{eqt:spr}
\mathbf{P_1}=
\begin{pmatrix}
+1 & 0 & 0 \\
0 & +1 & 0 \\
0 & 0 & -1
\end{pmatrix}
\end{equation}
Considering a rotation matrix whose axis is the $z$-axis
\[
\mathbf{R_1}=
\begin{pmatrix}
\cos\pi & -\sin\pi & 0 \\
\sin\pi & \cos\pi & 0 \\
0 & 0 & 1
\end{pmatrix}
=
\begin{pmatrix}
-1 & 0 & 0 \\
0 & -1 & 0 \\
0 & 0 & +1
\end{pmatrix}
\]
Then
\[
\mathbf{P}=\mathbf{P_1}\mathbf{R_1}=\mathbf{R_1}\mathbf{P_1}
\]
\end{proof}

\begin{theorem}
Each reflection is equivalent to performing a rotational transformation before or after a parity transformation or simple reflection.
\end{theorem}

\begin{proof}
Denote the matrix of reflection as $\mathbf{N}$, and the matrix of parity transformation or simple reflection as $\mathbf{\bar{P}}$, then $\mathbf{N\bar{P}}$ is an orthogonal matrix, which satisfies that $\mathrm{det}\ (\mathbf{N\bar{P}})=1$, i.e. $\mathbf{N\bar{P}}$ is a rotation matrix, as is denoted by $\mathbf{R}$. Likewise, $\mathbf{\bar{P}N}$ is a rotation matrix, denoted by $\mathbf{R'}$. So $\mathbf{N}=\mathbf{R\bar{P}}^{-1}=\mathbf{R\bar{P}}=\mathbf{\bar{P}}^{-1}\mathbf{R'}=\mathbf{\bar{P}R'}$.
\end{proof}
\section{Viewpoints and representations for rotations}\label{sec:view}
This section mainly discusses representations and viewpoints, which are very foundational but likely to be mistaken by starters.
\subsection{Notations of SO(3) elements}\label{sec:nota}
\begin{definition}
For a rotation by an angle $\phi$ about the axis $\boldsymbol{\hat{n}}$ $($one will see this is a general expression through Euler's Theorem in Section~\ref{sec:eig}$)$, the rotation matrix is denoted as
\[
\mathbf{R}(-\phi, \boldsymbol{\hat{n}})\footnote{This notation is used to keep in continuity with Liu (2019)\cite{liu2024}, which uses a different viewpoint from this article.}
\]
\end{definition}

Actually, the rotations about the axes are:
\begin{equation}\label{eqt:finrot}
\mathbf{R}(-\phi, \boldsymbol{\hat{x}})=
\begin{pmatrix}
1&0 &0 \\
0 & \cos\phi & -\sin\phi \\
0 & \sin\phi & \cos\phi
\end{pmatrix}
,\
\mathbf{R}(-\phi, \boldsymbol{\hat{y}})=
\begin{pmatrix}
\cos\phi&0 &\sin\phi\\
0&1&0 \\
-\sin\phi &0& \cos\phi
\end{pmatrix}
,\
\mathbf{R}(-\phi, \boldsymbol{\hat{z}})=
\begin{pmatrix}
\cos\phi& -\sin\phi &0 \\
\sin\phi & \cos\phi&0  \\
0&0&1
\end{pmatrix}
\end{equation}
\begin{definition}
For a rotation with Euler angles $(\varphi, \theta, \psi)$, the rotation matrix is also denoted as
\[
\mathbf{R}(-\theta,-\psi,-\varphi)\footnote{This notation is used to keep in continuity with L\"{u} (2025)\cite{lu2025}, which uses a different viewpoint from this article.}
\]
\end{definition}
\subsection{Relations between active viewpoint and passive viewpoint}\label{sec:relap}
Vectors are always represented in the inertial frame under active viewpoint, and in the body-attached frame under passive viewpoint.

Through space geometry, one can find the rotation of the frame about an axis $\boldsymbol{\hat{n}}$ by an angle $-\phi$ shares the same geometrical structure with that of vectors, about $\boldsymbol{\hat{n}}$ by $\phi$, like Section~\ref{sec:planar}. The rotation under passive viewpoint is
\[
\mathbf{R}(\phi, \boldsymbol{\hat{n}})
\]
It's natural to see that the first parameter is the angle and the other is the axis of coordinates' rotation. Actually, from plane geometry,
\begin{equation}\label{eqt:r-1r1org}
\mathbf{R}(\phi,\boldsymbol{\hat{n}})\mathbf{R}(-\phi,\boldsymbol{\hat{n}})=\mathbf{R}(-\phi,\boldsymbol{\hat{n}})\mathbf{R}(\phi,\boldsymbol{\hat{n}})=1
\end{equation}
Fixing the angle, the corresponding matrix under opposite viewpoints are inverse to each other.

Correspondingly, the rotations about the axes are:
\begin{equation}
\mathbf{R}(\phi, \boldsymbol{\hat{x}})=
\begin{pmatrix}
1&0 &0 \\
0 & \cos\phi & \sin\phi \\
0 & -\sin\phi & \cos\phi
\end{pmatrix}
,\
\mathbf{R}(\phi, \boldsymbol{\hat{y}})=
\begin{pmatrix}
\cos\phi&0 &-\sin\phi\\
0&1&0 \\
\sin\phi &0& \cos\phi
\end{pmatrix}
,\
\mathbf{R}(\phi,\boldsymbol{\hat{z}})=
\begin{pmatrix}
\cos\phi& \sin\phi &0 \\
-\sin\phi & \cos\phi&0  \\
0&0&1
\end{pmatrix}
\end{equation}

However, the relationship between viewpoints would be confusing and misleading if one imprudently extends the scope to successive rotations, as seen in Section~\ref{sec:mistk}.

Last but important, one can easily find that rotations under active viewpoint with those under passive viewpoint correspond one to one, i.e. each SO(3) element can be interpreted as a passive rotation. And because the passive viewpoint realize another representation to express any point in the 3D Euclidean space, any SO(3) matrix is an $\mathbb{R}^3\to\mathbb{R}^3$ linear mapping.
\subsection{Representation transformation}\label{sec:repiner1}
We can directly derive this statement by Theorem~\ref{eqt:similar}: If a transformation in the $A$ frame is represented by a matrix $\mathbf{F}$, while the rotation matrix of representation transformation for coordinates from $A$ to $B$ is $\mathbf{M}$, then that transformation in the $B$ frame is represented by 
\begin{equation}\label{eqt:rotsimiar}
\mathbf{B}=\mathbf{M}\mathbf{F}\mathbf{M}^\mathrm{T}
\end{equation}
Equation (\ref{eqt:rotsimiar}) will be applied in Part II to derive the two-ordered-tensor transformation. And in the cases that $\mathbf{F}$ is a rotation, the result can be applied to determine the effective axis and angle of SO(3) matrices in another representation.
\subsection{Successive rotations under active viewpoint}\label{sec:sruav}
We construct two frames, one inertial frame and one body-attached frame, like those in Section~\ref{sec:euler}.
\subsubsection{Rotations represented in the inertial frame}\label{sec:rotinerf}
If vectors experience successive rotations whose (axis and angle)-s are $(\boldsymbol{\hat{n}_1},\ \phi_1)$, $(\boldsymbol{\hat{n}_2},\ \phi_2)$, \dots, $(\boldsymbol{\hat{n}_n},\ \phi_n)$ in sequence represented in the inertial frame, then the total matrix of the active rotation is
\begin{equation}
\mathbf{M}=\mathbf{R}(-\phi_n,\boldsymbol{\hat{n}_n})\dots\mathbf{R}(-\phi_2,\boldsymbol{\hat{n}_2})\mathbf{R}(-\phi_1,\boldsymbol{\hat{n}_1})
\end{equation}
because the rotations act on vectors in series from the right to the left, i.e. the 1st acts on the original vector, and the 2nd acts on the vector after the 1st rotation, and so on.
\subsubsection{Rotations represented in the body-attached frame}\label{sec:avif}
Consider successive rotations whose (axis and angle)-s are $(\boldsymbol{\hat{n}_1},\ \phi_1)$, $(\boldsymbol{\hat{n}_2},\ \phi_2)$, \dots, $(\boldsymbol{\hat{n}_n},\ \phi_n)$ in sequence represented in the body-attached frame. First reduce the number of rotations to be 2. The matrix of the first rotation is $\mathbf{R}(-\phi_1,\boldsymbol{\hat{n}_1})$. For the second rotation, we apply (\ref{eqt:rotsimiar}). The matrix for the first active transformation is $\mathbf{R}(-\phi_1,\boldsymbol{\hat{n}_1})$, so the representation transformation matrix from the inertial frame to the body-attached is $\mathbf{R}^{-1}(-\phi_1,\boldsymbol{\hat{n}_1})$. Therefore, the representation matrix from the body-attached frame to the inertial is $\mathbf{R}(-\phi_1,\boldsymbol{\hat{n}_1})$. We can derive the second rotation represented in the inertial frame is $\mathbf{R}(-\phi_1,\boldsymbol{\hat{n}_1})\mathbf{R}(-\phi_2,\boldsymbol{\hat{n}_2})\mathbf{R}^{-1}(-\phi_1,\boldsymbol{\hat{n}_1})$, and the total matrix is $\mathbf{R}(-\phi_1,\boldsymbol{\hat{n}_1})\mathbf{R}(-\phi_2,\boldsymbol{\hat{n}_2})\mathbf{R}^{-1}(-\phi_1,\boldsymbol{\hat{n}_1})\cdot\mathbf{R}(-\phi_1,\boldsymbol{\hat{n}_1})=\mathbf{R}(-\phi_1,\boldsymbol{\hat{n}_1})\mathbf{R}(-\phi_2,\boldsymbol{\hat{n}_2})$.

With mathematical induction, one can know the composition of successive rotations is obtained by multiplying each rotation expressed in the body-attached frame, in the order of the rotation sequence from left to right, i.e.
\begin{equation}
\mathbf{M}=\mathbf{R}(-\phi_1,\boldsymbol{\hat{n}_1})\mathbf{R}(-\phi_2,\boldsymbol{\hat{n}_2})\dots\mathbf{R}(-\phi_n,\boldsymbol{\hat{n}_n})
\end{equation}
\subsection{Successive rotations under passive viewpoint}\label{sec:srupv}
\subsubsection{Matrices for basis transformation}\label{sec:subtl}
We have mentioned in Section~\ref{sec:defro} that passive rotations is caused by the change in the coordinate frame, i.e. the basis transformation. From linear algebra, the important result of transformation matrices for basis and coordinates is known: they are the inverses of each other, as is seen in (\ref{eqt:r-1r1org}). Consider a rotation by $\phi$ about the axis $\boldsymbol{\hat{n}}$. Due to the relationship between viewpoints, the corresponding matrix for coordinates' transformation under passive viewpoint is $\mathbf{R}(\phi,\boldsymbol{\hat{n}})$. Apply the relationship between basis and coordinate transformations, the rotation is expressed by 
\begin{equation}
\begin{pmatrix}
\boldsymbol{\hat{x}'}&\boldsymbol{\hat{y}'}&\boldsymbol{\hat{z}'}
\end{pmatrix}=
\begin{pmatrix}
\boldsymbol{\hat{x}}&\boldsymbol{\hat{y}}&\boldsymbol{\hat{z}}
\end{pmatrix}
\mathbf{R}^{-1}(\phi,\boldsymbol{\hat{n}})
\end{equation}
$\mathbf{R}^{-1}(\phi,\boldsymbol{\hat{n}})$ characterizes the relationship \textbf{only} between the bases, which depict the frames, before and after this rotation. 

Think about the circumstance that it is a single rotation, $\boldsymbol{\hat{n}}$ is the fixed axis seen in the frame before it. Correspondingly, if $\mathbf{R}^{-1}(\phi,\boldsymbol{\hat{n}})$ is one rotation in successive rotations, $\boldsymbol{\hat{n}}$ is the fixed axis seen in the frame just before this rotation, i.e. in the body-attached frame.

It's easy to see that the rotations act on basis in series from the left to the right, which the order of rotation matrices in the total rotation act on basis should follow.
\subsubsection{Rotations represented in the body-attached frame}
Recall the generalized situation mentioned in Section~\ref{sec:avif} and think about representation transformations through a routine similar with that in Section~\ref{sec:rotinerf}. The total matrix of the passive rotation should be $\mathbf{M}=\mathbf{R}(\phi_n,\boldsymbol{\hat{n}_n})\dots\mathbf{R}(\phi_2,\boldsymbol{\hat{n}_2})\mathbf{R}(\phi_1,\boldsymbol{\hat{n}_1})$
because the rotations act on vectors in series from the right to the left, in continuity with Section~\ref{sec:subtl}. And keep in mind the axes in the expression are represented in the body-attached frame, as is illustrated in Section~\ref{sec:subtl}.
\subsubsection{Rotations represented in the inertial frame}
First reduce the number of rotations to be 2, just like the thread in Section~\ref{sec:avif}. The matrix for the first active transformation is $\mathbf{R}(\phi_1,\boldsymbol{\hat{n}_1})$, which is the transformation matrix for representation. The second transformation represented in the body-attached frame is $\mathbf{R}(\phi_2,\boldsymbol{\hat{n}_2})$; therefore, it becomes $\mathbf{R}(\phi_1,\boldsymbol{\hat{n}_1})\mathbf{R}(\phi_2,\boldsymbol{\hat{n}_2})\mathbf{R}^{-1}(\phi_1,\boldsymbol{\hat{n}_1})$ in the inertial frame, and the total matrix goes to $\mathbf{R}(\phi_1,\boldsymbol{\hat{n}_1})\mathbf{R}(\phi_2,\boldsymbol{\hat{n}_2})$. So, in general case, the matrices should be multiplied from the left to the right, i.e. $\mathbf{M}=\mathbf{R}(\phi_1,\boldsymbol{\hat{n}_1})\mathbf{R}(\phi_2,\boldsymbol{\hat{n}_2})\dots\mathbf{R}(\phi_n,\boldsymbol{\hat{n}_n})$.
\subsection{Comparison between rotations under different viewpoints}\label{sec:comparison}
\subsubsection{A common mistake}\label{sec:mistk}
We can examine the correctness of the rules we have derived in Section~\ref{sec:sruav} and Section~\ref{sec:srupv}, e.g. for successive active rotations represented in the rest frame, the total matrix is $\mathbf{R}(-\phi_n,\boldsymbol{\hat{n}_n})\dots\mathbf{R}(-\phi_2,\boldsymbol{\hat{n}_2})\cdot\mathbf{R}(-\phi_1,\boldsymbol{\hat{n}_1})$; take its inverse and get $\mathbf{R}^{-1}(-\phi_1,\boldsymbol{\hat{n}_1})\mathbf{R}^{-1}(-\phi_2,\boldsymbol{\hat{n}_2})\dots\mathbf{R}^{-1}(-\phi_n,\boldsymbol{\hat{n}_n})=\mathbf{R}(\phi_1,\boldsymbol{\hat{n}_1})\cdot\mathbf{R}(\phi_2,\boldsymbol{\hat{n}_2})\dots\mathbf{R}(\phi_n,\boldsymbol{\hat{n}_n})$, which is just the corresponding ones under passive viewpoint.

Think about the equivalence between active and passive viewpoints in another way, e.g. see each passive rotation in a sequence represented in the body-attached frame under active viewpoint, and you will find that the total active rotation in the body-attached frame is $\mathbf{R}(-\phi_n,\boldsymbol{\hat{n}_n})\dots\mathbf{R}(-\phi_2,\boldsymbol{\hat{n}_2})\cdot\mathbf{R}(-\phi_1,\boldsymbol{\hat{n}_1})$, but actually that is in the inertial frame. In fact, the problem lies on the second rotation, because the $\boldsymbol{\hat{n}_2}$ is the axis viewing under the new basis, which characterizes the inertial frame. Since $\boldsymbol{\hat{n}_2}$ is an axis represented in the body-attached frame, there are only faults instead of paradoxes.
\subsubsection{Transformation of representations for one rotation in successive rotations}
A transformation of representations for rotation matrices is characterized by a similarity transformation, as is seen in Equation~\ref{eqt:rotsimiar}. For instance, for successive rotations whose (axis and angle)-s are $(\boldsymbol{\hat{n}_1},\ \phi_1)$, $(\boldsymbol{\hat{n}_2},\ \phi_2)$, \dots, $(\boldsymbol{\hat{n}_n},\ \phi_n)$ in sequence represented in the inertial frame under active viewpoint, the $i$-th rotation is represented by 
\begin{equation}\label{eqt:intfr}
\mathbf{R}(-\phi_i,\boldsymbol{\hat{n}_i})\ \to\ \mathbf{R}^{-1}(-\phi_1,\boldsymbol{\hat{n}_1})\dots\mathbf{R}^{-1}(-\phi_{i-1},\boldsymbol{\hat{n}_{i-1}})\mathbf{R}(-\phi_i,\boldsymbol{\hat{n}_i})\mathbf{R}(-\phi_{i-1},\boldsymbol{\hat{n}_{i-1}})\dots\mathbf{R}(-\phi_1,\boldsymbol{\hat{n}_1})
\end{equation}
in the body-attached frame, because the representation transformation matrix from the inertial frame to the body-attached is $\mathbf{R}^{-1}(-\phi_1,\boldsymbol{\hat{n}_1})\mathbf{R}^{-1}(-\phi_2,\boldsymbol{\hat{n}_2})\dots\mathbf{R}^{-1}(-\phi_{i-1},\boldsymbol{\hat{n}_{i-1}})$.

Readers can get the transformation relationships for another 3 scenarios if they are interested in those.
\subsection{Physical meanings of rotational parameters}
\subsubsection{Influences of a rotation on other rotations}\label{sec:influ}
For consideration of articular space, we only consider the scenario that the axes and angles are given in the body-attached representation, in which the expression method using Euler angles is. The axes are body-attached, i.e. they are vectors that need to experience active rotations. In conclusion, previous rotations can associate later rotations, but the reverse cannot. Specifically, any vector first rotates about $\boldsymbol{\hat{n}_1}$ by $\phi_1$, then about $\mathbf{R}(-\phi_1,\boldsymbol{\hat{n}_1})\boldsymbol{\hat{n}_2}$ by $\phi_2$, then about $\mathbf{R}(-\phi_2,\boldsymbol{\hat{n}_2})\mathbf{R}(-\phi_1,\boldsymbol{\hat{n}_1})\boldsymbol{\hat{n}_3}$ by $\phi_3$, and so on.
\subsubsection{Changing one angle in successive rotations}
Consider that the axes of rotations are fixed under a view from the body-attached frame, while the angles are changeable, seen as parameters of the composition of rotations. The total matrix should be $\mathbf{M}=\mathbf{R}(-\phi_1,\boldsymbol{\hat{n}_1})\mathbf{R}(-\phi_2,\boldsymbol{\hat{n}_2})\dots\mathbf{R}(-\phi_n,\boldsymbol{\hat{n}_n})$.
\begin{theorem}
The change in the angle of a constituent rotation equals to a rotation by the change about the axis corresponding to that parameter.
\end{theorem}
First consider the case the total rotation is composed by only two rotations following the notations where the change occurs on $\phi_2$. Try to derive the deformation of total transformation.
\begin{equation}
\mathbf{R}(-\phi_1,\boldsymbol{\hat{n}_1})\mathbf{R}(-\phi_2',\boldsymbol{\hat{n}_2})=\mathbf{R}(-\phi_1,\boldsymbol{\hat{n}_1})\mathbf{R}(-(\phi_2'-\phi_2),\boldsymbol{\hat{n}_2})\mathbf{R}^{-1}(-\phi_1,\boldsymbol{\hat{n}_1})\mathbf{R}(-\phi_1,\boldsymbol{\hat{n}_1})\mathbf{R}(-\phi_2,\boldsymbol{\hat{n}_2})
\end{equation}
In fact, $\mathbf{R}(-\phi_1,\boldsymbol{\hat{n}_1})\mathbf{R}(-(\phi_2'-\phi_2),\boldsymbol{\hat{n}_2})\mathbf{R}^{-1}(-\phi_1,\boldsymbol{\hat{n}_1})$ acting on the total rotation is that about the axis $\boldsymbol{\hat{n}_2}$ by $\phi_2'-\phi_2$ (viewed in inertial frame). That is the result, without surprise. Because, the representation transformation matrix from body-attached frame to the inertial is $\mathbf{R}(-\phi_1,\boldsymbol{\hat{n}_1})$, with which we can derive the expression. Then we generally give the proof below.
\begin{proof}
For the general case, i.e. the total transformation $\mathbf{M}$ equals to $\mathbf{R}(-\phi_1,\boldsymbol{\hat{n}_1})\mathbf{R}(-\phi_2,\boldsymbol{\hat{n}_2})\cdot$ $\dots\mathbf{R}(-\phi_i,\boldsymbol{\hat{n}_i})\dots\mathbf{R}(-\phi_n,\boldsymbol{\hat{n}_n})$ where the change occurs on $\phi_i$, the new total matrix is
\begin{align}
\mathbf{M'}&=\mathbf{R}(-\phi_1,\boldsymbol{\hat{n}_1})\mathbf{R}(-\phi_2,\boldsymbol{\hat{n}_2})\dots\mathbf{R}(-\phi_i',\boldsymbol{\hat{n}_i})\dots\mathbf{R}(-\phi_n,\boldsymbol{\hat{n}_n})\nonumber\\
&=\mathbf{R}(-\phi_1,\boldsymbol{\hat{n}_1})\dots\mathbf{R}(-\phi_{i-1},\boldsymbol{\hat{n}_{i-1}})\mathbf{R}(-(\phi_i'-\phi_i),\boldsymbol{\hat{n}_i})\mathbf{R}^{-1}(-\phi_{i-1},\boldsymbol{\hat{n}_{i-1}})\dots\mathbf{R}^{-1}(-\phi_1,\boldsymbol{\hat{n}_1})\cdot\nonumber\\
&\ \ \ \ \mathbf{R}(-\phi_1,\boldsymbol{\hat{n}_1})\dots\mathbf{R}(-\phi_1,\boldsymbol{\hat{n}_n})\nonumber\\
&=\mathbf{R}(-\phi_1,\boldsymbol{\hat{n}_1})\dots\mathbf{R}(-\phi_{i-1},\boldsymbol{\hat{n}_{i-1}})\mathbf{R}(-(\phi_i'-\phi_i),\boldsymbol{\hat{n}_i})\mathbf{R}^{-1}(-\phi_{i-1},\boldsymbol{\hat{n}_{i-1}})\dots\mathbf{R}^{-1}(-\phi_1,\boldsymbol{\hat{n}_1})\mathbf{M}
\end{align}
Because the representation transformation matrix for coordinates from the body-attached frame after the earliest $i-1$ rotations to the inertial frame is $\mathbf{R}(-\phi_1,\boldsymbol{\hat{n}_1})\mathbf{R}(-\phi_2,\boldsymbol{\hat{n}_2})\dots\mathbf{R}(-\phi_{i-1},\boldsymbol{\hat{n}_{i-1}})$ (derived from a change in the viewpoint and an inverse taken), the factors before $\mathbf{M}$ is a rotation by $\phi_i'-\phi_i$ about the axis $\boldsymbol{\hat{n}_i}$ in the representation of the inertial frame, i.e. the change of total rotation after the parameter change in $\phi_i$ can be expressed by a rotation about the axis corresponding to $\phi_i$ by $\phi_i'-\phi_i$.
\end{proof}
\section{Rotations expressed by parameters}\label{sec:repro}
From now on, we express active rotations in the inertial frame representation under active viewpoints unless specified, avoiding use equivalent but nonintuitive viewpoints and representations. To prepare for the relationship of different expressions, we will first analyze some global properties of SO(3) Group\textemdash the relationships between different SO(3) elements.
\subsection{Some global properties of SO(3) Group}\label{sec:algso3}
First, $\mathbf{R}(-\phi,\boldsymbol{\hat{n}})$ can be seen as a function of $\phi$ with period $2\pi$, i.e.
\begin{equation}\label{eqt:period}
\mathbf{R}(-(\phi+2k\pi),\boldsymbol{\hat{n}})=\mathbf{R}(-\phi,\boldsymbol{\hat{n}})\ \ \ (k\in\mathbb{Z})
\end{equation}

Second, with Rodrigues' rotation formula (\ref{eqt:rodrigues}), it is obvious that
\begin{equation}\label{eqt:glb1}
\mathbf{R}(\phi,-\boldsymbol{\hat{n}})=\mathbf{R}(-\phi,\boldsymbol{\hat{n}})
\end{equation}
i.e. a rotation about an axis by an angle is equivalent to about and by their opposites. With the two formulae, one can choose the convention that
\begin{equation}\label{eqt:convention}
\phi\in[0,\pi)
\end{equation}
This convention is \textbf{very important} because we can obtain a unique expression of the axis and angle with it, and will be taken in Sections~\ref{sec:axang} and~\ref{sec:eulertoaxang}.

Moreover, through an approach of plane geometry, if two rotations are about the same axis, then they commute with each other, written in the formula
\begin{equation}\label{eqt:glb2}
\mathbf{R}(-\phi_2,\boldsymbol{\hat{n}})\mathbf{R}(-\phi_1,\boldsymbol{\hat{n}})=\mathbf{R}(-(\phi_1+\phi_2),\boldsymbol{\hat{n}})=\mathbf{R}(-\phi_1,\boldsymbol{\hat{n}})\mathbf{R}(-\phi_2,\boldsymbol{\hat{n}})
\end{equation}
This equation can be concluded by a weaker theorem.
\begin{theorem}
$\mathbf{R}(-\phi,\boldsymbol{\hat{n}})$ with $\boldsymbol{\hat{n}}$ fixed constitute an Abelian group $\mathrm{(}$a group whose elements commute with each other$\mathrm{)}$.
\end{theorem}
A corollary is (\ref{eqt:r-1r1org}), i.e. 
\begin{equation}\label{eqt:glb3}
\mathbf{R}^{-1}(-\phi,\boldsymbol{\hat{n}})=\mathbf{R}(\phi,\boldsymbol{\hat{n}})
\end{equation}

Finally, we will get
\begin{equation}\label{eqt:lnindep}
\mathbf{R}(-\phi_1,\boldsymbol{\hat{n}_1})\neq\mathbf{R}(-\phi_2,\boldsymbol{\hat{n}_2})\ \ \ \ \ \mathrm{if\ }\boldsymbol{\hat{n}_1}\mathrm{\ and\ }\boldsymbol{\hat{n}_2}\mathrm{\ is\ linearly\ independent}
\end{equation}
in Section~\ref{sec:eig}.

Operate a simple variable substitution $\phi\to-\phi$, and we can precisely illustrate these laws under passive viewpoints.
\subsection{From Euler angles to matrix elements}\label{sec:eulertomat}
\subsubsection{Rotation matrices expressed by Euler angles}\label{sec:eulertomat1}
Express the matrices of Euler angles in the body-attached representation, i.e. the rotation about $\boldsymbol{\hat{x}_1}$ and $\boldsymbol{\hat{z}_2}$ are expressed as $\mathbf{R}(-\theta,\boldsymbol{\hat{x}})$ and $\mathbf{R}(-\psi,\boldsymbol{\hat{z}})$ instead of $\mathbf{R}(-\theta,\boldsymbol{\hat{x}_1})$ and $\mathbf{R}(-\psi,\boldsymbol{\hat{z}_2})$. According to the result mentioned in Section~\ref{sec:avif}, the total matrix of the composite rotation by those of Euler angles is
\begin{align}
\mathbf{R}(-\theta,-\psi,-\varphi)&= \mathbf{R}(-\varphi,\boldsymbol{\hat{z}})\mathbf{R}(-\theta,\boldsymbol{\hat{x}})\mathbf{R}(-\psi,\boldsymbol{\hat{z}})\label{eqt:3euler}\\
&= \begin{pmatrix}
\cos\varphi& -\sin\varphi &0 \\
\sin\varphi & \cos\varphi&0  \\
0&0&1
\end{pmatrix}
\begin{pmatrix}
1&0 &0 \\
0 & \cos\theta & -\sin\theta \\
0 & \sin\theta & \cos\theta
\end{pmatrix}
\begin{pmatrix}
\cos\psi& -\sin\psi &0 \\
\sin\psi & \cos\psi&0  \\
0&0&1
\end{pmatrix}\nonumber\\
&= \begin{pmatrix}
\cos\varphi\cos\psi-\cos\theta\sin\psi\sin\varphi& -\cos\varphi\sin\psi-\cos\theta\cos\psi\sin\varphi& \sin\varphi\sin\theta \\
\sin\varphi\cos\psi+\cos\theta\sin\psi\cos\varphi& -\sin\varphi\sin\psi+\cos\theta\cos\psi\cos\varphi& -\cos\varphi\sin\theta  \\
\sin\theta\sin\psi& \sin\theta\cos\psi& \cos\theta
\end{pmatrix}
\end{align}
\subsubsection{Further discussion on Euler-angle matrices}
The inverse transformation can be expressed as
\begin{align}
\mathbf{R}^{-1}(-\theta,-\psi,-\varphi)&=\mathbf{R}^{-1}(-\psi,\boldsymbol{\hat{z}})\mathbf{R}^{-1}(-\theta,\boldsymbol{\hat{x}})\mathbf{R}^{-1}(-\varphi,\boldsymbol{\hat{z}})=\mathbf{R}(\psi,\boldsymbol{\hat{z}})\mathbf{R}(\theta,\boldsymbol{\hat{x}})\mathbf{R}(\varphi,\boldsymbol{\hat{z}})=\mathbf{R}(\theta,\varphi,\psi)\nonumber\\
&=\begin{pmatrix}
\cos\psi\cos\varphi-\cos\theta\sin\varphi\sin\psi& \cos\psi\sin\varphi+\cos\theta\cos\varphi\sin\psi& \sin\psi\sin\theta \\
-\sin\psi\cos\varphi-\cos\theta\sin\varphi\cos\psi& -\sin\psi\sin\varphi+\cos\theta\cos\varphi\cos\psi& \cos\psi\sin\theta  \\
\sin\theta\sin\varphi& -\sin\theta\cos\varphi& \cos\theta
\end{pmatrix}
\end{align}
This formula under passive viewpoint also appears in Goldstein (2002), Liu (2019) and L\"{u} (2025) in those parts of Euler angles. 

We can see that for a rotation whose Euler angles are ($\varphi$, $\theta$, $\psi$), the Euler angles of its inverse are ($-\psi$, $-\theta$, $-\varphi$), not ($-\varphi$, $-\theta$, $-\psi$). This is just why the sequence of the matrices in successive rotations is reversed under opposite viewpoints.

Moreover, it is also interesting to view the rotation composed by Euler-angle rotations differently. The expression (\ref{eqt:3euler}) can be also interpreted as rotating by $\psi$ about $\boldsymbol{\hat{z}}$ at first, by $\theta$ about $\boldsymbol{\hat{x}}$ next, and by $\varphi$ about $\boldsymbol{\hat{z}}$ in the end, and the scenario is not the same with the original interpretation in general. 
\subsection{From matrix elements to Euler angles}\label{sec:mattoeuler}
In the matrix of Euler angles, whose elements denoted by $m_{ij}$
\[
\begin{cases}
\cos\theta=m_{33}\\
\sin\varphi\sin\theta=m_{13}\\
-\cos\varphi\sin\theta=m_{23}\\
\sin\theta\sin\psi=m_{31}\\
\sin\theta\cos\psi=m_{32}
\end{cases}
\]
which is equivalent to
\begin{equation}\label{eqt:eulerlaw}
\begin{cases}
\theta=\arccos m_{33}\\
\tan\varphi=-\frac{m_{13}}{m_{23}}\ \ \ \mathrm{with}\ \sin\varphi\ \mathrm{and}\ m_{13}\mathrm{\ sharing\ the\ same\ sign}\\
\tan\psi=\frac{m_{31}}{m_{32}}\ \ \ \mathrm{with}\ \sin\psi\ \mathrm{and}\ m_{31}\mathrm{\ sharing\ the\ same\ sign}\\
\end{cases}
\end{equation}

\noindent 1) If $m_{13},m_{23}$ are not both equal to zero, and $m_{31},m_{32}$ are not both equal to zero,

It can be solved from Euler matrix equations that
\begin{equation}\label{eqt:eulerlaw1}
\begin{cases}
\theta=\arccos m_{33}\\
\varphi=
\begin{cases}
\pi-\arctan\frac{m_{13}}{m_{23}}\ (m_{23}>0)\\
-\arctan\frac{m_{13}}{m_{23}}\ (m_{13}>0,m_{23}<0)\\
2\pi-\arctan\frac{m_{13}}{m_{23}}\ (m_{13}<0,m_{23}<0)\\
\pi\ (m_{13}=0,m_{23}>0)\\
0\ (m_{13}=0,m_{23}<0)\\
\frac{\pi}{2}\ (m_{13}>0,m_{23}=0)\\
\frac{3\pi}{2}\ (m_{13}<0,m_{23}=0)
\end{cases}\\
\psi=
\begin{cases}
\arctan\frac{m_{31}}{m_{32}}\ (m_{31}>0,m_{32}>0)\\
\pi+\arctan\frac{m_{31}}{m_{32}}\ (m_{32}<0)\\
2\pi+\arctan\frac{m_{31}}{m_{32}}\ (m_{31}<0,m_{32}>0)\\
0\ (m_{31}=0,m_{32}>0)\\
\pi\ (m_{31}=0,m_{32}<0)\\
\frac{\pi}{2}\ (m_{31}>0,m_{32}=0)\\
\frac{3\pi}{2}\ (m_{31}<0,m_{32}=0)
\end{cases}
\end{cases}
\end{equation}

\noindent 2) Otherwise,
$\theta =0$ or $\theta=\pi$

If $m_{33}=1$, then $\theta =0$, and the matrix goes to
\[
\begin{pmatrix}
\cos(\psi+\varphi) & -\sin(\psi+\varphi) & 0 \\
\sin(\psi+\varphi) & \cos(\psi+\varphi) & 0 \\
0&0&1
\end{pmatrix}
\]
Since $\psi+\varphi$ appears as a whole item, $\varphi$ and $\psi$ cannot be determined uniquely\textemdash the solutions are equivalent.

\noindent a. $m_{11}>0$ and $m_{21}>0$, $\psi+\varphi=\arcsin m_{21}=\arccos m_{11}$.

\noindent b. $m_{11}<0$ and $m_{21}>0$, $\psi+\varphi=\pi-\arcsin m_{21}=\arccos m_{11}$.

\noindent c. $m_{11}<0$ and $m_{21}<0$, $\psi+\varphi=\pi-\arcsin m_{21}=2\pi-\arccos m_{11}$.

\noindent d. $m_{11}>0$ and $m_{21}<0$, $\psi+\varphi=2\pi+\arcsin m_{21}=2\pi-\arccos m_{11}$.

\noindent e. $m_{11}=\pm 1$ and $m_{21}=0$, $\psi+\varphi=0(\pi)$.

\noindent f. $m_{11}=0$ and $m_{21}=\pm 1$, $\psi+\varphi=\frac{\pi}{2}(\frac{3\pi}{2})$.

If $m_{33}=-1$, then $\theta =\pi$, and the matrix goes to
\[
\begin{pmatrix}
\cos(\psi-\varphi) & -\sin(\psi-\varphi) & 0 \\
-\sin(\psi-\varphi) & -\cos(\psi-\varphi) & 0 \\
0&0&1
\end{pmatrix}
\]
$\psi-\varphi$ appears as a whole item similarly.

\noindent a. $m_{11}>0$ and $m_{21}>0$, $\psi-\varphi=2\pi-\arcsin m_{21}=2\pi-\arccos m_{11}$.

\noindent b. $m_{11}<0$ and $m_{21}>0$, $\psi-\varphi=\pi+\arcsin m_{21}=2\pi-\arccos m_{11}$.

\noindent c. $m_{11}<0$ and $m_{21}<0$, $\psi-\varphi=\pi+\arcsin m_{21}=\arccos m_{11}$.

\noindent d. $m_{11}>0$ and $m_{21}<0$, $\psi-\varphi=-\arcsin m_{21}=\arccos m_{11}$.

\noindent e. $m_{11}=\pm 1$ and $m_{21}=0$, $\psi-\varphi=0(\pi)$.

\noindent f. $m_{11}=0$ and $m_{21}=\pm 1$, $\psi-\varphi=\frac{3\pi}{2}(\frac{\pi}{2})$.
\subsection{From the axis and angle of rotation to matrix elements}\label{sec:axangtomat}
Let the rotation axis is $\boldsymbol{\hat{n}}=
\begin{pmatrix}
n_1&n_2&n_3
\end{pmatrix}
^\mathrm{T}$, one can write Rodrigues' rotation formula in the matrix form

\begin{align*}
\begin{pmatrix}
x' \\ y' \\ z'
\end{pmatrix}
&=\cos\phi
\begin{pmatrix}
x \\ y \\ z
\end{pmatrix}
+(1-\cos\phi)(n_1x+n_2y+n_3z)
\begin{pmatrix}
n_1 \\ n_2 \\n_3
\end{pmatrix}
+\sin\phi
\begin{pmatrix}
n_2z-n_3y\\ n_3x-n_1z \\ n_1y-n_2x
\end{pmatrix}\\
&=
\begin{pmatrix}
\cos\phi+{n_1}^2(1-\cos\phi)&n_1n_2(1-\cos\phi)-n_3\sin\phi&n_3n_1(1-\cos\phi)+n_2\sin\phi\\
n_1n_2(1-\cos\phi)+n_3\sin\phi&\cos\phi+{n_2}^2(1-\cos\phi)&n_2n_3(1-\cos\phi)-n_1\sin\phi\\
n_3n_1(1-\cos\phi)-n_2\sin\phi&n_2n_3(1-\cos\phi)+n_1\sin\phi&\cos\phi+{n_3}^2(1-\cos\phi)
\end{pmatrix}
\begin{pmatrix}
x\\y\\z
\end{pmatrix}
\end{align*}
So
\begin{equation}
\mathbf{R}(-\phi,\boldsymbol{\hat{n}})=
\begin{pmatrix}
\cos\phi+{n_1}^2(1-\cos\phi)&n_1n_2(1-\cos\phi)-n_3\sin\phi&n_3n_1(1-\cos\phi)+n_2\sin\phi\\
n_1n_2(1-\cos\phi)+n_3\sin\phi&\cos\phi+{n_2}^2(1-\cos\phi)&n_2n_3(1-\cos\phi)-n_1\sin\phi\\
n_3n_1(1-\cos\phi)-n_2\sin\phi&n_2n_3(1-\cos\phi)+n_1\sin\phi&\cos\phi+{n_3}^2(1-\cos\phi)
\end{pmatrix}
\end{equation}
Define $\mathbf{N}=
\begin{pmatrix}
0&-n_3&n_2\\n_3&0&n_1\\-n_2&n_1&0
\end{pmatrix}$ because for any vector $\boldsymbol{V}$, $\mathbf{N}\boldsymbol{V}=\boldsymbol{\hat{n}}\times\boldsymbol{V}$.
Readers can prove\cite{heard2008}
\begin{equation}
\mathbf{R}(-\phi,\boldsymbol{\hat{n}})=\mathbf{I}+\sin\phi\ \mathbf{N}+(1-\cos\phi)\ \mathbf{N}^2
\end{equation}

The rotation matrix under passive viewpoint is
\begin{align*}
\mathbf{R}(\phi,\boldsymbol{\hat{n}})&=
\begin{pmatrix}
\cos\phi+{n_1}^2(1-\cos\phi)&n_1n_2(1-\cos\phi)+n_3\sin\phi&n_3n_1(1-\cos\phi)-n_2\sin\phi\\
n_1n_2(1-\cos\phi)-n_3\sin\phi&\cos\phi+{n_2}^2(1-\cos\phi)&n_2n_3(1-\cos\phi)+n_1\sin\phi\\
n_3n_1(1-\cos\phi)+n_2\sin\phi&n_2n_3(1-\cos\phi)-n_1\sin\phi&\cos\phi+{n_3}^2(1-\cos\phi)
\end{pmatrix}\\
&=\mathbf{I}-\sin\phi\ \mathbf{N}+(1-\cos\phi)\ \mathbf{N}^2
\end{align*}
\subsection{From matrix elements to the axis and angle of rotation}\label{sec:mattoaxang}
We will see here in Section~\ref{sec:eig} the theorem we have repeatedly mentioned: \textbf{any rotation is about a single axis}.
\subsubsection{Eigenvalues and eigenvectors of orthogonal matrices}\label{sec:eig}
\begin{theorem}
$\mathbf{\ (}\textbf{Euler's Theorem}\mathbf{)}$ Any rotation matrix must have an eigenvalue $+1$, whose eigenvector corresponds to the axis of rotation.
\end{theorem}
\begin{proof}
Denote the rotation matrix $\mathbf{M}=
\begin{pmatrix}
m_{11} & m_{12} & m_{13} \\
m_{21} & m_{22} & m_{23} \\
m_{31} & m_{32} & m_{33}
\end{pmatrix}$
, then the characteristic equation is
\begin{align}
&\begin{vmatrix}
m_{11}-\lambda & m_{12} & m_{13} \\
m_{21} & m_{22}-\lambda & m_{23} \\
m_{31} & m_{32} & m_{33}-\lambda
\end{vmatrix} = 0 \nonumber\\
\Leftrightarrow\;& (m_{11}-\lambda)(m_{22}-\lambda)(m_{33}-\lambda)
+ m_{21}m_{32}m_{13} + m_{31}m_{12}m_{23} \nonumber\\
&\quad - m_{13}m_{31}(m_{22}-\lambda)
- m_{23}m_{32}(m_{11}-\lambda)
- m_{12}m_{21}(m_{33}-\lambda) = 0 \nonumber\\
\Leftrightarrow\;& -\lambda^3 + (m_{11}+m_{22}+m_{33})\lambda^2 \nonumber\\
&\quad -\big[(m_{11}m_{33}-m_{13}m_{31})
+ (m_{11}m_{22}-m_{12}m_{21})
+ (m_{22}m_{33}-m_{23}m_{32})\big]\lambda + \mathrm{det}\ \mathbf{M} = 0 \nonumber\\
\Leftrightarrow\;& -\lambda^3 + \lambda^2\,\mathrm{tr}\ \mathbf{M}
- \lambda\,\mathrm{tr}\ \mathbf{M} + 1 = 0 \nonumber\\
\Leftrightarrow\;& (\lambda-1)\big[\lambda^2 + (1-\mathrm{tr}\ \mathbf{M})\lambda + 1\big] = 0
\end{align}

Thus, $\mathbf{M}$ has an eigenvalue $+1$.
So, $\exists \boldsymbol{r}\ne\mathbf{0},\ \mathbf{M}\boldsymbol{r}=\boldsymbol{r}$. Because the vector $\boldsymbol{r}$ keeps unchanged after the transformation, the unit vector of this direction or its opposite must be the axis.\cite{liu2024}
\end{proof}

On the other hand, only this direction or its opposite can be the axis; thus, we can even know that rotations with linearly independent axes can't be the same, i.e. Equation (\ref{eqt:lnindep})
\[
\mathbf{R}(-\phi_1,\boldsymbol{\hat{n}_1})\neq\mathbf{R}(-\phi_2,\boldsymbol{\hat{n}_2})\ \ \ \ \ \mathrm{if\ }\boldsymbol{\hat{n}_1}\mathrm{\ and\ }\boldsymbol{\hat{n}_2}\mathrm{\ is\ linearly\ independent}
\]

In order to determine the rotation angle in principle, first prove a lemma (Theorem~\ref{eqt:lemma}).
\begin{theorem}\label{eqt:lemma}
A real orthogonal matrix is still inner-product-preserving for complex vectors.
\end{theorem}
\begin{proof}
$\forall \boldsymbol{r_1},\boldsymbol{r_2}$,
\[
(\mathbf{M}\boldsymbol{r_1})^\dagger(\mathbf{M}\boldsymbol{r_2})=\boldsymbol{r_1}^\dagger\mathbf{M}^\mathrm{T}\mathbf{M}\boldsymbol{r_2}=\boldsymbol{r_1}^\dagger\mathbf{M}^{-1}\mathbf{M}\boldsymbol{r_2}=\boldsymbol{r_1}^\dagger\boldsymbol{r_2}
\]
\end{proof}
\begin{theorem}
Any rotation matrix must have two conjugate eigenvalues $\mathrm{e}^{\pm\mathrm{i}\Theta}$, where $\Theta$ is the rotation angle. 
\end{theorem}
\begin{proof}

If $\lambda$ is an eigenvalue of a rotation matrix $\mathbf{M}$, i.e. $\exists\boldsymbol{r}$, $\mathbf{M}\boldsymbol{r}=\lambda\boldsymbol{r}$, then $||\mathbf{M}\boldsymbol{r}||=|\lambda|^2||\boldsymbol{r}||$, i.e. $|\lambda|^2=1$, so $\lambda=\mathrm{e}^{\mathrm{i}\Theta}$ for some $\Theta$. Substitute it into the characteristic equation and get $\mathrm{tr}\ \mathbf{M}=1+2\cos\Theta$.
Because $\mathrm{tr}\ \mathbf{M}$ is a real number, a $\lambda$ is a solution as long as its conjugate is a solution. Then one can write
\[
\lambda=\mathrm{e}^{\pm\mathrm{i}\Theta}
\]
with
\begin{equation}\label{eqt:ctnit}
\mathrm{tr}\ \mathbf{M}=1+2\cos\Theta
\end{equation}

Then we will prove that $\Theta$ is the angle of rotation. Make a passive rotation to let the axis of rotation to be the $z_1$-axis of the new coordinates, with the coordinate transformation matrix $\mathbf{M}$ (the transformation matrix back to the original frame is $\mathbf{M}^{-1}$), where $\mathbf{M}^\mathrm{T}=\mathbf{M}^{-1}$, so the rotation in the \textbf{original} representation is
\begin{equation}
\mathbf{A}=\mathbf{M}^\mathrm{T}\mathbf{R}(-\phi,\boldsymbol{\hat{z}})\mathbf{M}
\end{equation}
where $\phi$ is the actual angle of rotation.
In conclusion, $\mathbf{A}$ is similar to $\mathbf{R}(-\phi,\boldsymbol{\hat{z}})$, so their eigenvalues are the same. Solve the eigenvalues of $\mathbf{R}(-\phi,\boldsymbol{\hat{z}})$
\[
\begin{vmatrix}
\cos\phi-\lambda & -\sin\phi & 0 \\
\sin\phi & \cos\phi-\lambda & 0 \\
0 & 0 & 1-\lambda
\end{vmatrix} = 0
\]
The original equation is equivalent to $(1-\lambda)(\lambda^2-2\lambda\cos\phi+1)=0$, then $\lambda=1$ or $\mathrm{e}^{\pm\mathrm{i}\phi}$. So $\phi=\pm\Theta$. One can let $\phi=\Theta$.
\end{proof}

Likewise, there are some corresponding properties for reflections.
\begin{theorem}
Any reflection matrix must have three eigenvalues $-1$, $\mathrm{e}^{\pm\mathrm{i}\Theta}$ where $\Theta\in\mathbb{R}$.
\end{theorem}

\begin{proof}
Conducting a similar calculation, for reflection matrices,
\[-\lambda^3 + \lambda^2\,\mathrm{tr}\ \mathbf{M} + \lambda\,\mathrm{tr}\ \mathbf{M} - 1 = 0\]
which is equal to
\[
(\lambda+1)\big[\lambda^2 - (1+\mathrm{tr}\ \mathbf{M})\lambda + 1\big] = 0
\]
Thus, $+1$ is an eigenvalue. Speaking for the other two roots, $|\lambda|^2=1$ similarly, and there is also an $\lambda\to\lambda^*$ symmetry for the equation of them. Thus, $\mathrm{e}^{\pm\mathrm{i}\Theta}$ are the other two.
\end{proof}
So, for any reflection matrices $\mathbf{M}$, 
\begin{equation}
\mathrm{tr}\ \mathbf{M}=-1+2\cos\Theta
\end{equation}
which is immensely useful in Section~\ref{sec:indep}.

Note that not every reflection has a reflective surface. Actually, the trace of the transformation with reflective surface $xOy$ (\ref{eqt:spr}) is $+1$. With an arbitrary orthogonal diagonalization, one gets every simple reflection. But for each rotation $\mathbf{R}$,
\begin{equation}
\mathrm{tr}\ (\mathbf{R}\mathbf{P_1}\mathbf{R}^{-1})=\mathrm{tr}\ (\mathbf{P_1}\mathbf{R}\mathbf{R}^{-1})=\mathrm{tr}\ \mathbf{P_1}=+1
\end{equation}
and this expression can't include every reflection.
\subsubsection{Solution to the axis and angle}\label{sec:axang}
The next step is to derive the axis and the angle generally. Remember the conclusion in Section~\ref{sec:algso3} that the solution is unique. From Section~\ref{sec:axangtomat},
\begin{equation}\label{eqt:diagaxang}
\begin{cases}
\cos\phi+{n_1}^2(1-\cos\phi)=m_{11}\\
\cos\phi+{n_2}^2(1-\cos\phi)=m_{22}\\
\cos\phi+{n_3}^2(1-\cos\phi)=m_{33}
\end{cases}
\Leftrightarrow\ \ \ 
\begin{cases}
{n_1}^2=\frac{m_{11}-\cos\phi}{1-\cos\phi}\\
{n_2}^2=\frac{m_{22}-\cos\phi}{1-\cos\phi}\\
{n_3}^2=\frac{m_{33}-\cos\phi}{1-\cos\phi}
\end{cases}
\end{equation}
With ${n_1}^2+{n_2}^2+{n_3}^2=1$, we can easily get $\cos\phi=\frac{m_{11}+m_{22}+m_{33}-1}{2}$, in continuity with (\ref{eqt:ctnit}). Together with the convention $\phi\in[0,\pi)$, 
\begin{equation}
\phi=
\arccos\left(\frac{m_{11}+m_{22}+m_{33}-1}{2}\right)
\end{equation}

Then we consider $m_{23}$ and $m_{32}$.
\[
\begin{cases}
m_{23}=n_2n_3(1-\cos\phi)-n_1\sin\phi\\
m_{32}=n_2n_3(1-\cos\phi)+n_1\sin\phi
\end{cases}
\]

Subtract the second by the first, use cyclic symmetry and get
\begin{numcases}{}
n_1=\frac{m_{32}-m_{23}}{2\sin\phi}=\frac{m_{32}-m_{23}}{\sqrt{4-(m_{11}+m_{22}+m_{33}-1)^2}}\\
n_2=\frac{m_{13}-m_{31}}{2\sin\phi}=\frac{m_{13}-m_{31}}{\sqrt{4-(m_{11}+m_{22}+m_{33}-1)^2}}\\
n_3=\frac{m_{21}-m_{12}}{2\sin\phi}=\frac{m_{21}-m_{12}}{\sqrt{4-(m_{11}+m_{22}+m_{33}-1)^2}}
\end{numcases}
That's the final result.
\subsection{From Euler angles to the axis and angle of rotation}\label{sec:eulertoaxang}
Substitute
\[
\begin{cases}
m_{11}=\cos\varphi\cos\psi-\cos\theta\sin\psi\sin\varphi\\
m_{22}=-\sin\varphi\sin\psi+\cos\theta\cos\psi\cos\varphi\\
m_{33}=\cos\theta
\end{cases}
\]
then we get
\[
m_{11}+m_{22}+m_{33}-1=(1+\cos\theta)\cos(\varphi+\psi)-(1-\cos\theta)
\]
with which
\begin{numcases}{}
\phi=
\arccos\frac{(1+\cos\theta)\cos(\varphi+\psi)-(1-\cos\theta)}{2}
\\
n_1=
\frac{m_{32}-m_{23}}{\sqrt{4-[(1+\cos\theta)\cos(\varphi+\psi)-(1-\cos\theta)]^2}}
\\
n_2=
\frac{m_{13}-m_{31}}{\sqrt{4-[(1+\cos\theta)\cos(\varphi+\psi)-(1-\cos\theta)]^2}}
\\
n_3=
\frac{m_{21}-m_{12}}{\sqrt{4-[(1+\cos\theta)\cos(\varphi+\psi)-(1-\cos\theta)]^2}}
\end{numcases}

By the way,  by transforming the expressions, some interesting and insightful relationships can be derived. We calculate
\[
\cos^2\frac{\phi}{2}=\frac{1+\cos\phi}{2}=\frac{(1+\cos\theta)[1+\cos(\varphi+\psi)]}{4}=\cos^2\frac{\varphi+\psi}{2}\cos^2\frac{\theta}{2}
\]
Because we have chosen a convention that $\phi\in[0,\pi)$, meaning that $\cos\frac{\phi}{2}\geq0$. Also, for the same reason, $\cos\frac{\theta}{2}\geq0$. Therefore, the square root of LHS must be
\begin{equation}
\cos\frac{\phi}{2}=\left|\cos\frac{\varphi+\psi}{2}\right|\cos\frac{\theta}{2}
\end{equation}
Thus, $\cos\frac{\phi}{2}\geq\cos\frac{\theta}{2}$, meaning that
\begin{equation}
\phi\geq\theta
\end{equation}

Another phenomenon is when $\varphi+\psi$ goes from the left neighborhood of $\pi$ to the other side, the value of LHS may not be derivable. That can only be explained by that the axis flips in the meantime, meaning a discrete change in the expression in spite of a continuous change in the group element. 
\subsection{From the axis and angle of rotation to Euler angles}\label{sec:axangtoeuler}
From Section~\ref{sec:axangtomat}, one can know the matrix elements of a rotation with the axis $\boldsymbol{\hat{n}}$ and the angle $\phi$, including
\[
\begin{cases}
m_{13}=n_3n_1(1-\cos\phi)+n_2\sin\phi\\
m_{23}=n_2n_3(1-\cos\phi)-n_1\sin\phi\\
m_{33}=\cos\phi+{n_3}^2(1-\cos\phi)\\
m_{32}=n_2n_3(1-\cos\phi)+n_1\sin\phi\\
m_{31}=n_3n_1(1-\cos\phi)-n_2\sin\phi
\end{cases}
\]
Substitute all these relations into all the expressions in Section~\ref{sec:mattoeuler}, and we will get the final result. The article omits this process because it is as direct as possible. One example will be shown in Section~\ref{sec:successive}.
\subsection{Any rotation matrix expressed by independent elements}\label{sec:indep}
In the processes above we use some elements that are dependent to each other, more or less. A natural problem is how to use the least elements, i.e. independent ones, to recover the whole matrix, and get the axis and angle or Euler angles following this formulation. To solve this, we express any SO(3) matrices explicitly by the diagonal elements.

Consider an arbitrary rotation matrix $\mathbf{M}$ whose entry in the $i$-th row and $j$-th column is $m_{ij}$. With $\mathbf{M}\mathbf{M}^\mathrm{T}=\mathbf{I}$,
\begin{equation}
\begin{cases}
{m_{11}}^2+{m_{12}}^2+{m_{13}}^2=1 \\
{m_{21}}^2+{m_{22}}^2+{m_{23}}^2=1 \\
{m_{31}}^2+{m_{32}}^2+{m_{33}}^2=1
\end{cases}
\Leftrightarrow\ \ \ 
\begin{cases}
m_{13}=\pm\sqrt{1-{m_{11}}^2-{m_{12}}^2} \\
m_{21}=\pm\sqrt{1-{m_{22}}^2-{m_{23}}^2} \\
m_{32}=\pm\sqrt{1-{m_{31}}^2-{m_{33}}^2}
\end{cases}
\end{equation}
Meanwhile, with an equivalent form $\mathbf{M}^\mathrm{T}=\mathbf{M}^{-1}$,
\[
\begin{cases}
m_{11}=m_{22}m_{33}-m_{23}m_{32}\\
m_{22}=m_{33}m_{11}-m_{31}m_{13}\\
m_{33}=m_{11}m_{22}-m_{12}m_{21}
\end{cases}
\]
then
\begin{equation}\label{eqt:mnondrel}
\begin{cases}
m_{12}=\pm\frac{m_{11}m_{22}-m_{33}}{\sqrt{1-{m_{22}}^2-{m_{23}}^2}}\\
m_{23}=\pm\frac{m_{22}m_{33}-m_{11}}{\sqrt{1-{m_{31}}^2-{m_{33}}^2}}\\
m_{31}=\pm\frac{m_{33}m_{11}-m_{22}}{\sqrt{1-{m_{11}}^2-{m_{12}}^2}}
\end{cases}
\end{equation}
So
\begin{equation}
\mathbf{M}=
\begin{pmatrix}
m_{11}&\pm_2\frac{m_{11}m_{22}-m_{33}}{\sqrt{1-{m_{22}}^2-{m_{23}}^2}}&\pm_1\sqrt{1-{m_{11}}^2-{m_{12}}^2}\\[6pt]
\pm_2\sqrt{1-{m_{22}}^2-{m_{23}}^2}&m_{22}&\pm_3\frac{m_{22}m_{33}-m_{11}}{\sqrt{1-{m_{31}}^2-{m_{33}}^2}}\\[6pt]
\pm_1\frac{m_{33}m_{11}-m_{22}}{\sqrt{1-{m_{11}}^2-{m_{12}}^2}}&\pm_3\sqrt{1-{m_{31}}^2-{m_{33}}^2}&m_{33}
\end{pmatrix}
\end{equation}
where the sign correlation follows the subscript identity. It can be seen that $m_{12}$ and $m_{21}$ share the same sign if $m_{11}m_{22}-m_{33}>0$, and vice versa. For other non-diagonal elements, the law is similar. From Equation (\ref{eqt:mnondrel}), 
\begin{equation}\label{eqt:mnondrelsq}
\begin{cases}
{m_{12}}^2(1-{m_{22}}^2-{m_{23}}^2)=(m_{11}m_{22}-m_{33})^2\\
{m_{23}}^2(1-{m_{33}}^2-{m_{31}}^2)=(m_{22}m_{33}-m_{11})^2\\
{m_{31}}^2(1-{m_{11}}^2-{m_{12}}^2)=(m_{33}m_{11}-m_{22})^2
\end{cases}
\end{equation}
From the first equation of (\ref{eqt:mnondrelsq}), ${m_{12}}^2=\frac{(m_{11}m_{22}-m_{33})^2}{1-{m_{22}}^2-{m_{23}}^2}$; substitute it into the second, getting ${m_{31}}^2=1-{m_{33}}^2-\frac{(m_{22}m_{33}-m_{11})^2}{{m_{23}}^2}$; finally substitute the two equations into the third and get $\left(1-{m_{33}}^2-\frac{(m_{22}m_{33}-m_{11})^2}{{m_{23}}^2}\right)\left(1-{m_{11}}^2-\frac{(m_{11}m_{22}-m_{33})^2}{1-{m_{22}}^2-{m_{23}}^2}\right)=(m_{33}m_{11}-m_{22})^2$, which is equivalent to
\[
[(1-{m_{33}}^2){m_{23}}^2-(m_{22}m_{33}-m_{11})^2][(1-{m_{11}}^2)(1-{m_{22}}^2-{m_{23}}^2)]=(m_{33}m_{11}-m_{22})^2(1-{m_{22}}^2-{m_{23}}^2){m_{23}}^2
\]
Then we get a quadratic equation in ${m_{23}}^2$, denoted by $x$, from the formula above. After reducing the common factor $({m_{11}}^2+{m_{22}}^2+{m_{33}}^2-2m_{11}m_{22}m_{33}-1)$, we get
\begin{equation}
x^2+(-{m_{11}}^2+{m_{22}}^2+{m_{33}}^2-1)x+(m_{22}m_{33}-m_{11})^2=0
\end{equation}
The discriminant of the roots is
\begin{align}
\Delta&=(-{m_{11}}^2+{m_{22}}^2+{m_{33}}^2-1)^2-4(m_{22}m_{33}-m_{11})^2\nonumber\\
&=(m_{11}+m_{22}+m_{33}+1)(-m_{11}+m_{22}+m_{33}-1)(m_{11}+m_{22}-m_{33}-1)(-m_{11}+m_{22}-m_{33}+1)
\end{align}
It is worthwhile to emphasize that $\Delta$ is cyclic symmetrical.

Denote the reflection matrices about $n$-axes by $\mathbf{P}(\boldsymbol{\hat{n}})$. Considering the three matrices:
\[
\mathbf{M_1}=
\begin{pmatrix}
-m_{11} & -m_{12} & -m_{13} \\
m_{21} & m_{22} & m_{23} \\
m_{31} & m_{32} & m_{33}
\end{pmatrix},\ 
\mathbf{M_2}=
\begin{pmatrix}
m_{11} & m_{12} & m_{13} \\
m_{21} & m_{22} & m_{23} \\
-m_{31} & -m_{32} & -m_{33}
\end{pmatrix},\ 
\mathbf{M_3}=
\begin{pmatrix}
-m_{11} & -m_{12} & -m_{13} \\
m_{21} & m_{22} & m_{23} \\
-m_{31} & -m_{32} & -m_{33}
\end{pmatrix}
\]
Actually,
\[
\mathbf{M_1}=\mathbf{P}(\boldsymbol{\hat{x}})\mathbf{M},\ \mathbf{M_2}=\mathbf{P}(\boldsymbol{\hat{z}})\mathbf{M},\ \mathbf{M_3}=\mathbf{P}(\boldsymbol{\hat{x}})\mathbf{P}(\boldsymbol{\hat{z}})\mathbf{M}
\]
Due to $\mathbf{M}$, $\mathbf{P}(\boldsymbol{\hat{x}})$, $\mathbf{P}(\boldsymbol{\hat{z}})$ $\in\mathrm{O(3)}$, $\mathbf{M_1}$, $\mathbf{M_2}$, $\mathbf{M_3}$ $\in\mathrm{O(3)}$; and because $\mathrm{det}\ \mathbf{P}(\boldsymbol{\hat{x}})=\mathrm{det}\ \mathbf{P}(\boldsymbol{\hat{z}})=-1$, $\mathbf{M_1}$, $\mathbf{M_2}$ are reflection matrices while $\mathbf{M_3}\in\mathrm{SO(3)}$.

Thus, we express the discriminant
\begin{align}
\Delta&=(2+2\cos\Theta_\mathbf{M})(-2+2\cos\Theta_\mathbf{M_1})(-2+2\cos\Theta_\mathbf{M_2})(2+2\cos\Theta_\mathbf{M_3})\nonumber\\
&=8(1+\cos\Theta_\mathbf{M})(1-\cos\Theta_\mathbf{M_1})(1-\cos\Theta_\mathbf{M_2})(1+\cos\Theta_\mathbf{M_3})\\
&\geq0\nonumber
\end{align}
So, all SO(3) matrices can be only expressed by their diagonal elements. Specifically, $m_{23}=\pm_3\sqrt{\frac{1}{2}\left[({m_{11}}^2-{m_{22}}^2-{m_{33}}^2+1)\pm_4\sqrt{\Delta}\right]}$. Thus, $m_{32}=\pm_3\frac{m_{22}m_{33}-m_{11}}{\sqrt{\frac{1}{2}\left[({m_{11}}^2-{m_{22}}^2-{m_{33}}^2+1)\pm_4\sqrt{\Delta}\right]}}$. And due to cyclic symmetry, 
\begin{align*}
m_{21}&=\pm_2\frac{m_{11}m_{22}-m_{33}}{\sqrt{\frac{1}{2}\left[(-{m_{11}}^2-{m_{22}}^2+{m_{33}}^2+1)\pm\sqrt{\Delta}\right]}}\\
&=\pm\sqrt{\frac{1}{2}\left[(-{m_{11}}^2-{m_{22}}^2+{m_{33}}^2+1)\mp\sqrt{\Delta}\right]}
\end{align*}
The lower notation is removed because the sign depends on the relative relationship between $m_{11}m_{22}$ and $m_{33}$.
Furthermore, the solution of $m_{21}$ is constrained by ${m_{21}}^2+{m_{22}}^2+{m_{23}}^2=1$, i.e.
\begin{align*}
1&=\frac{1}{2}(-{m_{11}}^2-{m_{22}}^2+{m_{33}}^2+1)\mp\frac{1}{2}\sqrt{\Delta}+{m_{22}}^2+\frac{1}{2}({m_{11}}^2-{m_{22}}^2-{m_{33}}^2+1)\pm_4\frac{1}{2}\sqrt{\Delta}\\
&=1\mp\frac{1}{2}\sqrt{\Delta}\pm_4\frac{1}{2}\sqrt{\Delta}
\end{align*}
Thus, the minus-plus sign should be noted by 4. 

On the other hand, with $\mathbf{M}\mathbf{M}^\mathrm{T}=\mathbf{I}$, ${m_{13}}^2+{m_{23}}^2+{m_{33}}^2=1$. In a similar manner, $m_{13}=\pm\sqrt{\frac{1}{2}\left[(-{m_{11}}^2+{m_{22}}^2-{m_{33}}^2+1)\mp_4\sqrt{\Delta}\right]}$.

Use the cyclic symmetry and the method above again and again, we can conclude
\begin{adjustwidth}{-6.5em}{}
\begin{equation}
\begin{small}
\mathbf{M}=
\begin{pmatrix}
m_{11}&\pm\sqrt{\frac{1}{2}\left[(-{m_{11}}^2-{m_{22}}^2+{m_{33}}^2+1)\pm_4\sqrt{\Delta}\right]}&\pm\sqrt{\frac{1}{2}\left[(-{m_{11}}^2+{m_{22}}^2-{m_{33}}^2+1)\mp_4\sqrt{\Delta}\right]}\\[6pt]
\pm\sqrt{\frac{1}{2}\left[(-{m_{11}}^2-{m_{22}}^2+{m_{33}}^2+1)\mp_4\sqrt{\Delta}\right]}&m_{22}&\pm\sqrt{\frac{1}{2}\left[({m_{11}}^2-{m_{22}}^2-{m_{33}}^2+1)\pm_4\sqrt{\Delta}\right]}\\[6pt]
\pm\sqrt{\frac{1}{2}\left[(-{m_{11}}^2+{m_{22}}^2-{m_{33}}^2+1)\pm_4\sqrt{\Delta}\right]}&\pm\sqrt{\frac{1}{2}\left[({m_{11}}^2-{m_{22}}^2-{m_{33}}^2+1)\mp_4\sqrt{\Delta}\right]}&m_{33}
\end{pmatrix}
\end{small}
\end{equation}
\end{adjustwidth}
where $\Delta=(m_{11}+m_{22}+m_{33}+1)(-m_{11}+m_{22}+m_{33}-1)(m_{11}+m_{22}-m_{33}-1)(-m_{11}+m_{22}-m_{33}+1)$, and where for all totally different $i$, $j$, $k$, $m_{ij}$ and $m_{ji}$ share the same sign if $m_{ii}m_{jj}\geq m_{kk}$, and vice versa.

For each case, there are 16 solutions. Nevertheless, there may be extraneous roots within the solutions above because the constraint that $\mathbf{M}\in\mathrm{SO(3)}$ has not been fully used, so what is derived above is only a necessary condition. With the full constraint, some of them are likely to be excluded. Actually, for each case the real solutions totals 8, all of which are shown in Appendix~\ref{sec:appendix} (one can verify this statement with Mathematica).
\section{Continuous rotations in SO(3) Group}\label{sec:propso3}
This section aims to provide ways to describe continuous rotations, i.e. those through the integral of infinitesimal rotations. The first three subsections analyze and summarize some basic local properties of SO(3) elements expressed by the axis and angle of rotation. In it, we will first introduce Lie algebras in an intuitive way to highlight their physical meaning as a preparation, and then use them as an analytical paradigm of investigating local properties and apply them to SO(3) group and its correspondent angular velocities. After introducing Euler angles, the last three subsections constitute a systematic part about descriptions of continuous rotations, as understanding requires, beyond what appears in Section~\ref{sec:commt}, which is talked in a piecemeal manner.
\subsection{Basics of Lie algebras}
\subsubsection{Generators}\label{sec:generator}
See each element of a Lie group as a transformation. Considering an infinitesimal transformation $\hat{F}$; actually, it must be close to the identity transformation $\hat{I}$ infinitely. Then, it can \textbf{always} be written as
\begin{equation}
\hat{F}=\hat{I}+\varepsilon\hat{G}+\mathcal{O}(\varepsilon^2)
\end{equation}
where $\varepsilon$ is an infinitesimal number and $\hat{G}$ is a finite transformation.
\begin{definition}
Consider an infinitesimal transformation $\hat{F}$ in the group, if it can be expressed by the equation above, $\hat{G}$ is called a generator.
\end{definition}

\begin{theorem}\label{trm:timesgnr}
If $\hat{G}$ is a generator, then $\forall \lambda\in\mathbb{C}$, $\lambda\hat{G}$ is also a generator.
\end{theorem}
\begin{proof}
$\exists\textrm{ infinitesimal transformation }\hat{F}$, $\hat{F}=\hat{I}+\varepsilon\hat{G}+\mathcal{O}(\varepsilon^2)$, i.e. $\hat{F}=\hat{I}+\frac{\varepsilon}{\lambda}\lambda\hat{G}+\mathcal{O}(\varepsilon^2)$ where $\frac{\varepsilon}{\lambda}$ is also an infinitesimal number because $\lambda$ is finite.
\end{proof}
With a similar approach, any infinitesimal transformation can be reparametrized. Actually, $\forall$ infinitesimal number $\varepsilon$, any infinitesimal transformation $\hat{F}=\hat{I}+\varepsilon'\hat{G}+\mathcal{O}(\varepsilon'^2)$ can be expressed by $\hat{F}=\hat{I}+\varepsilon\frac{\varepsilon'}{\varepsilon}\hat{G}+\mathcal{O}(\varepsilon^2)$ where $\frac{\varepsilon'}{\varepsilon}\hat{G}$ is a finite transformation because both $\varepsilon$ and $\varepsilon'$ are first-order infinitesimal numbers. Thus, any infinitesimal transformation can be expressed as $\hat{F}=\hat{I}+\varepsilon\hat{G}+\mathcal{O}(\varepsilon^2)$ with $\varepsilon$ \textbf{fixed}. With this result, we can prove these theorems below in an easier way.
\begin{theorem}\label{trm:gnrt}
For any two infinitesimal transformation $\hat{F}_1=\hat{I}+\varepsilon\hat{G}_1+\mathcal{O}(\varepsilon^2)$ and $\hat{F}_2=\hat{I}+\varepsilon\hat{G}_2+\mathcal{O}(\varepsilon^2)$, $\hat{F}_2\hat{F}_1=\hat{I}+\varepsilon(\hat{G}_1+\hat{G}_2)+\mathcal{O}(\varepsilon^2)$.
\end{theorem}
\begin{proof}
\begin{equation}
\hat{F}_2\hat{F}_1=(\hat{I}+\varepsilon\hat{G}_2+\mathcal{O}(\varepsilon^2))(\hat{I}+\varepsilon\hat{G}_1+\mathcal{O}(\varepsilon^2))=\hat{I}+\varepsilon(\hat{G}_1+\hat{G}_2)+\mathcal{O}(\varepsilon^2)
\end{equation}
\end{proof}

We can see the product of group elements maps to the addition of their generators of that group.
\begin{theorem}
Any two infinitesimal transformations commute with each other.
\end{theorem}
\begin{proof}
\[
\hat{F}_2\hat{F}_1=\hat{I}+\varepsilon(\hat{G}_1+\hat{G}_2)+\mathcal{O}(\varepsilon^2)=\hat{I}+\varepsilon(\hat{G}_2+\hat{G}_1)+\mathcal{O}(\varepsilon^2)=\hat{F}_1\hat{F}_2+\mathcal{O}(\varepsilon^2)
\]
\end{proof}
We can see the theorem below from Theorem~\ref{trm:gnrt}, because $\hat{F}_2\hat{F}_1$ is also an infinitesimal transformation.
\begin{theorem}\label{trm:addgnr}
If $\hat{G}_1$, $\hat{G}_2$ are two generators, then $\hat{G}_1+\hat{G}_2$ is also a generator.
\end{theorem}
Due to the closure of any group, any infinitesimal transformation corresponding to a generator can be applied to elements everywhere in the group in a same manner, on which the following theorem is based.
\begin{theorem}\label{trm:expm}
Any-order self-product of a transformation in the group must belong to the group as well. $\mathrm{(}$Easy to prove$\mathrm{)}$
\end{theorem}
This can lead to the exponential mapping. The way to realizing the integral of infinitesimal rotations is to take the limit of a finite transformation:
\begin{equation}
\mathrm{e}^{\mu\hat{G}}=\lim\limits_{n\to\infty}(1+\frac{\mu}{n}\hat{G})^n
\end{equation}
where $\mu\in\mathbb{C}$, and $\frac{\mu}{n}$ goes to an infinitesimal angle, and $\hat{G}$ is a generator of the group $G$. Thus,
\begin{equation}
\forall\mu\in\mathbb{C},\ \mathrm{e}^{\mu\hat{G}}\in G
\end{equation}
\subsubsection{Generating set}\label{sec:genset}
From Theorems~\ref{trm:timesgnr} and~\ref{trm:addgnr}, $\forall\lambda,\mu\in\mathbb{C}$, $\lambda\hat{F}_1+\mu\hat{F}_2$ is also a generator. Then, one may conclude Theorem~\ref{trm:lnrspc} below.
\begin{theorem}\label{trm:lnrspc}
All generators of any Lie group constitute a linear space.
\end{theorem}
Thus, one can choose a set of non-linear generators to linearly express any generator.
\begin{definition}
The set of non-linear generators that can linearly express any generator is called a generating set denoted by $\mathfrak{g}$. The number of elements in the set is called the dimension of the generating set, denoted by $\mathrm{dim}\mathrm{(}\mathfrak{g}\mathrm{)}$.
\end{definition}

In fact, the choice for generating set need to meet the academic convention.

\begin{theorem}
The commutator of any two generators is also a generator.
\end{theorem}
To prove the theorem, we first introduce Baker-Campbell-Hausdorff (BCH) formula.
\begin{equation}
\mathrm{e}^{\hat{A}} \mathrm{e}^{\hat{B}}=\mathrm{e}^{\hat{A}+\hat{B}+\frac{1}{2}[\hat{A},\hat{B}]+\frac{1}{12}[\hat{A},[\hat{A},\hat{B}]]+\frac{1}{12}[[\hat{A},\hat{B}],\hat{B}]+\dots}
\end{equation}
\begin{proof}
Choose any two generators $\hat{X}$, $\hat{Y}$. Consider a series of exponential mapping, $\mathrm{e}^{s\hat{X}}\mathrm{e}^{t\hat{Y}}\mathrm{e}^{-s\hat{X}}\mathrm{e}^{-t\hat{Y}}$, a group element. When $s,\ t\to0$, the operation is an infinitesimal element in the group.
\[
\log(\mathrm{e}^{s\hat{X}}\mathrm{e}^{t\hat{Y}})=s\hat{X}+t\hat{Y}+\frac{1}{2}st[\hat{X},\hat{Y}]+\mathcal{O}((s,t)^3)
\]
\begin{align*}
\log(\mathrm{e}^{s\hat{X}}\mathrm{e}^{t\hat{Y}}\mathrm{e}^{-s\hat{X}})&=s\hat{X}+t\hat{Y}+\frac{1}{2}st[\hat{X},\hat{Y}]-s\hat{X}+\frac{1}{2}st[\hat{X},\hat{Y}]+\mathcal{O}((s,t)^3)\\
&=t\hat{Y}+st[\hat{X},\hat{Y}]+\mathcal{O}((s,t)^3)
\end{align*}
\begin{align*}
\log(\mathrm{e}^{s\hat{X}}\mathrm{e}^{t\hat{Y}}\mathrm{e}^{-s\hat{X}}\mathrm{e}^{-t\hat{Y}})&=t\hat{Y}+st[\hat{X},\hat{Y}]-t\hat{Y}+\mathcal{O}((s,t)^3)\\
&=st[\hat{X},\hat{Y}]+\mathcal{O}((s,t)^3)
\end{align*}
So,
\begin{equation}\label{sec:genliealg}
\mathrm{e}^{s\hat{X}}\mathrm{e}^{t\hat{Y}}\mathrm{e}^{-s\hat{X}}\mathrm{e}^{-t\hat{Y}}=\hat{I}+st[\hat{X},\hat{Y}]+\mathcal{O}((s,t)^3)
\end{equation}
Because LHS is an infinitesimal transformation of the group and $st$ is an infinitesimal number, $[\hat{X},\hat{Y}]$ must be a generator.
\end{proof}
\begin{definition}
Each commutator between two elements multiplied by $-\mathrm{i}$ in the generating set is called a Lie algebra, which is also a generator\cite{mazhongqi}.
\end{definition}

Lie algebras are not merely a mathematical trick; instead, it contains physical meanings and has many applications. With them, any path between any two group elements can be restored only by the generating set (with its Lie algebras) which appear solely as a linear term in the exponents. 

In fact, any exponential mapping can be expressed by
\begin{equation}
\mathrm{e}^{\mu\sum_{i=1}^{\mathrm{dim}\mathfrak{g}}c_i\hat{G}_i}
\end{equation}
where $c_i$-s are complex coefficients. Specifically, one can apply the Zassenhaus decomposition of BCH formula
\begin{equation}
\mathrm{e}^{\hat{A}+\hat{B}}=\mathrm{e}^{\hat{A}} \mathrm{e}^{\hat{B}} \mathrm{e}^{-\frac{1}{2}[\hat{A},\hat{B}]} \mathrm{e}^{\frac{1}{6}[\hat{A},[\hat{A},\hat{B}]]-\frac{1}{3}[[\hat{A},\hat{B}],\hat{B}]}\dots
\end{equation}

An example will be shown in Section~\ref{sec:countable}.
\subsection{Matrix representation of infinitesimal rotations}
\subsubsection[Generating set of SO(3) Group]{Generating set of SO(3) Group\cite{YRQ2024}}\label{sec:gensetso3}
The $\boldsymbol{J}$ in an infinitesimal transformation $\mathbf{M}=\mathbf{I}+\varepsilon\boldsymbol{J}+\mathcal{O}(\varepsilon^2)$ is an SO(3) generator if and only if
\[
\mathbf{I}=\mathbf{M}\mathbf{M}^\mathrm{T}=(\mathbf{I}+\varepsilon\boldsymbol{J}^\mathrm{T}+\mathcal{O}(\varepsilon^2))(\mathbf{I}+\varepsilon\boldsymbol{J}+\mathcal{O}(\varepsilon^2))=\mathbf{I}+\varepsilon(\boldsymbol{J}+\boldsymbol{J}^\mathrm{T})+\mathcal{O}(\varepsilon^2)
\]
Then
\begin{equation}
\boldsymbol{J}^\mathrm{T}=-\boldsymbol{J}
\end{equation}

So, all 3$\times$3 antisymmetric matrices are precisely the SO(3) generators, generally written as
\[
\boldsymbol{J}=
\begin{pmatrix}
0&a&b\\
-a&0&c\\
-b&-c&0
\end{pmatrix}
\]
One can choose the generating set as
\begin{equation}\label{eqt:gensetso3}
\boldsymbol{J_1}=
\begin{pmatrix}
0&0&0\\
0&0&-1\\
0&1&0
\end{pmatrix}
,\
\boldsymbol{J_2}=
\begin{pmatrix}
0&0&1\\
0&0&0\\
-1&0&0
\end{pmatrix}
,\
\boldsymbol{J_3}=
\begin{pmatrix}
0&-1&0\\
1&0&0\\
0&0&0
\end{pmatrix}
\footnote{In most literature\cite{mazhongqi,peskin}, the generating set is chosen to be 
$\boldsymbol{J_1}=\mathrm{i}
\begin{pmatrix}
0&0&0\\
0&0&-1\\
0&1&0
\end{pmatrix}
,\
\boldsymbol{J_2}=\mathrm{i}
\begin{pmatrix}
0&0&1\\
0&0&0\\
-1&0&0
\end{pmatrix}
,\
\boldsymbol{J_3}=\mathrm{i}
\begin{pmatrix}
0&-1&0\\
1&0&0\\
0&0&0
\end{pmatrix}$.
Rigorously, readers can understand $\boldsymbol{J_1}$, $\boldsymbol{J_2}$, $\boldsymbol{J_3}$ in (\ref{eqt:gensetso3}) as $-\mathrm{i}$ plus the elements in the generating set.}
\end{equation}
and any generator of SO(3) can be expressed as $\boldsymbol{J}=-c\boldsymbol{J_1}+b\boldsymbol{J_2}-a\boldsymbol{J_3}$.

\subsubsection{The angular velocity matrix of SO(3) Group}\label{sec:angv}
We have derived the time derivative of any position vector of a rotating rigid body in Section~\ref{sec:commt}
\begin{equation}
\frac{\mathrm{d}\boldsymbol{r}}{\mathrm{d}t}=\boldsymbol{\omega}\times\boldsymbol{r}=
\begin{pmatrix}
\omega_yz-\omega_zy\\
\omega_zx-\omega_xz\\
\omega_xy-\omega_yx
\end{pmatrix}
=\begin{pmatrix}
0 & -\omega_z & \omega_y \\
\omega_z & 0 & -\omega_x \\
-\omega_y & \omega_x & 0
\end{pmatrix}
\begin{pmatrix}
x\\
y\\
z
\end{pmatrix}
\end{equation}
Denote the angular velocity matrix $\begin{pmatrix}
0 & -\omega_z & \omega_y \\
\omega_z & 0 & -\omega_x \\
-\omega_y & \omega_x & 0
\end{pmatrix}$ by $\mathbf{\Omega}$, then
\begin{equation}
\boldsymbol{\dot{r}}=\mathbf{\Omega}\boldsymbol{r}
\end{equation}
and we find $\mathbf{\Omega}$ is an antisymmetrical matrix, i.e. 
\begin{equation}
\mathbf{\Omega}^\mathrm{T}=-\mathbf{\Omega}
\end{equation}

Characterize any vector fixed in the rigid body by an active coordinate transformation matrix $\mathbf{M}$, where $\boldsymbol{r}$ and $\mathbf{M}$ are functions of time $t$. And then we will write the relationship above expressed by $\mathbf{M}$ and its time-derivative.
On one hand, because $\boldsymbol{r}(t)=\mathbf{M}(t)\boldsymbol{r_0}$, $\boldsymbol{\dot{r}}(t)=\mathbf{\dot{M}}(t)\boldsymbol{r_0}$ where $\boldsymbol{r_0}$ is the coordinates represented in the body-attached frame. On the other hand, $\boldsymbol{\dot{r}}(t)=\mathbf{\Omega}\boldsymbol{r}(t)=\mathbf{\Omega}\mathbf{M}(t)\boldsymbol{r_0}$. Since $\boldsymbol{r_0}$ is arbitrary,
\begin{equation}\label{eqt:angv}
\mathbf{\dot{M}}=\mathbf{\Omega}\mathbf{M}
\end{equation}

Actually, the angular velocity matrix is an arbitrary antisymmetric matrix with dimension $\mathrm{T}^{-1}$, which corresponds to an arbitrary generator. That's inevitable and we will derive the relationship. Integrate the formula (\ref{eqt:angv}) and get $\mathrm{d}\mathbf{M}=\mathbf{\Omega} \mathrm{d}t\ \mathbf{M}$. Write the infinitesimal change acts on $\mathbf{M}$ as $\mathbf{M}(t+\mathrm{d}t)=(1+\varepsilon\boldsymbol{J})\mathbf{M}(t)$ (ignoring higher-order terms), then $\mathrm{d}\mathbf{M}=\varepsilon\boldsymbol{J}\mathbf{M}$, so
\begin{equation}\label{eqt:otej}
\mathbf{\Omega} \mathrm{d}t=\varepsilon\boldsymbol{J}
\end{equation}
This further shows that the angular velocity matrix is actually \textbf{the time-derivative of an infinitesimal quantity proportional to a generator} defined on the neighborhood of a moment.

Then compare the coefficients. Actually,
\begin{equation}\label{eqt:ooj}
\mathbf{\Omega}=\omega_x\boldsymbol{J_1}+\omega_y\boldsymbol{J_2}+\omega_z\boldsymbol{J_3}=\boldsymbol{\omega}\cdot\boldsymbol{\vec{J}}
\end{equation}
With the notation rule $x_1=x$, $x_2=y$, $x_3=z$, correspondingly, the angular velocity of coordinates is $\boldsymbol{\omega}=\omega_x\boldsymbol{\hat{x}_1}+\omega_y\boldsymbol{\hat{x}_2}+\omega_z\boldsymbol{\hat{x}_3}$. Any generator can be expressed by the form to emphasize its physical meaning
\begin{equation}\label{eqt:epj}
\varepsilon\boldsymbol{J}=
\begin{pmatrix}
0&-\mathrm{d}\phi_z&\mathrm{d}\phi_y\\
\mathrm{d}\phi_z&0&-\mathrm{d}\phi_x\\
-\mathrm{d}\phi_y&\mathrm{d}\phi_x&0
\end{pmatrix}
\end{equation}

\subsubsection{Normalization of generators}\label{sec:nomgen}
Calculate infinitesimal rotations about the Cartesian axes. From (\ref{eqt:finrot}),
\begin{equation}
\mathbf{R}(-\mathrm{d}\phi, \boldsymbol{\hat{x}})=
\begin{pmatrix}
1&0 &0 \\
0 & \cos\mathrm{d}\phi & -\sin\mathrm{d}\phi \\
0 & \sin\mathrm{d}\phi & \cos\mathrm{d}\phi
\end{pmatrix}
=
\begin{pmatrix}
1&0 &0 \\
0 & 1 & -\mathrm{d}\phi \\
0 & \mathrm{d}\phi & 1
\end{pmatrix}
+\mathcal{O}(\mathrm{d}\phi^2)
=\mathbf{I}+\mathrm{d}\phi\boldsymbol{J_1}+\mathcal{O}(\mathrm{d}\phi^2)
\end{equation}
\begin{equation}
\mathbf{R}(-\mathrm{d}\phi, \boldsymbol{\hat{y}})=
\begin{pmatrix}
\cos\mathrm{d}\phi&0 &\sin\mathrm{d}\phi\\
0&1&0 \\
-\sin\mathrm{d}\phi &0& \cos\mathrm{d}\phi
\end{pmatrix}
=
\begin{pmatrix}
1&0 &\mathrm{d}\phi\\
0&1&0 \\
-\mathrm{d}\phi &0&1
\end{pmatrix}
+\mathcal{O}(\mathrm{d}\phi^2)
=\mathbf{I}+\mathrm{d}\phi\boldsymbol{J_2}+\mathcal{O}(\mathrm{d}\phi^2)
\end{equation}
\begin{equation}
\mathbf{R}(-\mathrm{d}\phi, \boldsymbol{\hat{z}})=
\begin{pmatrix}
\cos\mathrm{d}\phi& -\sin\mathrm{d}\phi &0 \\
\sin\mathrm{d}\phi & \cos\mathrm{d}\phi&0  \\
0&0&1
\end{pmatrix}
=
\begin{pmatrix}
1& -\mathrm{d}\phi &0 \\
\mathrm{d}\phi & 1&0  \\
0&0&1
\end{pmatrix}
+\mathcal{O}(\mathrm{d}\phi^2)
=\mathbf{I}+\mathrm{d}\phi\boldsymbol{J_3}+\mathcal{O}(\mathrm{d}\phi^2)
\end{equation}
That's why we choose the convention of the generating set in Section~\ref{sec:gensetso3} because the infinitesimal parameter $\varepsilon$ is the angle $\mathrm{d}\phi$ as a result.

Second, we normalize any element in $\mathfrak{so}(3)$ (any infinitesimal rotation at any direction) as we choose $\varepsilon$ as $\mathrm{d}\phi$. We can see this from (\ref{eqt:epj})
\[
\varepsilon\boldsymbol{J}=\mathrm{d}\phi_x\boldsymbol{J_1}+\mathrm{d}\phi_y\boldsymbol{J_2}+\mathrm{d}\phi_z\boldsymbol{J_3}
\]

Actually, $\mathrm{d}\phi_x$, $\mathrm{d}\phi_y$ and $\mathrm{d}\phi_z$ can be seen as projections of the total angle $\mathrm{d}\phi$ on the Cartesian axes. Noticing ($n_1$, $n_2$, $n_3$) are the direction cosines, then
\begin{equation}
\varepsilon\boldsymbol{J}=\mathrm{d}\phi(n_1\boldsymbol{J_1}+n_2\boldsymbol{J_2}+n_3\boldsymbol{J_3})
\end{equation}
Let $\varepsilon=\mathrm{d}\phi$ and denote $\boldsymbol{\vec{J}}=
\begin{pmatrix}
\boldsymbol{J_1}&\boldsymbol{J_2}&\boldsymbol{J_3}
\end{pmatrix}^\mathrm{T}$, then
\begin{equation}
\boldsymbol{J}=\boldsymbol{\hat{n}}\cdot\boldsymbol{\vec{J}}
\end{equation}

With the norm, equation (\ref{eqt:otej}) goes to
\begin{equation}\label{eqt:ofj}
\mathbf{\Omega}=\dot{\phi}\boldsymbol{J}
\end{equation}
That's no surprise. Substitute the formula into (\ref{eqt:ooj}) and we will get $\dot{\phi}\boldsymbol{\hat{n}}\cdot\boldsymbol{\vec{J}}=\boldsymbol{\omega}\cdot\boldsymbol{\vec{J}}$, which actually corresponds to the definition of the angular velocity vector $\boldsymbol{\omega}=\dot{\phi}\boldsymbol{\hat{n}}$.

\subsection{Lie algebras of SO(3) Group}
\subsubsection{Matrix commutators of SO(3) Group}
Calculate commutators of SO(3) Group\footnote{Readers can know this is precise definition for Lie algebras according to the convention taken by most literature.}
\begin{equation}
[\boldsymbol{J_1},\boldsymbol{J_2}]=
\begin{pmatrix}
0&0&0\\0&0&-1\\0&1&0
\end{pmatrix}
\begin{pmatrix}
0&0&1\\0&0&0\\-1&0&0
\end{pmatrix}
-
\begin{pmatrix}
0&0&1\\0&0&0\\-1&0&0
\end{pmatrix}
\begin{pmatrix}
0&0&0\\0&0&-1\\0&1&0
\end{pmatrix}
=
\begin{pmatrix}
0&-1&0\\1&0&0\\0&0&0
\end{pmatrix}
=\boldsymbol{J_3}
\end{equation}
\begin{equation}
[\boldsymbol{J_2},\boldsymbol{J_3}]=
\begin{pmatrix}
0&0&1\\0&0&0\\-1&0&0
\end{pmatrix}
\begin{pmatrix}
0&-1&0\\1&0&0\\0&0&0
\end{pmatrix}
-
\begin{pmatrix}
0&-1&0\\1&0&0\\0&0&0
\end{pmatrix}
\begin{pmatrix}
0&0&1\\0&0&0\\-1&0&0
\end{pmatrix}
=
\begin{pmatrix}
0&0&0\\0&0&-1\\0&1&0
\end{pmatrix}
=\boldsymbol{J_1}
\end{equation}
\begin{equation}
[\boldsymbol{J_3},\boldsymbol{J_1}]=
\begin{pmatrix}
0&-1&0\\1&0&0\\0&0&0
\end{pmatrix}
\begin{pmatrix}
0&0&0\\0&0&-1\\0&1&0
\end{pmatrix}
-
\begin{pmatrix}
0&0&0\\0&0&-1\\0&1&0
\end{pmatrix}
\begin{pmatrix}
0&-1&0\\1&0&0\\0&0&0
\end{pmatrix}
=
\begin{pmatrix}
0&0&1\\0&0&0\\-1&0&0
\end{pmatrix}=\boldsymbol{J_2}
\end{equation}

\subsubsection{Exponential mapping of SO(3) Group}
As Section~\ref{sec:generator} shows, each generator corresponds to an exponential mapping. In order to highlight the meaning of the parameter $\mu$ correspondent in the mapping of SO(3), let's calculate this mapping again from the beginning.

Consider an element (able to express any rotation indeed) $\mathbf{R}(-\phi,\boldsymbol{\hat{n}})$. The expression include any rotation, indeed (See in~\ref{sec:eig}). Suppose dividing the finite rotation into $n$ identical portions\cite{gao2024}.

Take the limit of the process above, i.e.
\begin{equation}
\mathbf{R}(-\phi,\boldsymbol{\hat{n}})=\lim\limits_{n\to\infty}(\mathbf{I}+\frac{\phi}{n}\boldsymbol{\hat{n}}\cdot\boldsymbol{\vec{J}})^n=\mathrm{e}^{\phi\boldsymbol{\hat{n}}\cdot\boldsymbol{\vec{J}}}
\end{equation}
and one can write the exponential mapping under passive viewpoint
\begin{equation}
\mathbf{R}(\phi,\boldsymbol{\hat{n}})=\mathrm{e}^{-\phi\boldsymbol{\hat{n}}\cdot\boldsymbol{\vec{J}}}
\end{equation}
According to Euler's Theorem, SO(3) exponential mappings cover all its elements, which is a good property that makes expression of SO(3) Group convenient to readers.
\subsection{Euler's equations for rigid body kinematics}\label{sec:eerbk}
In this subsection, formulae are formulated not only in the inertial frame but also in the body-attached frame, because practically sometimes the torque is given in the body-attached frame, which makes writing Euler's equations in that frame convenient for further calculations.
\subsubsection{Formulation using Euler angles}\label{sec:formeuler}
Actually, as is mentioned in Section~\ref{sec:commt}, angular velocities are pseudo-vectors and commute with each other. Then a rigid body's angular velocity $\boldsymbol{\omega}$ is a sum can be expressed by the derivatives to time of Euler angles. With the geometric relationship shown in Figure~\ref{fig:eulerangles} analyzing projections (or with correspondent active rotation matrix for each axis), calculate the angular velocities caused by $\dot{\varphi}$, $\dot{\theta}$, and $\dot{\psi}$ respectively
\[
\begin{cases}
\boldsymbol{\omega}_\varphi=\dot{\varphi}\ \boldsymbol{\hat{z}}\\
\boldsymbol{\omega}_\theta=\dot{\theta}(\cos\varphi\ \boldsymbol{\hat{x}}+\sin\varphi\ \boldsymbol{\hat{y}})\\
\boldsymbol{\omega}_\psi=\dot{\psi}\ (\sin\varphi\sin\theta\ \boldsymbol{\hat{x}}-\cos\varphi\sin\theta\ \boldsymbol{\hat{y}}+\cos\theta\ \boldsymbol{\hat{z}})
\end{cases}
\]
where $\boldsymbol{\hat{x}}$, $\boldsymbol{\hat{y}}$ and $\boldsymbol{\hat{z}}$ are unit vectors of the axes of the \textbf{inertial} frame at that \textbf{fixed moment}. Because $\boldsymbol{\omega}=\boldsymbol{\omega}_\psi+\boldsymbol{\omega}_\theta+\boldsymbol{\omega}_\varphi$,
\begin{equation}
\begin{cases}
\omega_{x}=\dot{\theta}\cos\varphi+\dot{\psi}\sin\theta\sin\varphi\\
\omega_{y}=\dot{\theta}\sin\varphi-\dot{\psi}\sin\theta\cos\varphi\\
\omega_{z}=\dot{\psi}\cos\theta+\dot{\varphi}
\end{cases}
\end{equation}

Likewise, in the body-attached frame\cite{goldstein} (readers can derive it in a similar way or just with the basis transformation matrix),
\begin{equation}
\begin{cases}
\omega_{x'}=\dot{\varphi}\sin\theta\sin\psi+\dot{\theta}\cos\psi\\
\omega_{y'}=\dot{\varphi}\sin\theta\cos\psi-\dot{\theta}\sin\psi\\
\omega_{z'}=\dot{\varphi}\cos\theta+\dot{\psi}
\end{cases}
\end{equation}

\subsubsection{Formulation using Tait-Bryan angles}
We can also derive the angular velocities caused by time-derivatives of Tait-Bryan angles $\dot{\varphi}$, $\dot{\theta}$, and $\dot{\psi}$
\[
\begin{cases}
\boldsymbol{\omega}_\varphi=\dot{\varphi}\ \boldsymbol{\hat{z}}\\
\boldsymbol{\omega}_\theta=\dot{\theta}(-\sin\varphi\ \boldsymbol{\hat{x}}+\cos\varphi\ \boldsymbol{\hat{y}})\\
\boldsymbol{\omega}_\psi=\dot{\psi}(\cos\theta\cos\varphi\ \boldsymbol{\hat{x}}+\cos\theta\sin\varphi\ \boldsymbol{\hat{y}}-\sin\theta\ \boldsymbol{\hat{z}})
\end{cases}
\]

So,
\begin{equation}
\begin{cases}
\omega_{x}=\dot{\psi}\cos\theta\cos\varphi-\dot{\theta}\sin\varphi\\
\omega_{y}=\dot{\psi}\cos\theta\sin\varphi+\dot{\theta}\cos\varphi\\
\omega_{z}=\dot{\varphi}-\dot{\psi}\sin\theta
\end{cases}
\end{equation}

In the body-attached frame\cite{goldstein},
\begin{equation}
\begin{cases}
\omega_{x'}=-\dot{\varphi}\sin\theta+\dot{\psi}\\
\omega_{y'}=\dot{\varphi}\cos\theta\sin\psi+\dot{\theta}\cos\psi\\
\omega_{z'}=\dot{\varphi}\cos\theta\cos\psi-\dot{\theta}\sin\psi
\end{cases}
\end{equation}

\subsection{Further physical meaning of Euler angles and Tait-Bryan angles}

\subsubsection{Relationship between Euler angles and spherical coordinates}\label{sec:rbeaasc}
Imagine a top whose coordinate frame coincides with the inertial frame. Calculate its $z'$-axis in the inertial frame.
\begin{equation}
\boldsymbol{\hat{z}'}=\mathbf{R}(-\theta,-\psi,-\varphi)\boldsymbol{\hat{z}}=
\begin{pmatrix}
\sin\varphi\sin\theta\\
-\cos\varphi\sin\theta\\
\cos\theta
\end{pmatrix}
=
\begin{pmatrix}
\sin\theta\cos(\varphi-\frac{\pi}{2})\\
\sin\theta\sin(\varphi-\frac{\pi}{2})\\
\cos\theta
\end{pmatrix}
\end{equation}
On the other hand, expressed by spherical coordinates mentioned in Section~\ref{sec:degfr}, $\boldsymbol{\hat{z}'}=
\begin{pmatrix}
\sin\Theta\cos\Phi\\
\sin\Theta\sin\Phi\\
\cos\Theta
\end{pmatrix}$.
Thus,
\begin{equation}
\Theta=\theta,\ \Phi=\varphi-\frac{\pi}{2}
\end{equation}
so the first two Euler angles is directly associated with the spherical coordinates of the $z'$-axis of the top, and the method of Euler angles is a realization of the idea mentioned to derive the number of degrees of freedom in Section~\ref{sec:degfr}. These two relationships are also useful when solving problems of rigid body dynamics.

\subsubsection{Gyroscope and Tait-Bryan angles}
Consider a gyroscope, with the initial state below. The rigid body rotation caused by an axis is attached by all the outer axes, but the reverse cannot. Establish a coordinate frame so that the $z$-axis corresponds to the outer axis, and $y$-axis to the middle, $x$-axis to the inner. One can easily deduce that the ($\varphi$, $\theta$, $\psi$) in the picture is actually the ones of Tait-Bryan angles.

\begin{figure}[htbp]
  \centering
  \includegraphics[width=0.5\linewidth]{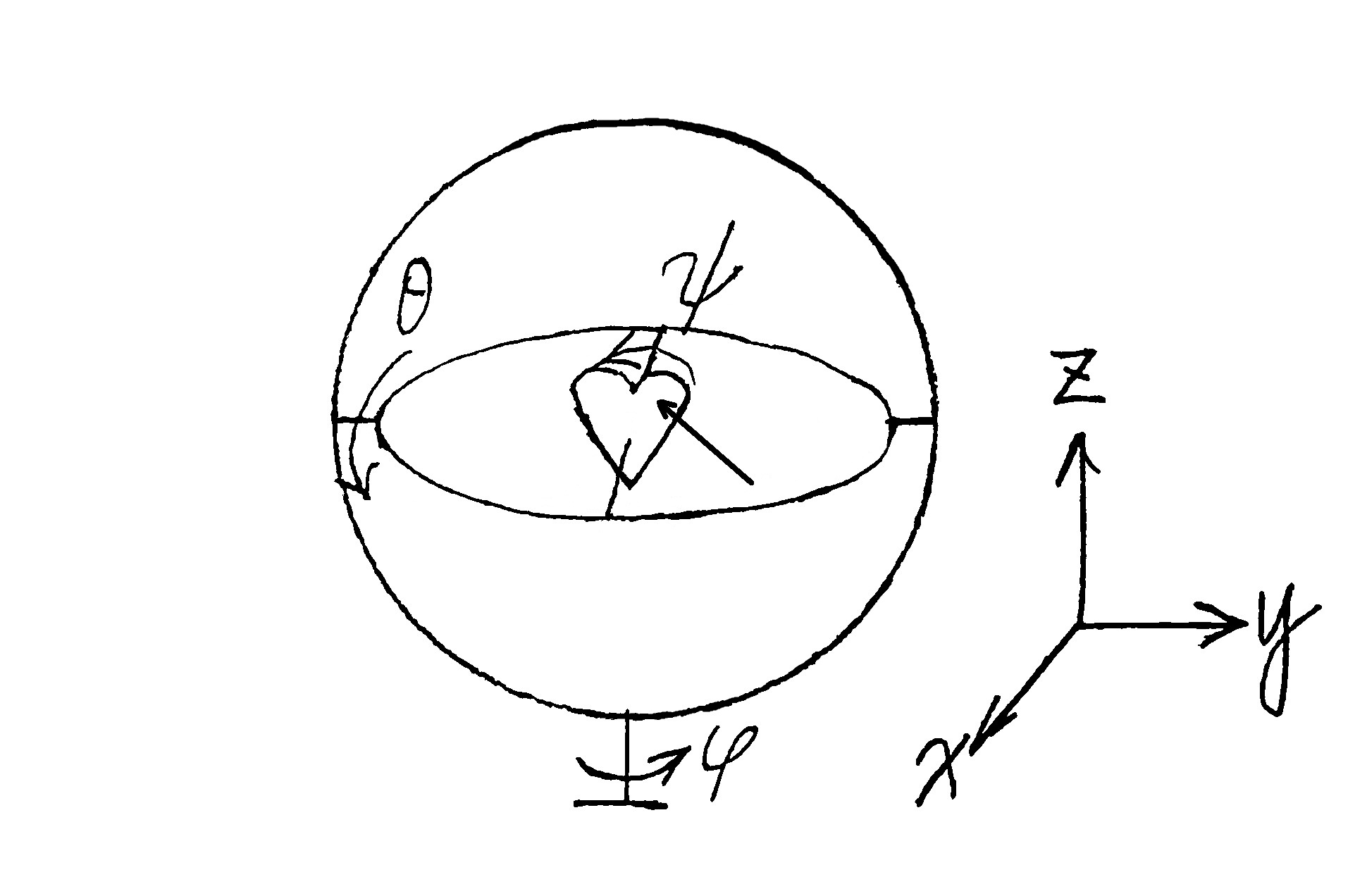}
  \caption{A gyroscope.}
  \label{fig:gyroscope}
\end{figure}

\subsubsection{Origin of names of Euler angles and Tait-Bryan angles}
From Section~\ref{sec:rbeaasc}, the angle $\varphi$ characterize the orientation because it and the spherical coordinate $\Theta$ of the top differ by a constant, so it is called the \textbf{precession angle}; $\theta$ is precisely the spherical coordinate $\Phi$ of the top, characterizing its tilt, thus is called the \textbf{notation angle}; the $z'$-axis of the top is independent of the angle $\psi$, so $\psi$ is called the \textbf{revolution angle}.

Similarly, the projection of orientation of the $x'$-axis on the $xOy$-plane is independent of $\theta$ and $\psi$, so it is called the \textbf{orientation angle}; the angle between the $x'$-axis and $xOy$-plane is $\theta$, so $\theta$ is named the \textbf{pitch angle}; the change of $\psi$ does not change the orientation of the $x'$-axis, so the angle is called the \textbf{yaw angle}.
\subsection{The gimbal lock problem}
Consider a possibility for the gyroscope shown in Figure~\ref{fig:gyroscope} that two axes become parallel. Then, the dimension of possible angular velocity reduces to two. Nevertheless, the dimension of all angular velocities under the configuration is still three since this configuration is no different with others. This phenomenon is called gimbal lock.

For any 3 successive rotations, denoted as $\mathbf{R}(-\phi_1,\boldsymbol{\hat{n}_1})$ first, $\mathbf{R}(-\phi_2,\boldsymbol{\hat{n}_2})$ next, and $\mathbf{R}(-\phi_3,\boldsymbol{\hat{n}_3})$ final. The conditions for gimbal lock, i.e. that the time-derivatives of the angle parameters fail to generate some infinitesimal rotations, is two of the constituent rotations share the same or the opposite axes. If the axes of adjacent rotations are't parallel, the condition is that the axes of the first and third rotations are parallel to each other, i.e.
\begin{equation}
\mathbf{R}(-\phi_2,\boldsymbol{\hat{n}_2})\mathbf{R}(-\phi_1,\boldsymbol{\hat{n}_1})\boldsymbol{\hat{n}_3}\ //\ \boldsymbol{\hat{n}_1}
\end{equation}
The gimbal lock always exist regardless of the convention. For Euler angles, the necessary and sufficient condition is $\theta=0$ or $\pi$; for Tait-Bryan angles, that is $\theta=\pm\frac{\pi}{2}$.
\section{Compositions of rotations}\label{sec:compo}
For intuition, rotations will be expressed by the axis and angle in the inertial frame representation under active viewpoint. Readers can easily derive the correspondent formulae in the other 3 scenarios.
\subsection{Composition of finite rotations}
\subsubsection{Composition of two rotations}\label{sec:compo2}
Consider two successive arbitrary rotations $\mathbf{R}(-\phi_1,\boldsymbol{\hat{n}}_1)$ and $\mathbf{R}(-\phi_2,\boldsymbol{\hat{n}}_2)$. The angle between the axes of the rotations is denoted by $\theta$, while the total rotation by $\mathbf{R}(-\Theta,\boldsymbol{\hat{n}})$. From Section~\ref{sec:algso3}, we can choose a convention that $\phi_1,\phi_2\in[0,\pi)$ and $\theta\in[0,\pi]$. In order to derive the expression of $\Theta$ and $\boldsymbol{\hat{n}}$, establish two coordinate frames: the first, $
\begin{pmatrix}
x&y&z
\end{pmatrix}
^\mathrm{T}$, whose $z$-axis coincides with $\boldsymbol{\hat{n}}_1$ and $zOx$-plane contains $\boldsymbol{\hat{n}}_2$; the second, $
\begin{pmatrix}
x'&y'&z'
\end{pmatrix}
^\mathrm{T}$, whose $z'$-axis coincides with $\boldsymbol{\hat{n}}_2$ and $y'$-axis coincides with the $y$-axis. This assumption is general.
\begin{figure}[htbp]
  \centering
  \includegraphics[width=0.4\linewidth]{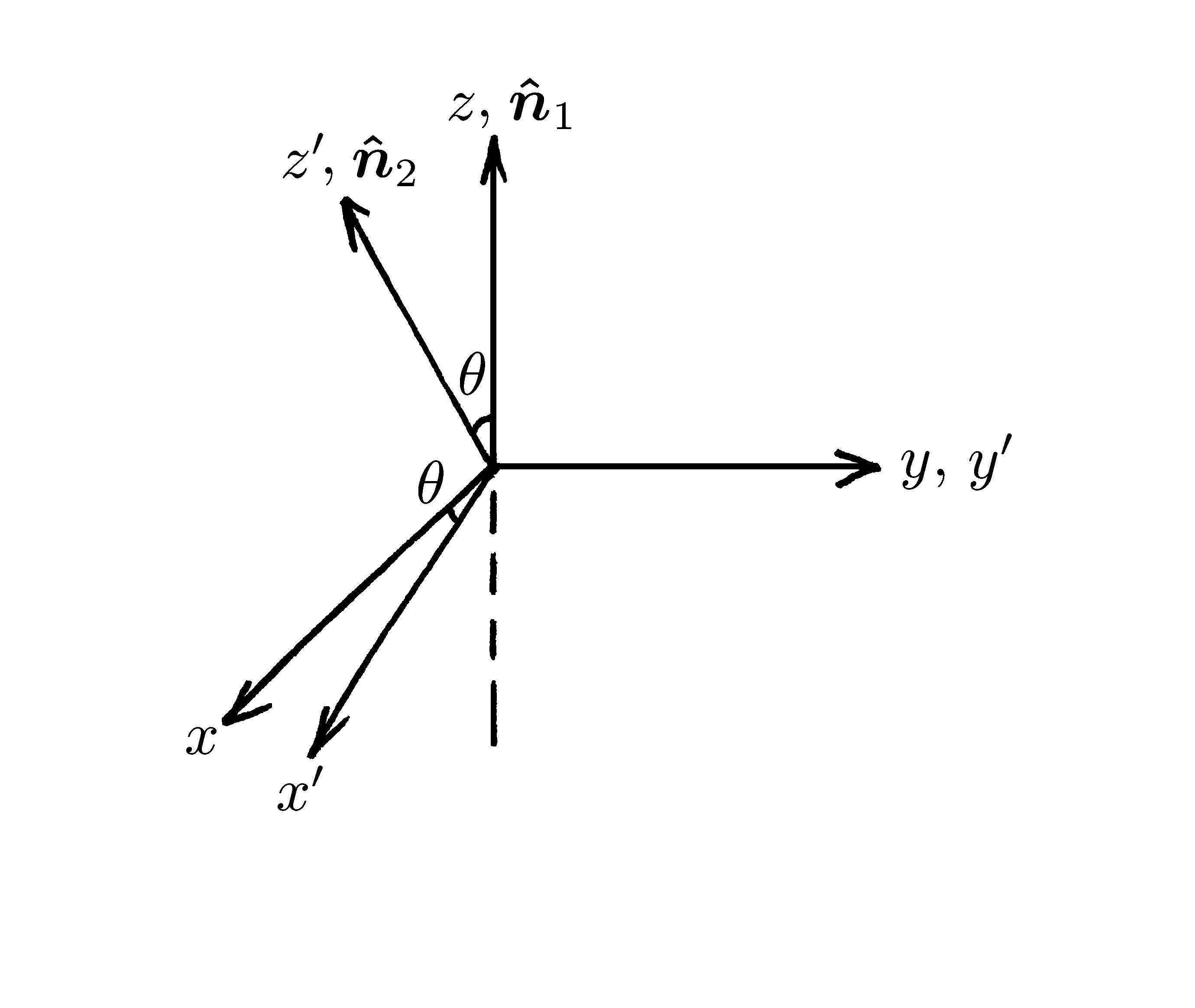}
  \caption{Composition of rotations.}
  \label{fig:compo}
\end{figure}

Written in mathematical language, $\boldsymbol{\hat{n}}_1=\boldsymbol{\hat{z}}$, $\boldsymbol{\hat{n}}_2=\boldsymbol{\hat{z}'}=\cos\theta\ \boldsymbol{\hat{z}}+\sin\theta\ \boldsymbol{\hat{x}}$. On the other hand, $\boldsymbol{\hat{y}'}=\boldsymbol{\hat{y}}$, so $\boldsymbol{\hat{x}'}=-\sin\theta\ \boldsymbol{\hat{z}}+\cos\theta\ \boldsymbol{\hat{x}}$. We can thus determine the basis transformation matrix.
\begin{equation}
\begin{pmatrix}
\boldsymbol{\hat{x}'} & \boldsymbol{\hat{y}'} & \boldsymbol{\hat{z}'}
\end{pmatrix}
=
\begin{pmatrix}
\boldsymbol{\hat{x}} & \boldsymbol{\hat{y}} & \boldsymbol{\hat{z}}
\end{pmatrix}
\begin{pmatrix}
\cos\theta&0&\sin\theta\\0&1&0\\-\sin\theta&0&\cos\theta
\end{pmatrix}
=
\begin{pmatrix}
\boldsymbol{\hat{x}} & \boldsymbol{\hat{y}} & \boldsymbol{\hat{z}}
\end{pmatrix}
\mathbf{R}(-\theta,\boldsymbol{\hat{y}})
\end{equation}
i.e. 
\[
\begin{pmatrix}
x \\ y \\ z
\end{pmatrix}
=\mathbf{R}(-\theta,\boldsymbol{\hat{y}})
\begin{pmatrix}
x' \\ y' \\ z'
\end{pmatrix}
\]
So, $\mathbf{R}(-\phi_1,\boldsymbol{\hat{n}}_1)=\mathbf{R}(-\phi_1,\boldsymbol{\hat{z}})$, $\mathbf{R}(-\phi_2,\boldsymbol{\hat{n}}_2)=\mathbf{R}(-\theta,\boldsymbol{\hat{y}})\mathbf{R}(-\phi_2,\boldsymbol{\hat{z}})\mathbf{R}^{-1}(-\theta,\boldsymbol{\hat{y}})\mathbf{R}(-\phi_1,\boldsymbol{\hat{z}})$. Then
\begin{adjustwidth}{-5.5em}{}
{\fontsize{7pt}{9pt}\selectfont
\begin{align*}
&\ \ \ \ \mathbf{R}(-\Theta,\boldsymbol{\hat{n}})=\mathbf{R}(-\phi_2,\boldsymbol{\hat{n}}_2)\mathbf{R}(-\phi_1,\boldsymbol{\hat{n}}_1)=\mathbf{R}(-\theta,\boldsymbol{\hat{y}})\mathbf{R}(-\phi_2,\boldsymbol{\hat{z}})\mathbf{R}(\theta,\boldsymbol{\hat{y}})\mathbf{R}(-\phi_1,\boldsymbol{\hat{z}})\\
&=\begin{pmatrix}
\cos\theta&0 &\sin\theta\\
0&1&0 \\
-\sin\theta &0& \cos\theta
\end{pmatrix}
\begin{pmatrix}
\cos\phi_2& -\sin\phi_2 &0 \\
\sin\phi_2 & \cos\phi_2&0  \\
0&0&1
\end{pmatrix}
\begin{pmatrix}
\cos\theta&0 &-\sin\theta\\
0&1&0 \\
\sin\theta &0& \cos\theta
\end{pmatrix}
\begin{pmatrix}
\cos\phi_1& -\sin\phi_1 &0 \\
\sin\phi_1 & \cos\phi_1&0  \\
0&0&1
\end{pmatrix}\\
&=
\begin{pmatrix}
\cos^2\theta\cos\phi_2\cos\phi_1-\cos\theta\sin\phi_2\sin\phi_1+\sin^2\theta\cos\phi_1& -\cos^2\theta\cos\phi_2\sin\phi_1-\cos\theta\sin\phi_2\cos\phi_1-\sin^2\theta\sin\phi_1 &-\cos\theta\sin\theta\cos\phi_2+\cos\theta\sin\theta\\
\cos\theta\sin\phi_2\cos\phi_1+\cos\phi_2\sin\phi_1 & -\cos\theta\sin\phi_2\sin\phi_1+\cos\phi_2\cos\phi_1&-\sin\theta\sin\phi_2\\
-\cos\theta\sin\theta\cos\phi_2\cos\phi_1+\sin\theta\sin\phi_2\sin\phi_1+\cos\theta\sin\theta\cos\phi_1 & \cos\theta\sin\theta\cos\phi_2\sin\phi_1+\sin\theta\sin\phi_2\cos\phi_1-\cos\theta\sin\theta\sin\phi_1 &\sin^2\theta\cos\phi_2+\cos^2\theta
\end{pmatrix}
\end{align*}
}
\end{adjustwidth}
\begin{equation}
\relax
\end{equation}

The characteristic equation with the eigenvalue 1 is
\begin{adjustwidth}{-6em}{}
{\fontsize{7pt}{9pt}\selectfont
\[
\begin{pmatrix}
\cos^2\theta\cos\phi_2\cos\phi_1-\cos\theta\sin\phi_2\sin\phi_1+\sin^2\theta\cos\phi_1-1& -\cos^2\theta\cos\phi_2\sin\phi_1-\cos\theta\sin\phi_2\cos\phi_1-\sin^2\theta\sin\phi_1 &-\cos\theta\sin\theta\cos\phi_2+\cos\theta\sin\theta\\
\cos\theta\sin\phi_2\cos\phi_1+\cos\phi_2\sin\phi_1 & -\cos\theta\sin\phi_2\sin\phi_1+\cos\phi_2\cos\phi_1-1&-\sin\theta\sin\phi_2\\
-\cos\theta\sin\theta\cos\phi_2\cos\phi_1+\sin\theta\sin\phi_2\sin\phi_1+\cos\theta\sin\theta\cos\phi_1 & \cos\theta\sin\theta\cos\phi_2\sin\phi_1+\sin\theta\sin\phi_2\cos\phi_1-\cos\theta\sin\theta\sin\phi_1 &\sin^2\theta\cos\phi_2-\sin^2\theta
\end{pmatrix}
\boldsymbol{r}=\boldsymbol{0}
\]
}
\end{adjustwidth}
\begin{equation}
\relax
\end{equation}
Solve this equation with Mathematica's help
\begin{equation}
\boldsymbol{r}=r_0
\begin{pmatrix}\label{eqt:compoaxis}
1&-\tan\frac{\phi_1}{2}&\frac{\cos\theta+\cot\frac{\phi_2}{2}\tan\frac{\phi_1}{2}}{\sin\theta}
\end{pmatrix}^\mathrm{T}
\end{equation}
where $r_0$ is an arbitrary real number. Then normalize it to get $\boldsymbol{\hat{n}}$, the unit vector of the axis.

We choose a convention that satisfy $\boldsymbol{\hat{n}}=\boldsymbol{\hat{z}}$ when $\theta=0$ because the meaning of the convention is natural, and then
\begin{equation}
\boldsymbol{\hat{n}}=
\begin{pmatrix}
\frac{1}{\sqrt{\frac{1}{\cos^2\frac{\phi_1}{2}}+\left(\frac{\cos\theta+\cot\frac{\phi_2}{2}\tan\frac{\phi_1}{2}}{\sin\theta}\right)^2}}&
-\frac{\tan\frac{\phi_1}{2}}{\sqrt{\frac{1}{\cos^2\frac{\phi_1}{2}}+\left(\frac{\cos\theta+\cot\frac{\phi_2}{2}\tan\frac{\phi_1}{2}}{\sin\theta}\right)^2}}&
\frac{\frac{\cos\theta+\cot\frac{\phi_2}{2}\tan\frac{\phi_1}{2}}{\sin\theta}}{\sqrt{\frac{1}{\cos^2\frac{\phi_1}{2}}+\left(\frac{\cos\theta+\cot\frac{\phi_2}{2}\tan\frac{\phi_1}{2}}{\sin\theta}\right)^2}}
\end{pmatrix}
^\mathrm{T}
\end{equation}

With 
\begin{numcases}{}
\boldsymbol{\hat{z}}=\boldsymbol{\hat{n}}_1\\ \boldsymbol{\hat{y}}=\frac{\boldsymbol{\hat{n}}_1\times\boldsymbol{\hat{n}}_2}{||\boldsymbol{\hat{n}}_1\times\boldsymbol{\hat{n}}_2||}=\frac{1}{\sin\theta}(\boldsymbol{\hat{n}}_1\times\boldsymbol{\hat{n}}_2)\\ \boldsymbol{\hat{x}}=\boldsymbol{\hat{y}}\times\boldsymbol{\hat{z}}=\frac{1}{\sin\theta}(\boldsymbol{\hat{n}}_1\times\boldsymbol{\hat{n}}_2)\times\boldsymbol{\hat{n}}_1=\frac{1}{\sin\theta}[\boldsymbol{\hat{n}}_2-(\boldsymbol{\hat{n}}_1\cdot\boldsymbol{\hat{n}}_2)\boldsymbol{\hat{n}}_1]=\frac{1}{\sin\theta}\boldsymbol{\hat{n}}_2-\cot\theta\ \boldsymbol{\hat{n}}_1
\end{numcases}
we can write the expression in the covariant form
\begin{adjustwidth}{-4em}{}
\begin{align}
\boldsymbol{\hat{n}}&=\frac{1}{\sqrt{\frac{1}{\cos^2\frac{\phi_1}{2}}+\left(\frac{\cos\theta+\cot\frac{\phi_2}{2}\tan\frac{\phi_1}{2}}{\sin\theta}\right)^2}}\left(\frac{1}{\sin\theta}\boldsymbol{\hat{n}}_2-\cot\theta\boldsymbol{\hat{n}}_1\right)-\frac{\tan\frac{\phi_1}{2}}{\sqrt{\frac{1}{\cos^2\frac{\phi_1}{2}}+\left(\frac{\cos\theta+\cot\frac{\phi_2}{2}\tan\frac{\phi_1}{2}}{\sin\theta}\right)^2}}\frac{1}{\sin\theta}\boldsymbol{\hat{n}}_1\times\boldsymbol{\hat{n}}_2\nonumber\\
&\ \ \ +\frac{\frac{\cos\theta+\cot\frac{\phi_2}{2}\tan\frac{\phi_1}{2}}{\sin\theta}}{\sqrt{\frac{1}{\cos^2\frac{\phi_1}{2}}+\left(\frac{\cos\theta+\cot\frac{\phi_2}{2}\tan\frac{\phi_1}{2}}{\sin\theta}\right)^2}}\boldsymbol{\hat{n}_1}\nonumber\\
&=\frac{1}{\sqrt{\frac{\sin^2\theta}{\cos^2\frac{\phi_1}{2}}+\left(\cos\theta+\cot\frac{\phi_2}{2}\tan\frac{\phi_1}{2}\right)^2}}\left[\cot\frac{\phi_2}{2}\tan\frac{\phi_1}{2}\boldsymbol{\hat{n}}_1+\boldsymbol{\hat{n}}_2-\tan\frac{\phi_1}{2}(\boldsymbol{\hat{n}}_1\times\boldsymbol{\hat{n}}_2)\right]\nonumber\\
&=\frac{1}{\sqrt{\tan^2\frac{\phi_1}{2}+\tan^2\frac{\phi_2}{2}+2\tan\frac{\phi_2}{2}\tan\frac{\phi_1}{2}\cos\theta+\tan^2\frac{\phi_2}{2}\tan^2\frac{\phi_1}{2}\sin^2\theta}}\left[\tan\frac{\phi_1}{2}\boldsymbol{\hat{n}}_1+\tan\frac{\phi_2}{2}\boldsymbol{\hat{n}}_2-\tan\frac{\phi_2}{2}\tan\frac{\phi_1}{2}(\boldsymbol{\hat{n}}_1\times\boldsymbol{\hat{n}}_2)\right]\label{eqt:compoangle0}
\end{align}
\end{adjustwidth}

Then talk about the direction of the axis of the total transformation. From the expression, the rotation veers towards the direction $\frac{\boldsymbol{\hat{n}}_2\times\boldsymbol{\hat{n}}_1}{||\boldsymbol{\hat{n}}_2\times\boldsymbol{\hat{n}}_1||}\ne0$ from the plane spanned by the axes of the two rotations when $\boldsymbol{\hat{n}}_1$ and $\boldsymbol{\hat{n}}_2$ are not parallel. If the two are parallel, the expression degenerates into (\ref{eqt:glb2}), and there is no veer. (Actually, composition of two rotations is analogous to Wigner rotation in this phenomenon\cite{lu2025}, and readers can consider why.) Then analyze the direction of the projection on the plane spanned by the axes of the two rotations. Apply the formulae $\frac{\tan x}{x}>0$ and $\frac{\mathrm{d}}{\mathrm{d}x}(\frac{\tan x}{x})\geq0$ when $0 \leq x<\frac{\pi}{2}$\footnote{It is proved that if the derivative don't change its sign with only a finite number of zero points on an interval, then the function must be monotonic on that.}. Due to the convention, when $\phi_1>\phi_2$,
$
\frac{\tan\frac{\phi_1}{2}}{\tan\frac{\phi_2}{2}}\Big/\frac{\phi_1}{\phi_2}=\frac{\tan\frac{\phi_1}{2}}{\frac{\phi_1}{2}}\Big/\frac{\tan\frac{\phi_2}{2}}{\frac{\phi_2}{2}}>1
$, and vice versa. Thus, the projection leans more towards the rotation axis corresponds to the larger angle, compared with the simplistic equating of the angle to a vector with $\boldsymbol{\phi}=\phi\boldsymbol{\hat{n}}$, which is satisfied regarding infinitesimal rotations.

Furthermore, one can even see the difference, of the composition of two rotations in reverse order, in the axis: the projections on the plane are same while the veers are opposite.

Speaking of the angle, $1+2\cos\Theta=\mathrm{tr}\ \mathbf{R}(-\Theta,\boldsymbol{\hat{n}})=(1+\cos^2\theta)\cos\phi_2\cos\phi_1-2\cos\theta\sin\phi_2\sin\phi_1+\sin^2\theta(\cos\phi_2+\cos\phi_1)+\cos^2\theta$, and then
\begin{equation}\label{eqt:costh}
\cos\Theta=\frac{1}{2}\bigl[(1+\cos^2\theta)\cos\phi_2\cos\phi_1-2\cos\theta\sin\phi_2\sin\phi_1+\sin^2\theta(\cos\phi_2+\cos\phi_1-1)\bigr]
\end{equation}

We have chosen the convention that $\phi_1,\phi_2\in[0,\pi)$, so $\phi_1+\phi_2\in[0,2\pi)$, which will be used next. In order to discuss the range of $\Theta$ under certain circumstances, first calculate the derivative of $\cos\Theta$, from (\ref{eqt:costh})
\begin{align}
\frac{\mathrm{d}\cos\Theta}{\mathrm{d}\theta}&=-\sin\theta\cos\theta\cos\phi_2\cos\phi_1+\sin\theta\sin\phi_2\sin\phi_1+\sin\theta\cos\theta(\cos\phi_2+\cos\phi_1-1)\nonumber\\
&=\sin\theta[\sin\phi_2\sin\phi_1-\cos\theta(1-\cos\phi_2)(1-\cos\phi_1)]
\end{align}
So, if $\cos\Theta$ has any extremum as $\theta$ is increasing, its point must satisfy $\cos\theta=\frac{\sin\phi_2\sin\phi_1}{(1-\cos\phi_2)(1-\cos\phi_1)}$. Solve the inequations $-1<\frac{\sin\phi_2\sin\phi_1}{(1-\cos\phi_2)(1-\cos\phi_1)}<1$, which means the existence of $\cos\Theta$'s extremum. The inequations is equivalent to
\begin{equation}
\cos(\phi_1\pm\phi_2)-\cos\phi_2-\cos\phi_1+1>0
\end{equation}
Make an identity transformation on LHS of the equation
\begin{align}
\mathrm{LHS}&=-2\sin\frac{\phi_1\pm2\phi_2}{2}\sin\frac{\phi_1}{2}-\cos\phi_1+1=2\sin\frac{\phi_1}{2}(\sin\frac{\phi_1}{2}-\sin\frac{\phi_1\pm2\phi_2}{2})\nonumber\\
&=2\sin\frac{\phi_1}{2}(\mp2\cos\frac{\phi_1\pm\phi_2}{2}\sin\frac{\phi_2}{2})=\mp4\sin\frac{\phi_1}{2}\sin\frac{\phi_2}{2}\cos\frac{\phi_1\pm \phi_2}{2}
\end{align}

Considering the domain of $\phi_1+\phi_2$ that $\phi_1+\phi_2\in[0,2\pi)$ and $\phi_1-\phi_2\in(-\pi,\pi)$, the equation for the lower sign must be satisfied, and for the upper sign, the solution is $\phi_1+\phi_2>\pi$, so we analyze $\Theta$ in two cases. Regarding $\cos\Theta=\frac{1}{2}[(1+\cos^2\theta)\cos\phi_2\cos\phi_1-2\cos\theta\sin\phi_2\sin\phi_1+\sin^2\theta(\cos\phi_2+\cos\phi_1-1)]$, one can always let $\Theta\in[0,2\pi)$, then possible $\Theta$-s form 2 branches, $\Theta_1\in[0,\pi)$ and $\Theta_2=2\pi-\Theta_1\in[\pi,2\pi)$.

One can easily find the continuity of the rotation axis, and out of consideration of that of the rotation matrix (we give up the convention (\ref{eqt:convention})), $\Theta$ must keep continuous as $\theta$ is continuously changing, and then $\Theta$'s solution may change the branch only when $\Theta=\pi$, i.e. $\cos\Theta=-1$.

If $\phi_1+\phi_2\leq\pi$, one can see that when $\theta=0$, $\Theta=\phi_1+\phi_2=\arccos\{\frac{1}{2}[(1+\cos^2\theta)\cos\phi_2\cos\phi_1-2\cos\theta\sin\phi_2\sin\phi_1+\sin^2\theta(\cos\phi_2+\cos\phi_1-1)]\}$. The law of composition for $\theta=\pi$ is similar, as we have already known directly. In fact, $\cos\Theta$ is monotonic and cannot change the branch. Thus, for all $\theta$,
\begin{equation}\label{eqt:compoangle}
\Theta=\arccos\Big\{\frac{1}{2}\left[(1+\cos^2\theta)\cos\phi_2\cos\phi_1-2\cos\theta\sin\phi_2\sin\phi_1+\sin^2\theta(\cos\phi_2+\cos\phi_1-1)\right]\Big\}
\end{equation}
and $\Theta$ and $\theta$ constitute a bijection. Substituting $\phi_1,\ \phi_2,\ \theta\to0^+$, we can derive $\frac{\mathrm{d}\cos\Theta}{\mathrm{d}\theta}>0$ generally under $\phi_1+\phi_2<\pi$, so $\Theta$ decreases monotonically with $\theta$ because $\cos\Theta$ decreases monotonically with $\Theta$ when $\Theta\in[0,\pi)$ from this expression. Specifically, from (\ref{eqt:compoangle}), $\Theta$ goes from $\phi_1+\phi_2$ to $|\phi_1-\phi_2|$ as $\theta$ goes from 1 to $-1$, because in (\ref{eqt:compoaxis}), $\Theta$ does not flip when $\theta=\pi$ if $\theta_1>\theta_2$, and does when $\theta_1<\theta_2$. 

If $\phi_1+\phi_2>\pi$, $\Theta=\phi_1+\phi_2\in(\pi,2\pi)$ also satisfies when $\theta=0$; and for $\theta=\pi$, as we mentioned in the discussion for the previous case, $\Theta=|\phi_1-\phi_2|\in[0,\pi)$. Then $\Theta$ must change the branch of its value at some $\theta$. For the continuity, this must happen at when $\cos\Theta=-1$, i.e. $\Theta=\pi$. Because there exists only one extremum, only one possibility is remained\textemdash $\Theta$ is monotonic by $\theta$, and $\cos\Theta$ changes its monotonicity at $\theta=\frac{\sin\phi_2\sin\phi_1}{(1-\cos\phi_2)(1-\cos\phi_1)}$ where $\cos\Theta=-1$. Thus,
\begin{adjustwidth}{-5.5em}{}
\begin{equation}
\Theta=
\begin{cases}
2\pi-\arccos\{\frac{1}{2}[(1+\cos^2\theta)\cos\phi_2\cos\phi_1-2\cos\theta\sin\phi_2\sin\phi_1+\sin^2\theta(\cos\phi_2+\cos\phi_1-1)]\},\ \theta<\arccos\frac{\sin\phi_2\sin\phi_1}{(1-\cos\phi_2)(1-\cos\phi_1)}\\
\arccos\{\frac{1}{2}[(1+\cos^2\theta)\cos\phi_2\cos\phi_1-2\cos\theta\sin\phi_2\sin\phi_1+\sin^2\theta(\cos\phi_2+\cos\phi_1-1)]\}\ \ \ \ \ \ \ ,\ \theta\geq\arccos\frac{\sin\phi_2\sin\phi_1}{(1-\cos\phi_2)(1-\cos\phi_1)}
\end{cases}
\end{equation}
\end{adjustwidth}

The fourth-order finite difference between the real value of $\Theta$ and the vectorial sum of $\phi_1$ and $\phi_2$ is indefinite. Thus, the numerical magnitude relationship between them is indefinite.

Because of the exchange symmetry of $\phi_1$ and $\phi_2$ in the expression of $\Theta$, the rotation angle of composition of two rotations is independent of the order of the two rotations; however, the rotation axis is not. Thus, the associativity of rotation is generally not satisfied. Furthermore, for successive rotations total more than two, the axis and angle of the total transformation are both subjective to the order of the rotations. In result, the angle of a rotation is neither a real vector nor a pseudo-vector.

When $\phi_1,\ \phi_2\to0^+$, one can see
\begin{numcases}{}
\boldsymbol{\hat{n}}=\frac{1}{\sqrt{{\phi_1}^2+{\phi_2}^2}}\left(\phi_1\boldsymbol{\hat{n}_1}+\phi_2\boldsymbol{\hat{n}_2}+\mathcal{O}((\phi_1,\phi_2)^2)\right)\\
\Theta^2={\phi_1}^2+{\phi_2}^2+2\phi_1\phi_2\cos\theta+\mathcal{O}((\phi_1,\phi_2)^4)
\end{numcases}
That's the vectorial character of infinitesimal angles.

Write the results together as

\noindent 1)For $\phi_1+\phi_2\leq\pi$,
\begin{align}
&\mathbf{R}\Big(-\arccos\Big\{\frac{1}{2}[(1+\cos^2\theta)\cos\phi_2\cos\phi_1-2\cos\theta\sin\phi_2\sin\phi_1+\sin^2\theta(\cos\phi_2+\cos\phi_1-1)]\Big\},\nonumber\\
&\ \ \ \ \ \ \ \ \ \ \ \ \ \ \ \ \ \ \ \ \ \ \ \ \ \ \ \ \ \frac{\tan\frac{\phi_1}{2}\boldsymbol{\hat{n}}_1+\tan\frac{\phi_2}{2}\boldsymbol{\hat{n}}_2-\tan\frac{\phi_2}{2}\tan\frac{\phi_1}{2}(\boldsymbol{\hat{n}}_1\times\boldsymbol{\hat{n}}_2)}{\sqrt{\tan^2\frac{\phi_1}{2}+\tan^2\frac{\phi_2}{2}+2\tan\frac{\phi_2}{2}\tan\frac{\phi_1}{2}\cos\theta+\tan^2\frac{\phi_2}{2}\tan^2\frac{\phi_1}{2}\sin^2\theta}}\Bigr)
\end{align}
2)For $\phi_1+\phi_2>\pi$,
\begin{adjustwidth}{-4em}{}
\begin{equation}
\mathbf{R}(-\Theta,\boldsymbol{\hat{n}})=
\begin{cases}
\mathbf{R}\Big(-\Big[2\pi-\arccos\{\frac{1}{2}[(1+\cos^2\theta)\cos\phi_2\cos\phi_1-2\cos\theta\sin\phi_2\sin\phi_1+\sin^2\theta(\cos\phi_2+\cos\phi_1-1)]\}\Big],\\
\ \ \ \ \ \ \ \ \ \ \ \ \ \ \ \ \ \ \ \ \ \frac{\tan\frac{\phi_1}{2}\boldsymbol{\hat{n}}_1+\tan\frac{\phi_2}{2}\boldsymbol{\hat{n}}_2-\tan\frac{\phi_2}{2}\tan\frac{\phi_1}{2}(\boldsymbol{\hat{n}}_1\times\boldsymbol{\hat{n}}_2)}{\sqrt{\tan^2\frac{\phi_1}{2}+\tan^2\frac{\phi_2}{2}+2\tan\frac{\phi_2}{2}\tan\frac{\phi_1}{2}\cos\theta+\tan^2\frac{\phi_2}{2}\tan^2\frac{\phi_1}{2}\sin^2\theta}}\Big),\ \theta<\arccos\frac{\sin\phi_2\sin\phi_1}{(1-\cos\phi_2)(1-\cos\phi_1)}\\
\mathbf{R}\Big(-\arccos\{\frac{1}{2}[(1+\cos^2\theta)\cos\phi_2\cos\phi_1-2\cos\theta\sin\phi_2\sin\phi_1+\sin^2\theta(\cos\phi_2+\cos\phi_1-1)]\},\\
\ \ \ \ \ \ \ \ \ \ \ \ \ \ \ \ \ \ \ \ \ \frac{\tan\frac{\phi_1}{2}\boldsymbol{\hat{n}}_1+\tan\frac{\phi_2}{2}\boldsymbol{\hat{n}}_2-\tan\frac{\phi_2}{2}\tan\frac{\phi_1}{2}(\boldsymbol{\hat{n}}_1\times\boldsymbol{\hat{n}}_2)}{\sqrt{\tan^2\frac{\phi_1}{2}+\tan^2\frac{\phi_2}{2}+2\tan\frac{\phi_2}{2}\tan\frac{\phi_1}{2}\cos\theta+\tan^2\frac{\phi_2}{2}\tan^2\frac{\phi_1}{2}\sin^2\theta}}\Bigr),\ \theta\geq\arccos\frac{\sin\phi_2\sin\phi_1}{(1-\cos\phi_2)(1-\cos\phi_1)}
\end{cases}
\end{equation}
\end{adjustwidth}
With $\theta_1$ and $\theta_2$ fixed and $\theta$ increasing, $\Theta$ decreases monotonically, from $\phi_1+\phi_2$ at $\theta=0$ to $|\phi_1-\phi_2|$ at $\theta=\pi$.
\begin{example}
We can consider a specific example that rotations don't commute to each other.

\noindent Consider two rotations\cite{zhaokaihua}: the first is composed by a initial rotation about the $x$-axis and a following one about the $y$-axis; the second is of the opposite order. The angles are the same:
\begin{align*}
\Theta&=\arccos\left\{\frac{1}{2}\left[(1+\cos^2\frac{\pi}{2})\cos\frac{\pi}{2}\cos\frac{\pi}{2}-2\cos\frac{\pi}{2}\sin\frac{\pi}{2}\sin\frac{\pi}{2}+\sin^2\theta(\cos\frac{\pi}{2}+\cos\frac{\pi}{2}-1)\right]\right\}\\
&=\arccos\left\{\frac{1}{2}\left[(1+0^2)\times0^2-2\times0\times1^2+1^2\times(2\times0-1)\right]\right\}\\
&=\arccos\left(-\frac{1}{2}\right)=\frac{2\pi}{3}
\end{align*}
Speaking of the axes, for the circumstances,
\begin{align*}
\boldsymbol{\hat{n}}&\propto\tan\frac{\pi}{4}\boldsymbol{\hat{x}}+\tan\frac{\pi}{4}\boldsymbol{\hat{y}}\mp\tan\frac{\pi}{4}\tan\frac{\pi}{4}(\boldsymbol{\hat{x}}\times\boldsymbol{\hat{y}})\\
&=\boldsymbol{\hat{x}}+\boldsymbol{\hat{y}}\mp\boldsymbol{\hat{z}}
\end{align*}
Normalize the axis and we get
\[
\boldsymbol{\hat{n}}=\frac{1}{\sqrt{3}}(\boldsymbol{\hat{x}}+\boldsymbol{\hat{y}}\mp\boldsymbol{\hat{z}})
\]
where the minus sign is for the first circumstance and the plus one for the other.

\noindent So,
\[
\begin{cases}
\mathbf{R}\left(-\frac{2\pi}{3},\frac{1}{\sqrt{3}}(\boldsymbol{\hat{x}}+\boldsymbol{\hat{y}}+\boldsymbol{\hat{z}})\right)=\mathbf{R}(-\frac{\pi}{2},\boldsymbol{\hat{x}})\mathbf{R}(-\frac{\pi}{2},\boldsymbol{\hat{y}})\\
\mathbf{R}\left(-\frac{2\pi}{3},\frac{1}{\sqrt{3}}(\boldsymbol{\hat{x}}+\boldsymbol{\hat{y}}-\boldsymbol{\hat{z}})\right)=\mathbf{R}(-\frac{\pi}{2},\boldsymbol{\hat{y}})\mathbf{R}(-\frac{\pi}{2},\boldsymbol{\hat{x}})
\end{cases}
\]
Readers can examine them with matrix multiplications. These expressions can be useful for the analysis for the symmetry of crystals. 
\end{example}
\begin{figure}[htbp]
  \centering
  \includegraphics[width=1\linewidth]{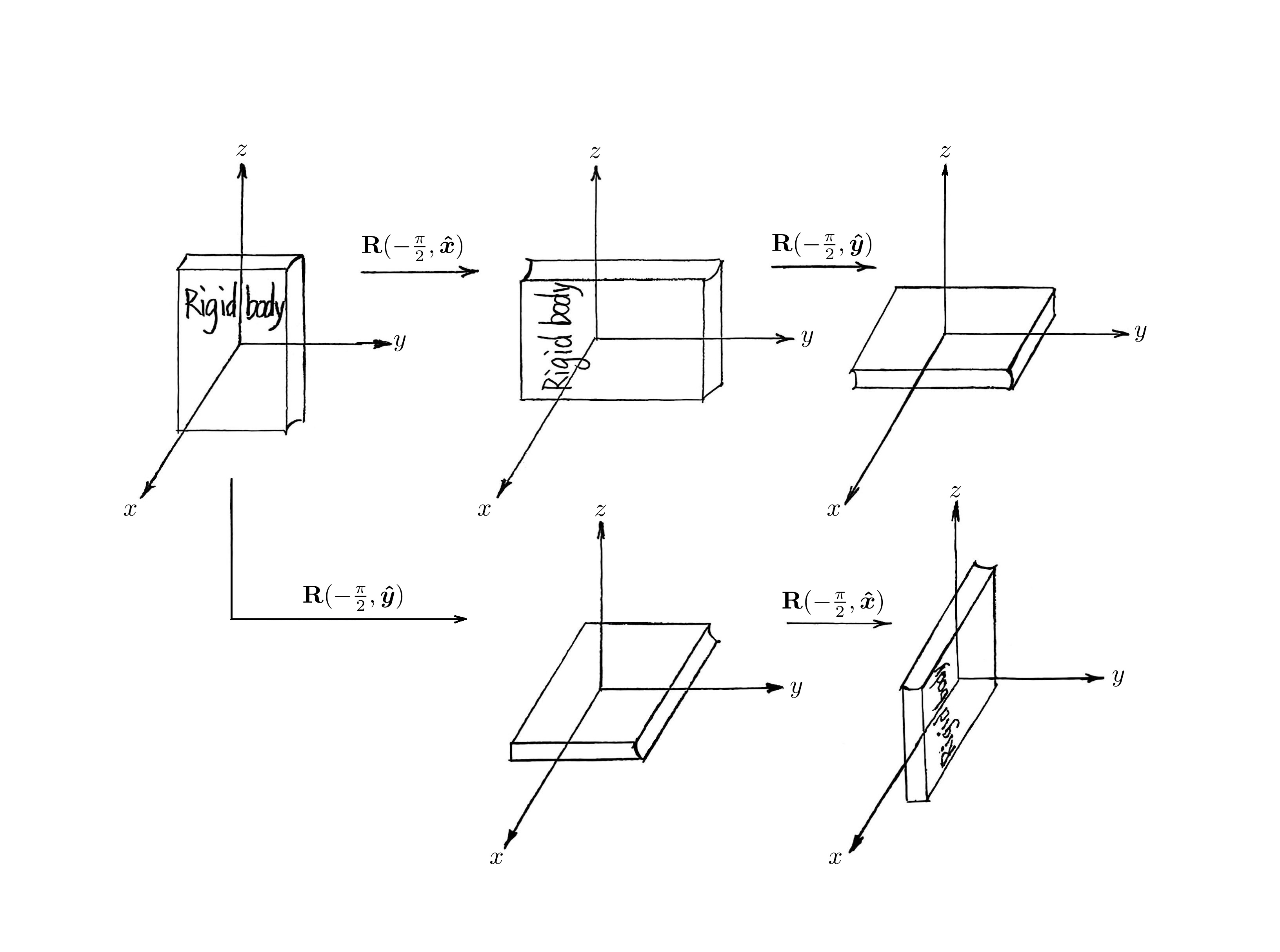}
  \caption{The non-commutative property.}
  \label{fig:noncommt}
\end{figure}
\subsubsection{Composition of several successive rotations}\label{sec:successive}
As Section~\ref{sec:compo2} said, the axis and angle of the total transformation are both subjective to the order of the rotations. It's no wonder that the axis and angle of the total rotation composed by successive rotations have no general formulae. But we can know some properties from the discussion in Section~\ref{sec:compo2}. For instance, from (\ref{eqt:compoangle0}), we can know the rotation with higher angle plays a more significant role compared to vectorial composition, as is mentioned in that subsection.

The most classic example about composition of successive rotations is that of 3 rotations for sure because SO(3) has 3 degrees of freedom.
\begin{example}
Consider using Euler angles to express the rotation about $y$-axis by $\frac{\pi}{2}$. From $\mathrm{(\ref{eqt:eulerlaw})}$ or $\mathrm{(\ref{eqt:eulerlaw1})}$, $\theta=\frac{\pi}{2}$, $\varphi=\frac{\pi}{2}$, $\psi=-\frac{\pi}{2}$, i.e. 
\[
\mathbf{R}(-\frac{\pi}{2},\boldsymbol{\hat{y}})=\mathbf{R}(-\frac{\pi}{2},\boldsymbol{\hat{z}})\mathbf{R}(-\frac{\pi}{2},\boldsymbol{\hat{x}})\mathbf{R}(\frac{\pi}{2},\boldsymbol{\hat{z}})
\]
The expression can be also interpreted as the matrix of total rotation composed by rotations about $z$-axis by $-\frac{\pi}{2}$, about $x$-axis by $\frac{\pi}{2}$, about $z$-axis by $\frac{\pi}{2}$, in sequence.
\end{example}
\subsection{Composition of infinitesimal rotations}
\subsubsection{Composition of countably infinite rotations}\label{sec:countable}
Apply the BCH formula (See in Section~\ref{sec:genset}) to any SO(3) rotation. For $-1<\phi<1$, ignoring terms whose order is higher than $\phi^3$ for $\log(\mathrm{e}^{\phi\boldsymbol{\hat{n}}\cdot\boldsymbol{\vec{J}}})$, so
\begin{adjustwidth}{-3em}{}
\begin{align*}
\mathbf{R}(-\phi,\boldsymbol{\hat{n}})&=\mathrm{e}^{\phi\boldsymbol{\hat{n}}\cdot\boldsymbol{\vec{J}}}\\
&=\mathrm{e}^{\phi n_1\boldsymbol{J_1}+\phi n_2\boldsymbol{J_2}+\phi n_3\boldsymbol{J_3}}\\
&=\mathrm{e}^{\phi n_1\boldsymbol{J_1}+\phi n_2\boldsymbol{J_2}}\mathrm{e}^{\phi n_3\boldsymbol{J_3}}\mathrm{e}^{-\frac{1}{2}[\phi n_1\boldsymbol{J_1}+\phi n_2\boldsymbol{J_2},\phi n_3\boldsymbol{J_3}]}\mathrm{e}^{\frac{1}{6}[\phi n_1\boldsymbol{J_1}+\phi n_2\boldsymbol{J_2},[\phi n_1\boldsymbol{J_1}+\phi n_2\boldsymbol{J_2},\phi n_3\boldsymbol{J_3}]]-\frac{1}{3}[[\phi n_1\boldsymbol{J_1}+\phi n_2\boldsymbol{J_2},\phi n_3\boldsymbol{J_3}],\phi n_3\boldsymbol{J_3}]}\dots
\end{align*}
\end{adjustwidth}
Similarly,
\begin{align*}
\mathrm{e}^{\phi n_1\boldsymbol{J_1}+\phi n_2\boldsymbol{J_2}}&=\mathrm{e}^{\phi n_1\boldsymbol{J_1}}\mathrm{e}^{\phi n_2\boldsymbol{J_2}}\mathrm{e}^{-\frac{1}{2}[\phi n_1\boldsymbol{J_1},\phi n_2\boldsymbol{J_2}]}\mathrm{e}^{\frac{1}{6}[\phi n_1\boldsymbol{J_1},[\phi n_1\boldsymbol{J_1},\phi n_2\boldsymbol{J_2}]]-\frac{1}{3}[[\phi n_1\boldsymbol{J_1},\phi n_2\boldsymbol{J_2}],\phi n_2\boldsymbol{J_2}]}\dots\\
&=\mathrm{e}^{\phi n_1\boldsymbol{J_1}}\mathrm{e}^{\phi n_2\boldsymbol{J_2}}\mathrm{e}^{-\frac{1}{2}\phi^2 n_1n_2\boldsymbol{J_3}}\mathrm{e}^{-\frac{1}{6}\phi^3{n_1}^2n_2\boldsymbol{J_2}}\mathrm{e}^{-\frac{1}{3}\phi^3n_1{n_2}^2\boldsymbol{J_1}}\dots
\end{align*}
Next, it can be derived that
\begin{equation}\label{eqt:commt}
[\phi n_1\boldsymbol{J_1}+\phi n_2\boldsymbol{J_2},\phi n_3\boldsymbol{J_3}]=\phi^2(-n_1n_3\boldsymbol{J_2}+n_2n_3\boldsymbol{J_1})
\end{equation}
with which
\[
\mathrm{e}^{-\frac{1}{2}[\phi n_1\boldsymbol{J_1}+\phi n_2\boldsymbol{J_2},\phi n_3\boldsymbol{J_3}]}=\mathrm{e}^{-\frac{1}{2}(\phi^2n_1n_3[\boldsymbol{J_1},\boldsymbol{J_3}]+\phi^2n_2n_3[\boldsymbol{J_2},\boldsymbol{J_3}])}
=\mathrm{e}^{\frac{1}{2}\phi^2n_1n_3\boldsymbol{J_2}}\mathrm{e}^{-\frac{1}{2}\phi^2n_2n_3\boldsymbol{J_1}}\dots
\]
Last but not least, with (\ref{eqt:commt})
\[
\begin{cases}
[\phi n_1\boldsymbol{J_1}+\phi n_2\boldsymbol{J_2},[\phi n_1\boldsymbol{J_1}+\phi n_2\boldsymbol{J_2},\phi n_3\boldsymbol{J_3}]]=-\phi^3({n_1}^2+{n_2}^2)n_3\boldsymbol{J_3}\\
[[\phi n_1\boldsymbol{J_1}+\phi n_2\boldsymbol{J_2},\phi n_3\boldsymbol{J_3}],\phi n_3\boldsymbol{J_3}]=-\phi^3{n_3}^2(n_1\boldsymbol{J_1}+n_2\boldsymbol{J_2})
\end{cases}
\]
Then,
\begin{align*}
&\ \ \ \ \mathrm{e}^{\frac{1}{6}[\phi n_1\boldsymbol{J_1}+\phi n_2\boldsymbol{J_2},[\phi n_1\boldsymbol{J_1}+\phi n_2\boldsymbol{J_2},\phi n_3\boldsymbol{J_3}]]-\frac{1}{3}[[\phi n_1\boldsymbol{J_1}+\phi n_2\boldsymbol{J_2},\phi n_3\boldsymbol{J_3}],\phi n_3\boldsymbol{J_3}]}\\
&=\mathrm{e}^{-\frac{1}{6}\phi^3({n_1}^2+{n_2}^2)n_3\boldsymbol{J_3}}\mathrm{e}^{\frac{1}{3}\phi^3{n_3}^2n_1\boldsymbol{J_1}}\mathrm{e}^{\frac{1}{3}\phi^3{n_3}^2n_2\boldsymbol{J_2}}\dots
\end{align*}
In conclusion,
\begin{align}
&\mathbf{R}(-\phi,\boldsymbol{\hat{n}})=\mathrm{e}^{\phi n_1\boldsymbol{J_1}}\mathrm{e}^{\phi n_2\boldsymbol{J_2}}\mathrm{e}^{-\frac{1}{2}\phi^2 n_1n_2\boldsymbol{J_3}}\mathrm{e}^{-\frac{1}{6}\phi^3{n_1}^2n_2\boldsymbol{J_2}}\mathrm{e}^{-\frac{1}{3}\phi^3n_1{n_2}^2\boldsymbol{J_1}}\dots\mathrm{e}^{\phi n_3\boldsymbol{J_3}}\mathrm{e}^{\frac{1}{2}\phi^2n_1n_3\boldsymbol{J_2}}\mathrm{e}^{-\frac{1}{2}\phi^2n_2n_3\boldsymbol{J_1}}\cdot\nonumber\\
&\quad\quad\quad\quad\quad\quad\quad\dots\mathrm{e}^{-\frac{1}{6}\phi^3({n_1}^2+{n_2}^2)n_3\boldsymbol{J_3}}\mathrm{e}^{\frac{1}{3}\phi^3{n_3}^2n_1\boldsymbol{J_1}}\mathrm{e}^{\frac{1}{3}\phi^3{n_3}^2n_2\boldsymbol{J_2}}\dots
\end{align}

These ``\dots'' represent terms the order of whose logarithm is higher than 3. Ignore them and we will get an approximation of any rotation whose angle is in $(-1,1)$; or take the order to infinity in a similar approach so we will get a precise expression with countable infinite matrices in principle. In fact, any rotation can be decomposed into those with $\phi\in(-1,1)$; thus, this method is general.

This formulation serves as an addition to a rotation's generation with the generating set's elements.

\subsubsection{Integral of continuous infinitesimal rotations}\label{sec:intinf}
Review formula (\ref{eqt:angv})
\begin{equation}\label{eqt:angv2}
\frac{\mathrm{d}\mathbf{M}(t)}{\mathrm{d}t}=\mathbf{\Omega}(t)\mathbf{M}(t)
\end{equation}
Integrate both sides by $\mathrm{d}t$ and express $\mathbf{M}(t)$ explicitly
\begin{align}
\mathbf{M}(t)&=\mathbf{M}(t_0)+\int_{t_0}^{t}\mathbf{\Omega}(t_1)\mathbf{M}(t_1)\mathrm{d}t_1\nonumber\\
&=\mathbf{M}(t_0)+\int_{t_0}^{t}\mathbf{\Omega}(t_1)\left\{\mathbf{M}(t_0)+\int_{t_0}^{t_1}\mathbf{\Omega}(t_2)\left[\mathbf{M}(t_0)+\int_{t_0}^{t_2}\mathbf{\Omega}(t_3)\mathbf{M}(t_3)\mathrm{d}t_3\right]\mathrm{d}t_2\right\}\mathrm{d}t_1\nonumber\\
&=\dots\nonumber\\
&=\mathbf{M}(t_0)+\sum_{{n=1}}^{\infty}\int_{t_0}^{t}\mathrm{d}t_1\int_{t_0}^{t_1}\mathrm{d}t_2 \dots\int_{t_0}^{t_{n-1}}\mathrm{d}t_n\mathbf{\Omega}(t_1)\mathbf{\Omega}(t_2)\dots\mathbf{\Omega}(t_n)\mathbf{M}(t_0)\nonumber\\
&=\mathbf{M}(t_0)+\sum_{{n=1}}^{\infty}\int_{t_0}^{t}\mathrm{d}t_1\int_{t_0}^{t}\mathrm{d}t_2 \dots\int_{t_0}^{t}\mathrm{d}t_n\Theta(t_1-t_2)\Theta(t_2-t_3)\dots\Theta(t_{n-1}-t_n)\mathbf{\Omega}(t_1)\mathbf{\Omega}(t_2)\dots\mathbf{\Omega}(t_n)\mathbf{M}(t_0)\label{eqt:bianshi}
\end{align}
where $\Theta(t)$ is the Heaviside function\footnote{This kind of method can be seen in \cite{QM}.}. As the number of iterations goes to the infinity, $\mathbf{M}(t_i)$, where $i$ is the maximal possible value of $n$, goes to $\mathbf{M}(t_0)$, thus vanishes. As a result, we have derived an explicit expression of $\mathbf{M}(t)$.

In a special case where $\mathbf{\Omega}(t)$ is a constant matrix $\mathbf{\Omega}$, do Taylor expansion on both sides of (\ref{eqt:angv2}) and get
\begin{equation}
\mathbf{M}(t)=\mathrm{e}^{(t-t_0)\mathbf{\Omega}}
\end{equation}
This equation is equivalent to an exponential mapping since $\mathbf{\Omega} \Delta t=\boldsymbol{J} \Delta \phi$ (Integrated from (\ref{eqt:ofj}) in Section~\ref{sec:nomgen}).

For real equations that need to be solved, the situation is more complex, partly because the angular velocity matrix is determined not only by time but also by the configuration. Furthermore, the equations even turns to second-order equations following Euler's equations for rigid body dynamics. So, generally it is not convenient to use formula (\ref{eqt:bianshi}) or its counterpart in the body-attached frame directly. Physicists tend to transform the integral to those of the parameters to solve continuous rotations, by Euler's equations for rigid body kinematics, as is seen in Section~\ref{sec:eerbk}; they then integrate the equations together with Euler's equations for rigid body dynamics. There are only a few situations where the motion has explicit analytical solutions, which include the motions for asymmetrical free tops, symmetrical tops, Kovalevskaya tops and Goryachev-Chaplygin tops.
\subsection{Diversity of routes to the same destination}
From the examples shown all above, a single rotation can be generated by the multiplication of finite number of finite rotations, e.g. formulation of Euler angles or Tait-Bryan angles, or a single one with the effective axis and angle. It can also be generated by countable infinite number of rotations, e.g. formulation using the BCH formula. Moreover, it can also be generated by continuous rotations, e.g. the ones mentioned in the first possibility, interpreted as continuous ones. All of these show the diversity of generating a certain finite rotation.
\subsection{Composition of a finite rotation and an infinitesimal rotation}
In this subsection we will investigate the time-derivative of a finite rotation, which can be interpreted as an exploration of the composition of a finite rotation and an infinitesimal rotation. The aim is to derive the relationship between the angular velocity and the axis and angle.

Introduce the notation $\phi\in[0,\pi)$ again. Consider an infinitesimal rotation denoted by $\mathbf{R}(-\mathrm{d}\varphi,\boldsymbol{\hat{n}}_\mathrm{i})$ acting on the rotation $\mathbf{R}(-\phi,\boldsymbol{\hat{n}})$. From (\ref{eqt:costh}), $-\sin\phi\mathrm{d}\phi=\mathrm{d}\cos\phi=-\cos\theta\sin\phi\mathrm{d}\varphi+\mathcal{O}(\mathrm{d}\varphi^2)$, i.e. 
\[
\mathrm{d}\phi=\cos\theta\ \mathrm{d}\varphi+\mathcal{O}(\mathrm{d}\varphi^2)
\]
which is equivalent to
\begin{equation}
\dot{\phi}=\dot{\varphi}\cos\theta=\boldsymbol{\omega}\cdot\boldsymbol{\hat{n}}
\end{equation}

Differentiate both sides of (\ref{eqt:compoangle0}) and get 
\[
\mathrm{d}\boldsymbol{\hat{n}}=\Big(\frac{1}{2}\cot\frac{\phi}{2}\Big[\boldsymbol{\hat{n}}_\mathrm{i}-(\boldsymbol{\hat{n}}\cdot\boldsymbol{\hat{n}}_\mathrm{i})\boldsymbol{\hat{n}}\Big]+\frac{1}{2}\boldsymbol{\hat{n}}_\mathrm{i}\times\boldsymbol{\hat{n}}\Big)\mathrm{d}\varphi
\]
The formula is equivalent to 
\begin{equation}
\boldsymbol{\dot{\hat{n}}}=\frac{1}{2}\cot\frac{\phi}{2}\Big[\boldsymbol{\omega}-(\boldsymbol{\omega}\cdot\boldsymbol{\hat{n}})\boldsymbol{\hat{n}}\Big]+\frac{1}{2}\boldsymbol{\omega}\times\boldsymbol{\hat{n}}
\end{equation}

Because $\boldsymbol{\hat{n}}$ is a vector with constant length, there must be $\boldsymbol{\dot{\hat{n}}}\cdot\boldsymbol{\hat{n}}=0$. We can verify that.
\begin{align}
\boldsymbol{\dot{\hat{n}}}\cdot\boldsymbol{\hat{n}}&=\Big\{\frac{1}{2}\cot\frac{\phi}{2}\Big[\boldsymbol{\omega}-(\boldsymbol{\omega}\cdot\boldsymbol{\hat{n}})\boldsymbol{\hat{n}}\Big]+\frac{1}{2}\boldsymbol{\omega}\times\boldsymbol{\hat{n}}\Big\}\cdot\boldsymbol{\hat{n}}\nonumber\\
&=\frac{1}{2}\Big\{\cot\frac{\phi}{2}\Big[\boldsymbol{\omega}\cdot\boldsymbol{\hat{n}}-(\boldsymbol{\omega}\cdot\boldsymbol{\hat{n}})(\boldsymbol{\hat{n}}\cdot\boldsymbol{\hat{n}})\Big]+(\boldsymbol{\omega}\times\boldsymbol{\hat{n}})\cdot\boldsymbol{\hat{n}}\Big\}\nonumber\\
&=0
\end{align}
Similarly,
\begin{align}
\boldsymbol{\dot{\hat{n}}}\times\boldsymbol{\hat{n}}&=\frac{1}{2}\Big[\cot\frac{\phi}{2}(\boldsymbol{\omega}\times\boldsymbol{\hat{n}})+(\boldsymbol{\omega}\times\boldsymbol{\hat{n}})\times\boldsymbol{\hat{n}}\Big]\nonumber\\
&=\frac{1}{2}\Big[\cot\frac{\phi}{2}(\boldsymbol{\omega}\times\boldsymbol{\hat{n}})+(\boldsymbol{\omega}\cdot\boldsymbol{\hat{n}})\boldsymbol{\hat{n}}-\boldsymbol{\omega}\Big]
\end{align}

Then we conversely use $\boldsymbol{\dot{\hat{n}}}$ and $\boldsymbol{\hat{n}}$ to express $\boldsymbol{\omega}$, which will be extremely useful in Section~\ref{sec:angvsu2g}. The result is
\begin{equation}\label{eqt:quaternion}
\boldsymbol{\omega}=\dot{\phi}\boldsymbol{\hat{n}}+\sin\phi\boldsymbol{\dot{\hat{n}}}+(1-\cos\phi)\boldsymbol{\hat{n}}\times\boldsymbol{\dot{\hat{n}}}
\end{equation}
To prove that, we should keep in mind that
\begin{equation}
\cot\frac{\phi}{2}=\frac{\sin\phi}{1-\cos\phi}=\frac{1+\cos\phi}{\sin\phi}
\end{equation}
The RHS of (\ref{eqt:quaternion}) is
\begin{align*}
\mathrm{RHS}&=(\boldsymbol{\omega}\cdot\boldsymbol{\hat{n}})\boldsymbol{\hat{n}}+\frac{1}{2}[(1+\cos\phi)\boldsymbol{\omega}-(1+\cos\phi)(\boldsymbol{\omega}\cdot\boldsymbol{\hat{n}})\boldsymbol{\hat{n}}+\sin\phi(\boldsymbol{\omega}\times\boldsymbol{\hat{n}})]\\
&\ \ \ -\frac{1}{2}[\sin\phi(\boldsymbol{\omega}\times\boldsymbol{\hat{n}})+(1-\cos\phi)(\boldsymbol{\omega}\cdot\boldsymbol{\hat{n}})\boldsymbol{\hat{n}}-(1-\cos\phi)\boldsymbol{\omega}]\\
&=\boldsymbol{\omega}=\mathrm{LHS}
\end{align*}
\section{Description of rotations using SU(2)}\label{sec:descsu2}
In this section, we will try another way to express spatial rotations. We will write coordinates in a matrix form, and derive the corresponding rotation matrices. All of the rotation matrices constitute SU(2) Group. Spinors turn out naturally from SU(2) representation of rotations, whose properties will be explored in Part II. Unlike most literature, this section of the article includes derivations as complete as possible.
\subsection{Matrix of coordinates}
In the previous discussion, one writes position as the coordinate form like $\boldsymbol{r}=
\begin{pmatrix}
x&y&z
\end{pmatrix}
^\mathrm{T}
$. In this section we will explore another equivalent representation of position\cite{liu2024}
\begin{equation}
\boldsymbol{X}=
\begin{pmatrix}
z&x-\mathrm{i}y\\
x+\mathrm{i}y&-z
\end{pmatrix}
\end{equation}

\begin{definition}
An Hermitian matrix $\mathbf{H}$ is one that satisfies $\mathbf{H}^\dagger=\mathbf{H}$.
\end{definition}

\begin{theorem}
There is a one-to-one correspondence between every 3D real vector and every traceless $2\times 2$ Hermitian matrix.
\end{theorem}
\begin{proof}
One denotes the entry in the $i$-th row and $j$-th column of $\mathbf{H}$ by $h_{ij}$.
$\mathbf{H}^\dagger=\mathbf{H}$ equals to
\begin{equation}
\begin{cases}
h_{11}=h_{11}^*\\
h_{22}=h_{22}^*\\
h_{12}=h_{21}^*
\end{cases}
\end{equation}
That is to say that the diagonal elements are real, and the two non-diagonal elements are conjugate to each other. What's more, with the traceless condition, the two diagonal elements are opposite to each other.

Under these conditions, $h_{11}$ and $h_{12}$ are independent. The most general expression is $h_{11}=z,\ h_{12}=x-\mathrm{i}y$, and then $h_{12}=x+\mathrm{i}y$ and $h_{22}=-z$ where $x,\ y,\ z$ are real, and that's the same with the matrix of coordinates $\boldsymbol{X}$.
\end{proof}

\begin{definition}
The tensor of Pauli matrices $\boldsymbol{\vec{\sigma}}=
\begin{pmatrix}
\boldsymbol{\sigma}_1&\boldsymbol{\sigma}_2&\boldsymbol{\sigma}_3
\end{pmatrix}
^\mathrm{T}$
. The components of it is
\begin{equation}
\boldsymbol{\sigma}_1=
\begin{pmatrix}
0&1\\
1&0
\end{pmatrix}
,\
\boldsymbol{\sigma}_2=
\begin{pmatrix}
0&-\mathrm{i}\\
\mathrm{i}&0
\end{pmatrix}
,\
\boldsymbol{\sigma}_3=
\begin{pmatrix}
1&0\\
0&-1
\end{pmatrix}
\end{equation}
\end{definition}

\begin{theorem}
The three Pauli matrices are Hermitian. $($Easy to prove$)$
\end{theorem}

Then

\begin{equation}
\boldsymbol{X}=\boldsymbol{r}\cdot\boldsymbol{\vec{\sigma}}
\end{equation}
Introduce the notation rule $x_1=x$, $x_2=y$, $x_3=z$. Conversely\cite{liu2024},
\begin{equation}
x_i=\frac{1}{2}\mathrm{tr}\ (\boldsymbol{X}\boldsymbol{\sigma}_i)
\end{equation}

There is a bijection between $\boldsymbol{X}$ and $\boldsymbol{r}$. Also, the square of determinant of the coordinates matrix $\boldsymbol{X}$ connects directly with the length of a vector\cite{liu2024}

\begin{equation}
\mathrm{det}\ \boldsymbol{X}=-||\boldsymbol{r}||^2
\end{equation}

The product of vectors corresponding to $\boldsymbol{A}$ and $\boldsymbol{B}$ is
\begin{equation}
\langle \boldsymbol{A},\boldsymbol{B}\rangle=\frac{1}{2}\mathrm{tr}\ (\boldsymbol{A}\boldsymbol{B})
\end{equation}
That's because
\begin{align*}
\frac{1}{2}\operatorname{tr}(\boldsymbol{A}\boldsymbol{B})
&=\frac{1}{2}\operatorname{tr}\left[
\begin{pmatrix}
z_A & x_A-\mathrm{i}y_A \\
x_A+\mathrm{i}y_A & z_A
\end{pmatrix}
\begin{pmatrix}
z_B & x_B-\mathrm{i}y_B \\
x_B+\mathrm{i}y_B & z_B
\end{pmatrix}
\right] \\
&=\frac{1}{2}\Big[\,z_Az_B+x_Ax_B+y_Ay_B
   +\mathrm{i}(x_Ay_B-x_By_A) \\
&\qquad\quad +\,x_Ax_B+y_Ay_B
   +\mathrm{i}(x_By_A-x_Ay_B)+z_Az_B \Big] \\
&=x_Ax_B+y_Ay_B+z_Az_B \\
&=\langle \boldsymbol{A},\boldsymbol{B}\rangle
\end{align*}

\subsection{Rotations about Cartesian axes acting on traceless-Hermitian space}\label{sec:racaithr}
We postulate the existence of an operation which maps the coordinate matrix to the matrix of new coordinates after rotation. Let this matrix $\mathbf{Q}$ be of the form
\begin{equation}
\boldsymbol{X'}=\mathbf{Q}\boldsymbol{X}\mathbf{Q}^\dagger
\end{equation}
This choice of definition is motivated by that $\boldsymbol{X}$ shall fulfill the definition of a complex tensor. Denote $\mathbf{Q}=
\begin{pmatrix}
r_1\mathrm{e}^{\mathrm{i}\theta_1} & r_2\mathrm{e}^{\mathrm{i}\theta_2} \\
r_3\mathrm{e}^{\mathrm{i}\theta_3} & r_4\mathrm{e}^{\mathrm{i}\theta_4}
\end{pmatrix}
$
where $\forall$ $j$, $r_j$ and $\theta_j$ are real.
\begin{definition}
The solution $\mathbf{Q}$ corresponding to $\mathbf{R}(-\phi,\boldsymbol{\hat{n}})$ is denoted as $\mathbf{A}(-\phi,\boldsymbol{\hat{n}})$. Denote the Euler angles of this rotation by $(\varphi,\ \theta,\ \psi)$, then $\mathbf{Q}$ is also denoted as $\mathbf{A}(-\theta, -\psi, -\varphi)$.
\end{definition}
Out of simplicity, first solve the expressions of some simple rotations.
For $\mathbf{R}(-\phi, \boldsymbol{\hat{z}})=
\begin{pmatrix}
\cos\phi& -\sin\phi &0 \\
\sin\phi & \cos\phi&0  \\
0&0&1
\end{pmatrix}
$, $\begin{pmatrix}
x' \\ y' \\ z'
\end{pmatrix}
=\mathbf{R}(-\phi, \boldsymbol{\hat{z}})
\begin{pmatrix}
x \\ y \\ z
\end{pmatrix}
$, so on one hand,
\begin{align*}
\begin{pmatrix}
z'&x'-\mathrm{i}y'\\
x'+\mathrm{i}y'&-z'
\end{pmatrix}
&=
\begin{pmatrix}
z&(x\cos\phi-y\sin\phi)-\mathrm{i}(x\sin\phi+y\cos\phi)\\
(x\cos\phi-y\sin\phi)+\mathrm{i}(x\sin\phi+y\cos\phi)&-z
\end{pmatrix}\\
&=
\begin{pmatrix}
z&\mathrm{e}^{-\mathrm{i}\phi}(x-\mathrm{i}y)\\
\mathrm{e}^{\mathrm{i}\phi}(x+\mathrm{i}y)&-z
\end{pmatrix}
\end{align*}
On the other hand,
\[
\begin{pmatrix}
z'&x'-\mathrm{i}y'\\
x'+\mathrm{i}y'&-z'
\end{pmatrix}
=
\boldsymbol{X'}
=
\mathbf{Q}\boldsymbol{X}\mathbf{Q}^\dagger=
\begin{pmatrix}
r_1\mathrm{e}^{\mathrm{i}\theta_1} & r_2\mathrm{e}^{\mathrm{i}\theta_2} \\
r_3\mathrm{e}^{\mathrm{i}\theta_3} & r_4\mathrm{e}^{\mathrm{i}\theta_4}
\end{pmatrix}
\begin{pmatrix}
z&x-\mathrm{i}y\\
x+\mathrm{i}y&-z
\end{pmatrix}
\begin{pmatrix}
r_1\mathrm{e}^{-\mathrm{i}\theta_1} & r_3\mathrm{e}^{-\mathrm{i}\theta_3} \\
r_2\mathrm{e}^{-\mathrm{i}\theta_2} & r_4\mathrm{e}^{-\mathrm{i}\theta_4}
\end{pmatrix}
=
\begin{pmatrix}
q_{11}&q_{12}\\
q_{21}&q_{22}
\end{pmatrix}
\]
whose elements are
\[
\begin{cases}
q_{11}=({r_1}^2-{r_2}^2)z+2r_1r_2[x\cos(\theta_2-\theta_1)+y\sin(\theta_2-\theta_1)]\\
q_{12}=r_1r_3\mathrm{e}^{\mathrm{i}(\theta_1-\theta_3)}z+r_2r_3\mathrm{e}^{\mathrm{i}(\theta_2-\theta_3)}(x+\mathrm{i}y)+r_1r_4\mathrm{e}^{\mathrm{i}(\theta_1-\theta_4)}(x-\mathrm{i}y)-r_2r_4\mathrm{e}^{\mathrm{i}(\theta_2-\theta_4)}z\\
q_{21}=r_1r_3\mathrm{e}^{\mathrm{i}(\theta_3-\theta_1)}z+r_2r_3\mathrm{e}^{\mathrm{i}(\theta_3-\theta_2)}(x-\mathrm{i}y)+r_1r_4\mathrm{e}^{\mathrm{i}(\theta_4-\theta_1)}(x+\mathrm{i}y)-r_2r_4\mathrm{e}^{\mathrm{i}(\theta_4-\theta_2)}z\\
q_{22}=({r_3}^2-{r_4}^2)z+2r_3r_4[x\cos(\theta_4-\theta_3)+y\sin(\theta_4-\theta_3)]
\end{cases}
\]
By coefficient comparison, $q_{11}$ satisfies
\[
\begin{cases}
{r_1}^2-{r_2}^2=1\\
r_1r_2\cos(\theta_2-\theta_1)=0\\
r_1r_2\sin(\theta_2-\theta_1)=0
\end{cases}
\]
$r_1r_2$ must be zero, then $r_1=0$ or $r_2=0$. If $r_1=0$, ${r_1}^2-{r_2}^2=1$ cannot be true because ${r_1}^2-{r_2}^2<0$. So $r_2=0$ and $r_1=1$.
$q_{22}$ satisfies
\[
\begin{cases}
{r_3}^2-{r_4}^2=-1\\
r_3r_4\cos(\theta_4-\theta_3)=0\\
r_3r_4\sin(\theta_4-\theta_3)=0
\end{cases}
\]
Similarly, $r_3=0$ and $r_4=1$.

It is easy to find that $q_{12}$ and $q_{21}$ are conjugate to each other. Moreover, the non-diagonal elements of $\boldsymbol{X'}$ are conjugate too. Then the $q_{12}$-equation and $q_{21}$-equation are equivalent. So, one can just substitute the non-diagonal elements to the $q_{12}$-equation and derive that
\[
\mathrm{e}^{\mathrm{i}(\theta_1-\theta_4)}(x-\mathrm{i}y)=\mathrm{e}^{-\mathrm{i}\phi}(x-\mathrm{i}y)
\]
Since $x-\mathrm{i}y$ is an arbitrary number, $\mathrm{e}^{\mathrm{i}(\theta_1-\theta_4)}=\mathrm{e}^{-\mathrm{i}\phi}$. Then $\theta_4-\theta_1=\phi+2k\pi$ $(k \in\mathbb{Z})$, and one can let $\theta_4-\theta_1=\phi$. That's the general solution.

However, there is one more degree of freedom due to a global phase difference caused by $\theta_1$. We can choose a gauge
\[
\mathrm{det}\ \mathbf{Q}=1
\]
This means $\mathrm{e}^{\mathrm{i}(2\theta_4-\phi)}=1$, which is equivalent to $2\theta_4-\phi=2k\pi$ $(k\in\mathbb{Z})$. One can choose a convention $\phi\in[0,2\pi)$ so that $\theta_4=\frac{\phi}{2}$ or $\frac{\phi}{2}+\pi$. The correspondent formula of $\theta_1$ is $\theta_1=-\frac{\phi}{2}$ and $\theta_1=-\frac{\phi}{2}+\pi$, respectively. In conclusion, the remaining solutions sum only two
\[
\mathbf{Q}=
\begin{pmatrix}
\mathrm{e}^{-\mathrm{i}\frac{\phi}{2}}&0\\
0&\mathrm{e}^{\mathrm{i}\frac{\phi}{2}}
\end{pmatrix}
\ \mathrm{or}\
\begin{pmatrix}
-\mathrm{e}^{-\mathrm{i}\frac{\phi}{2}}&0\\
0&-\mathrm{e}^{\mathrm{i}\frac{\phi}{2}}
\end{pmatrix}
\]
With $\mathbf{A}(-\phi,\boldsymbol{\hat{z}})\rightarrow\mathbf{I}$ when $\phi\rightarrow0$, $\mathbf{A}(-\phi,\boldsymbol{\hat{z}})=
\begin{pmatrix}
\mathrm{e}^{-\mathrm{i}\frac{\phi}{2}}&0\\
0&\mathrm{e}^{\mathrm{i}\frac{\phi}{2}}
\end{pmatrix}
$.

For $\mathbf{R}(-\phi,\boldsymbol{\hat{x}})=\begin{pmatrix}
1&0 &0 \\
0 & \cos\phi & -\sin\phi \\
0 & \sin\phi & \cos\phi
\end{pmatrix}
$, do coefficient comparison as well. Suppose $\{r_j\}$, $\{\theta_j\}$ are not necessarily positive. With $x'=x$, $y'=y\cos\phi-z\sin\phi$, $z'=y\sin\phi+z\cos\phi$,
\[
\boldsymbol{X'}=
\begin{pmatrix}
y\sin\phi+z\cos\phi&x-\mathrm{i}(y\cos\phi-z\sin\phi)\\
x+\mathrm{i}(y\cos\phi-z\sin\phi)&-(y\sin\phi+z\cos\phi)
\end{pmatrix}
\]
and the $q_{11}$-equations are
\[
\begin{cases}
{r_1}^2-{r_2}^2=\cos\phi\\
2r_1r_2\cos(\theta_2-\theta_1)=0\\
2r_1r_2\sin(\theta_2-\theta_1)=-\sin\phi
\end{cases}
\]
One can choose the convention to let $\theta_2-\theta_1=-\frac{\pi}{2}$, and $2r_1r_2=\sin\phi$. The solution is ${r_1}^2=\cos^2\frac{\phi}{2}$ and ${r_2}^2=\sin^2\frac{\phi}{2}$. One can choose a convention about $\{\theta_j\}$ to let $r_1=\cos\frac{\phi}{2}$, then $r_2=\sin\frac{\phi}{2}$. The $q_{22}$-equations are
\[
\begin{cases}
{r_3}^2-{r_4}^2=-\cos\phi\\
2r_3r_4\cos(\theta_4-\theta_3)=0\\
2r_3r_4\sin(\theta_4-\theta_3)=\sin\phi
\end{cases}
\]
Similarly, $r_4=\cos\frac{\phi}{2}$, $r_3=\sin\frac{\phi}{2}$, $\theta_3-\theta_4=-\frac{\pi}{2}$. Substitute these results into the $q_{12}$-equation and get $\mathrm{e}^{\mathrm{i}(\theta_1-\theta_4)}=1$, so one can choose that $\theta_1=\theta_4$. Follow the gauge $\mathrm{det}\ \mathbf{Q}=1$: $\mathrm{e}^{2\mathrm{i}\theta_1}=1$, so $\mathrm{e}^{\mathrm{i}\theta_1}=\pm1$. Then
\[
\mathbf{Q}=
\begin{pmatrix}
\cos\frac{\phi}{2}&-\mathrm{i}\sin\frac{\phi}{2}\\
-\mathrm{i}\sin\frac{\phi}{2}&\cos\frac{\phi}{2}
\end{pmatrix}
\ \mathrm{or}\
\begin{pmatrix}
-\cos\frac{\phi}{2}&\mathrm{i}\sin\frac{\phi}{2}\\
\mathrm{i}\sin\frac{\phi}{2}&-\cos\frac{\phi}{2}
\end{pmatrix}
\]
With $\mathbf{A}(-\phi,\boldsymbol{\hat{x}})\rightarrow\mathbf{I}$ when $\phi\rightarrow0$, $\mathbf{A}(-\phi,\boldsymbol{\hat{x}})=
\begin{pmatrix}
\cos\frac{\phi}{2}&-\mathrm{i}\sin\frac{\phi}{2}\\
-\mathrm{i}\sin\frac{\phi}{2}&\cos\frac{\phi}{2}
\end{pmatrix}
$.

For $\mathbf{R}(-\phi,\boldsymbol{\hat{y}})=\begin{pmatrix}
\cos\phi&0 &\sin\phi\\
0&1&0 \\
-\sin\phi &0& \cos\phi
\end{pmatrix}
$, with $x'=x\cos\phi+z\sin\phi$, $y'=y$, $z'=-x\sin\phi+z\cos\phi$,
\[
\boldsymbol{X'}=
\begin{pmatrix}
-x\sin\phi+z\cos\phi&x\cos\phi+z\sin\phi-\mathrm{i}y\\
x\cos\phi+z\sin\phi+\mathrm{i}y&x\sin\phi-z\cos\phi
\end{pmatrix}
\]
The $q_{11}$-equations are
\[
\begin{cases}
{r_1}^2-{r_2}^2=\cos\phi\\
2r_1r_2\cos(\theta_2-\theta_1)=-\sin\phi\\
2r_1r_2\sin(\theta_2-\theta_1)=0
\end{cases}
\]
and the $q_{22}$-equations are
\[
\begin{cases}
{r_1}^2-{r_2}^2=-\cos\phi\\
2r_1r_2\cos(\theta_2-\theta_1)=\sin\phi\\
2r_1r_2\sin(\theta_2-\theta_1)=0
\end{cases}
\]
By similar calculation, $r_1=\cos\frac{\phi}{2}$, $r_2=-\sin\frac{\phi}{2}$, $r_3=\sin\frac{\phi}{2}$, $r_4=\cos\frac{\phi}{2}$, $\theta_2=\theta_1$, $\theta_4=\theta_3$. Substitute these equations into the $q_{12}$-equation and get $\mathrm{e}^{\mathrm{i}(\theta_1-\theta_4)}=1$. Similarly, with the gauge $\mathrm{det}\ \mathbf{Q}=1$, $\mathrm{e}^{\mathrm{i}\theta_1}=\mathrm{e}^{\mathrm{i}\theta_4}=\pm1$.
\[
\mathbf{Q}=
\begin{pmatrix}
\cos\frac{\phi}{2}&-\sin\frac{\phi}{2}\\
\sin\frac{\phi}{2}&\cos\frac{\phi}{2}
\end{pmatrix}
\ \mathrm{or}\
\begin{pmatrix}
-\cos\frac{\phi}{2}&\sin\frac{\phi}{2}\\
-\sin\frac{\phi}{2}&-\cos\frac{\phi}{2}
\end{pmatrix}
\]
Then $\mathbf{A}(-\phi,\boldsymbol{\hat{y}})=
\begin{pmatrix}
\cos\frac{\phi}{2}&-\sin\frac{\phi}{2}\\
\sin\frac{\phi}{2}&\cos\frac{\phi}{2}
\end{pmatrix}
$.

One can easily discover that $\mathbf{A}(-\phi,\boldsymbol{\hat{x}})$, $\mathbf{A}(-\phi,\boldsymbol{\hat{y}})$, $\mathbf{A}(-\phi,\boldsymbol{\hat{z}})$ are periodic functions of $\phi$ with period $4\pi$. And we have known
\begin{equation}
\mathbf{A}(-\phi,\boldsymbol{\hat{x}})=
\begin{pmatrix}
\cos\frac{\phi}{2}&-\mathrm{i}\sin\frac{\phi}{2}\\
-\mathrm{i}\sin\frac{\phi}{2}&\cos\frac{\phi}{2}
\end{pmatrix}
,\
\mathbf{A}(-\phi,\boldsymbol{\hat{y}})=
\begin{pmatrix}
\cos\frac{\phi}{2}&-\sin\frac{\phi}{2}\\
\sin\frac{\phi}{2}&\cos\frac{\phi}{2}
\end{pmatrix}
,\
\mathbf{A}(-\phi,\boldsymbol{\hat{z}})=
\begin{pmatrix}
\mathrm{e}^{-\mathrm{i}\frac{\phi}{2}}&0\\
0&\mathrm{e}^{\mathrm{i}\frac{\phi}{2}}
\end{pmatrix}
\end{equation}
For these three matrices, with $\phi$ goes to $\phi+2\pi$, the new coordinates remain the same, while the rotation matrices differ in a minus sign. And one can easily testify that the ``abandoned'' $\mathbf{Q}$-s are actually
\begin{adjustwidth}{-3em}{}
\begin{equation}
\mathbf{A}(-(\phi+2\pi),\boldsymbol{\hat{x}})=
\begin{pmatrix}
-\cos\frac{\phi}{2}&\mathrm{i}\sin\frac{\phi}{2}\\
\mathrm{i}\sin\frac{\phi}{2}&-\cos\frac{\phi}{2}
\end{pmatrix}
,\
\mathbf{A}(-(\phi+2\pi),\boldsymbol{\hat{y}})=
\begin{pmatrix}
-\cos\frac{\phi}{2}&\sin\frac{\phi}{2}\\
-\sin\frac{\phi}{2}&-\cos\frac{\phi}{2}
\end{pmatrix}
,\
\mathbf{A}(-(\phi+2\pi),\boldsymbol{\hat{z}})=
\begin{pmatrix}
-\mathrm{e}^{-\mathrm{i}\frac{\phi}{2}}&0\\
0&-\mathrm{e}^{\mathrm{i}\frac{\phi}{2}}
\end{pmatrix}
\end{equation}
\end{adjustwidth}
And one can write those matrices of passive viewpoint
\begin{equation}
\mathbf{A}(\phi,\boldsymbol{\hat{x}})=
\begin{pmatrix}
\cos\frac{\phi}{2}&\mathrm{i}\sin\frac{\phi}{2}\\
\mathrm{i}\sin\frac{\phi}{2}&\cos\frac{\phi}{2}
\end{pmatrix}
,\
\mathbf{A}(\phi,\boldsymbol{\hat{y}})=
\begin{pmatrix}
\cos\frac{\phi}{2}&\sin\frac{\phi}{2}\\
-\sin\frac{\phi}{2}&\cos\frac{\phi}{2}
\end{pmatrix}
,\
\mathbf{A}(\phi,\boldsymbol{\hat{z}})=
\begin{pmatrix}
\mathrm{e}^{\mathrm{i}\frac{\phi}{2}}&0\\
0&\mathrm{e}^{-\mathrm{i}\frac{\phi}{2}}
\end{pmatrix}
\end{equation}

\subsection{Definition of U(2) Group and SU(2) Group}
\begin{definition}
$\mathbf{A}$ is unitary if $\mathbf{A}^\dagger\mathbf{A}=\mathbf{A}\mathbf{A}^\dagger=\mathbf{I}$.
\end{definition}
\begin{theorem}
A unitary matrix $\mathbf{A}$ satisfies $\mathrm{det}\ \mathbf{A}=\mathrm{e}^{\mathrm{i}\lambda}$ for some real $\lambda$.
\end{theorem}
\begin{proof}
\[
1=\mathrm{det}\ (\mathbf{A}^\dagger\mathbf{A})=(\mathrm{det}\ \mathbf{A})^*\cdot\mathrm{det}\ \mathbf{A}
\]
Then
\[
|\mathrm{det}\ \mathbf{A}|=1
\]
So
\[
\mathrm{det}\ \mathbf{A}=\mathrm{e}^{\mathrm{i}\lambda}
\]
for some real $\lambda$\cite{lu2025}.
\end{proof}
\begin{theorem}
All 2D unitary matrices constitutes a group.
\end{theorem}
\begin{proof}
\
\begin{enumerate}
\item \textbf{Closure:} If $\mathbf{A}$ and $\mathbf{B}$ are both unitary matrices, then
\[
(\mathbf{A}\mathbf{B})^\dagger (\mathbf{A}\mathbf{B})
= \mathbf{B}^\dagger \mathbf{A}^\dagger \mathbf{A} \mathbf{B}
= \mathbf{B}^{-1} \mathbf{A}^{-1} \mathbf{A} \mathbf{B}
= \mathbf{I}
\]
so $\mathbf{A}\mathbf{B}$ is also a unitary matrix.
\item \textbf{Associativity:} By the associativity of matrix multiplication,
\[
(\mathbf{A}\mathbf{B}) \mathbf{C} = \mathbf{A} (\mathbf{B}\mathbf{C})
\]

\item \textbf{Identity element:} By the properties of matrices, there exists a unique identity matrix $\mathbf{I}$.

\item \textbf{Inverse element:} For an orthogonal matrix $\mathbf{A}$, $\mathbf{A}^\dagger$ serves as its inverse, as $\mathbf{A}^\dagger=\mathbf{A}^{-1}$ and $\mathbf{A}^{-1}$ is defined such that
\[
\mathbf{A}^{-1} \mathbf{A} = \mathbf{A} \mathbf{A}^{-1} = \mathbf{I}
\]
\end{enumerate}
\end{proof}
The group is named U(2) Group.

\begin{theorem}
All 2D unitary matrices whose determinant equals to $+1$ constitute a group.
\end{theorem}
\begin{proof}
\
\item \textbf{Closure:} If unit matrices $\mathbf{A},\mathbf{B}$ satisfy $\mathrm{det}\ \mathbf{A}=\mathrm{det}\ \mathbf{B}=1$, then $\mathrm{det}\ (\mathbf{AB})=\mathrm{det}\ \mathbf{A} \cdot \mathrm{det}\ \mathbf{B}=1$. Based on the closure of U(2), unit matrices whose determinant equals to 1 is also closed.
\item \textbf{Other properties:} Following the ones of the U(2) Group.
\end{proof}

The group is named SU(2) Group.

\subsection{Any rotation acting on traceless-Hermitian space}\label{sec:arithr}
\subsubsection{Rotation of spinors}
Consider a special kind of quantity, where a single spatial configuration transforms like vectors to some degree but corresponds to two distinct states (inspired by that a plane has two surfaces), which are characterized by two ``quantum numbers''. Under a 2$\pi$ rotation, this quantity transforms into the other quantum state that corresponds to its initial configuration; it only restores its original state after a complete 4$\pi$ rotation. We will demonstrate that any rotation of this quantity can be described using Euler angles.

First consider any final configuration of coordinates with the initial one. There are two possibilities of state. If the quantum number is the same, the whole rotation can be generated in the way like vectors. For another circumstance, one can give any Euler angle an extra $2\pi$ and keep the others the same. In result, any rotation can be generated by Euler angles.

We have proved that any rotation for vectors can be described this way, and so can that for this quantity, because each Euler angle that belongs to $[2\pi,4\pi)$ can also be described by a matrix uniquely, and the total effect of two matrices of this kind is equivalent to that of the ones whose corresponding Euler angles belong to those minus $2\pi$, in $[0,2\pi)$, i.e. changing the quantum number twice equals no change in it, in which $\mathbf{A}(-(\phi+2\pi),\boldsymbol{\hat{x}})=-\mathbf{A}(-\phi,\boldsymbol{\hat{x}})$, $\mathbf{A}(-(\phi+2\pi),\boldsymbol{\hat{y}})=-\mathbf{A}(-\phi,\boldsymbol{\hat{y}})$, $\mathbf{A}(-(\phi+2\pi),\boldsymbol{\hat{z}})=-\mathbf{A}(-\phi,\boldsymbol{\hat{z}})$ fit.

In addition, for this kind of quantities, each rotation has its axis as well. For the rotations whose final frame share the same quantum number with the initial one, the statement holds true. And the other is equivalent to the former one followed by an $2\pi$ rotation about an arbitrary axis. Then, one adds $2\pi$ to a rotation angle to get the required rotation.

Then we can choose $\varphi\in[0,2\pi)$, $\theta\in[0,\pi)$, $\psi\in[0,4\pi)$.

An SU(2) rotation matrix, which is defined by acting on the traceless-Hermitian space, is capable of representing vectors. Furthermore, the representation on the $\mathcal{H}_0$ space, also brings out a kind of quantities that transforms like that demonstrated in this subsection, named spinors. 
\subsubsection{Result of rotation}\label{sec:anrottrlsrep}
We continue the gauge $\mathrm{det}\ \mathbf{Q}=1$.
\begin{theorem}\label{trm:timesq}
If $\mathbf{Q_1}$, $\mathbf{Q_2}$ respectively correspond to rotation matrices $\mathbf{M_1}$ which maps $(x,\ y,\ z)$ to $(x',\ y',\ z')$, and $\mathbf{M_2}$ which maps $(x',\ y',\ z')$ to $(x'',\ y'',\ z'')$, then $\mathbf{Q_2}\mathbf{Q_1}$ maps $(x,\ y,\ z)$ to $(x'',\ y'',\ z'')$.
\end{theorem}
\begin{proof}
Because $\boldsymbol{X'}=\mathbf{Q_1}\boldsymbol{X}\mathbf{Q_1}^\dagger$, and $\boldsymbol{X''}=\mathbf{Q_2}\boldsymbol{X'}\mathbf{Q_2}^\dagger$
\[
\boldsymbol{X''}=\mathbf{Q_2}(\mathbf{Q_1}\boldsymbol{X}\mathbf{Q_1}^\dagger)\mathbf{Q_2}^\dagger\\
=(\mathbf{Q_2}\mathbf{Q_1})\boldsymbol{X}(\mathbf{Q_2}\mathbf{Q_1})^\dagger
\]
\end{proof}

Any rotation matrix $\mathbf{A}(-\phi, \boldsymbol{\hat{n}})$ can be expressed by the composition of rotations of Euler angles. For an arbitrary rotation, since $\mathbf{R}(-\phi, \boldsymbol{\hat{n}})= \mathbf{R}(-\varphi,\boldsymbol{\hat{z}})\mathbf{R}(-\theta,\boldsymbol{\hat{x}})\mathbf{R}(-\psi,\boldsymbol{\hat{z}})$,
\begin{align}
\mathbf{A}(-\phi, \boldsymbol{\hat{n}})
&=\mathbf{A}(-\theta,-\psi,-\varphi)\nonumber\\&= \mathbf{A}(-\varphi,\boldsymbol{\hat{z}})\mathbf{A}(-\theta,\boldsymbol{\hat{x}})\mathbf{A}(-\psi,\boldsymbol{\hat{z}})\nonumber\\
&=\begin{pmatrix}
\mathrm{e}^{-\mathrm{i}\frac{\varphi}{2}}&0\\
0&\mathrm{e}^{\mathrm{i}\frac{\varphi}{2}}
\end{pmatrix}
\begin{pmatrix}
\cos\frac{\theta}{2}&-\mathrm{i}\sin\frac{\theta}{2}\\
-\mathrm{i}\sin\frac{\theta}{2}&\cos\frac{\theta}{2}
\end{pmatrix}
\begin{pmatrix}
\mathrm{e}^{-\mathrm{i}\frac{\psi}{2}}&0\\
0&\mathrm{e}^{\mathrm{i}\frac{\psi}{2}}
\end{pmatrix}\nonumber\\
&=\begin{pmatrix}
\cos\frac{\theta}{2}\mathrm{e}^{-\mathrm{i}\frac{\varphi+\psi}{2}}&-\mathrm{i}\sin\frac{\theta}{2}\mathrm{e}^{-\mathrm{i}\frac{\varphi-\psi}{2}}\\
-\mathrm{i}\sin\frac{\theta}{2}\mathrm{e}^{\mathrm{i}\frac{\varphi-\psi}{2}}&\cos\frac{\theta}{2}\mathrm{e}^{\mathrm{i}\frac{\varphi+\psi}{2}}
\end{pmatrix}\label{eqt:eulangsu2}
\end{align}
Under passive viewpoint, the rotation matrix is
\begin{equation}
\mathbf{A}(\phi, \boldsymbol{\hat{n}})=\mathbf{A}^{-1}(-\phi, \boldsymbol{\hat{n}})=
\begin{pmatrix}
\cos\frac{\theta}{2}\mathrm{e}^{\mathrm{i}\frac{\psi+\varphi}{2}}&\mathrm{i}\sin\frac{\theta}{2}\mathrm{e}^{\mathrm{i}\frac{\psi-\varphi}{2}}\\
\mathrm{i}\sin\frac{\theta}{2}\mathrm{e}^{-\mathrm{i}\frac{\psi-\varphi}{2}}&\cos\frac{\theta}{2}\mathrm{e}^{-\mathrm{i}\frac{\psi+\varphi}{2}}
\end{pmatrix}
\end{equation}
\begin{theorem}
Each rotation matrix is an $\mathrm{SU(2)}$ matrix.\label{trm:roteqsu2}
\end{theorem}
\begin{proof}
It's known that $\det\ \mathbf{A}(-\varphi,\boldsymbol{\hat{z}})=\det\ \mathbf{A}(-\theta,\boldsymbol{\hat{x}})=\det\ \mathbf{A}(-\psi,\boldsymbol{\hat{z}})=1$ and easy to verify that $\mathbf{A}(-\varphi,\boldsymbol{\hat{z}})\mathbf{A^\dagger}(-\varphi,\boldsymbol{\hat{z}})=\mathbf{I}$, $\mathbf{A}(-\theta,\boldsymbol{\hat{x}})\mathbf{A^\dagger}(-\theta,\boldsymbol{\hat{x}})=\mathbf{I}$, $\mathbf{A}(-\psi,\boldsymbol{\hat{z}})\mathbf{A^\dagger}(-\psi,\boldsymbol{\hat{z}})=\mathbf{I}$, so the 3 matrices belong to SU(2). As a result, the product of them $\mathbf{A}(-\theta,-\psi,-\varphi)$ is an SU(2) matrix.
\end{proof}
With this theorem and properties of matrix operations, we can derive a corollary.
\begin{theorem}
Any rotation matrix $\mathbf{A}(-\phi, \boldsymbol{\hat{n}})$ has its unique inverse.
\end{theorem}

The contents above in this subsubsection talk about rotation matrices in terms of their effect on vectors, and we will distinguish the matrices with the same effect below.
\begin{theorem}\label{trm:pm}
For any vector rotation, if $\mathbf{Q}$ is a rotation matrix, the one and only one another equivalent matrix is $-\mathbf{Q}$.
\end{theorem}
\begin{proof}
Consider two matrices $\mathbf{Q_1}$, $\mathbf{Q_2}$ satisfy
\[
\mathbf{Q_1}\boldsymbol{X}\mathbf{Q_1}^\dagger=\boldsymbol{X'}
\]
\[
\mathbf{Q_2}\boldsymbol{X}\mathbf{Q_2}^\dagger=\boldsymbol{X'}
\]
Then
\[
(\mathbf{Q_2}^{-1}\mathbf{Q_1})\boldsymbol{X}(\mathbf{Q_2}^{-1}\mathbf{Q_1})^\dagger=\boldsymbol{X}
\]
This transformation corresponds to $\mathbf{M}=\mathbf{R}(0,\boldsymbol{\hat{z}})$,
and substitute $\phi=0$ to $\mathbf{A}(-\phi,\boldsymbol{\hat{z}})$ to get the solutions
\[
\mathbf{Q_2}^{-1}\mathbf{Q_1}=
\begin{pmatrix}
  1&0 \\
  0&1
\end{pmatrix}
\ \mathrm{or}\
\begin{pmatrix}
  -1&0 \\
  0&-1
\end{pmatrix}
\]
Because rotation matrices are invertible,
\[
\mathbf{Q_1}=\mathbf{Q_2}\ \mathrm{or}\ \mathbf{Q_1}=-\mathbf{Q_2}
\]
\end{proof}

This means that each SO(3) rotation corresponds to two SU(2) rotations. For vectors, the identity of a rotation is defined by the transformation of ($x$, $y$, $z$), so each rotation can be expressed by two equivalent matrices. However, for spinors, the identity of a rotation is defined by the transformation of ($x$, $y$, $z$) and the sign, so it is characterized by the transformation matrix, as will be shown in Part II.

Then with Theorem~\ref{trm:timesq}, for composite transformation which is composed by $\mathbf{Q_1}$ first and $\mathbf{Q_2}$ next, the total matrix is $\mathbf{Q}=\pm\mathbf{Q_2}\mathbf{Q_1}$. However, for spinors the two cases differ, so it is needed to determine which represents the total transformation.

For small rotations there must be $\mathbf{Q}=\mathbf{Q_2}\mathbf{Q_1}$ considering that a small transformation must be close to the identity. Apply mathematical induction and we will get the expression of any total transformation composed by successive rotations: $\mathbf{Q}=\mathbf{Q_n}\mathbf{Q_{n-1}}\dots\mathbf{Q_2}\mathbf{Q_1}$ where $\mathbf{Q_i}$ is the $i$-th rotation. Then, for the composition of any two finite rotations $\mathbf{Q_1}=\mathbf{A}(-\phi_1,\boldsymbol{\hat{n}_1})$ and $\mathbf{Q_2}=\mathbf{A}(-\phi_2,\boldsymbol{\hat{n}_2})$, we can view both of them as the composition of infinitesimal rotations, thus $
\mathbf{Q_1}=\lim\limits_{n\to\infty} \left[\mathbf{A}(-\frac{\phi_1}{n},\boldsymbol{\hat{n}_1})\right]^n$ and
$
\mathbf{Q_2}=\lim\limits_{n\to\infty} \left[\mathbf{A}(-\frac{\phi_2}{n},\boldsymbol{\hat{n}_2})\right]^n
$. So,
\[
\mathbf{Q}=\lim\limits_{n\to\infty} \left[\mathbf{A}(-\frac{\phi_2}{n},\boldsymbol{\hat{n}_2})\right]^n \lim\limits_{n\to\infty} \left[\mathbf{A}(-\frac{\phi_1}{n},\boldsymbol{\hat{n}_1})\right]^n=\mathbf{Q_2}\mathbf{Q_1}
\]
for any rotation acting on traceless-Hermitian space.

We turn back and emphasize this once again: $\psi$ and $\psi\pm2\pi$ in (\ref{eqt:eulangsu2}) should be regarded as different angles.
\subsection{Physical meanings and properties of U(2) Group and SU(2) Group}\label{sec:phymn}
\subsubsection[Caylay-Klein parameters]{Caylay-Klein parameters\cite{lu2025}}\label{sec:caylay}
We will express any SU(2) Group explicitly. Write an arbitrary SU(2) matrix as
\[
\mathbf{Q}=
\begin{pmatrix}
\alpha&\beta\\
\gamma&\delta
\end{pmatrix}
\]
And $\mathbf{Q}$ must satisfy
\[
1=\mathbf{Q}^\dagger\mathbf{Q}=
\begin{pmatrix}
\alpha^*&\gamma^*\\
\beta^*&\delta^*
\end{pmatrix}
\begin{pmatrix}
\alpha&\beta\\
\gamma&\delta
\end{pmatrix}
=
\begin{pmatrix}
\alpha^*\alpha+\gamma^*\gamma&\alpha^*\beta+\gamma^*\delta\\
\beta^*\alpha+\delta^*\gamma&\beta^*\beta+\delta^*\delta
\end{pmatrix}
\]
Then
\begin{numcases}{}
\alpha^*\alpha+\gamma^*\gamma=1\label{eqt:ck1}\\
\beta^*\beta+\delta^*\delta=1\label{eqt:ck2}\\
\alpha^*\beta+\gamma^*\delta=0\label{eqt:ck3}\\
\alpha\delta-\beta\gamma=1\label{eqt:ck4}
\end{numcases}
From (\ref{eqt:ck3}), $\delta=-\frac{\alpha^*\beta}{\gamma^*}$. Substitute it into (\ref{eqt:ck4}) and multiply by $\gamma$ and get $-\beta(\alpha^*\alpha+\gamma^*\gamma)=\gamma^*$. With (\ref{eqt:ck1}),
\[
\gamma=-\beta^*
\]
Then $\delta=\alpha^*$, and equation (\ref{eqt:ck1})(\ref{eqt:ck2}) goes to $\alpha^*\alpha+\beta^*\beta=1$, so
\begin{equation}\label{eqt:caylayklein}
\mathbf{Q}=
\begin{pmatrix}
\alpha&\beta\\
-\beta^*&\alpha^*
\end{pmatrix}
\ \ \ \ \ \ \ \ \mathrm{subject\ to}\ \ \ \ \ \ \ \ \alpha^*\alpha+\beta^*\beta=1
\end{equation}
Because there are 2 independent complex parameters subject to 1 real condition, there exists $2\times2-1=3$ real parameters, which are called Caylay-Klein parameters.

\subsubsection{The equivalence between SU(2) Group and rotations}
\begin{theorem}\label{trm:su2eqrot}
Each $\mathrm{SU(2)}$ matrix is a rotation matrix.
\end{theorem}
\begin{proof}
Write an arbitrary rotation matrix
\[
\mathbf{A}(-\theta,-\psi,-\varphi)=
\begin{pmatrix}
\cos\frac{\theta}{2}\mathrm{e}^{-\mathrm{i}\frac{\varphi+\psi}{2}}&-\mathrm{i}\sin\frac{\theta}{2}\mathrm{e}^{-\mathrm{i}\frac{\varphi-\psi}{2}}\\
-\mathrm{i}\sin\frac{\theta}{2}\mathrm{e}^{\mathrm{i}\frac{\varphi-\psi}{2}}&\cos\frac{\theta}{2}\mathrm{e}^{\mathrm{i}\frac{\varphi+\psi}{2}}
\end{pmatrix}
\]
On the other hand, for each SU(2) element, denote its entries: $q_{11}=r_1\mathrm{e}^{\mathrm{i}\theta_1}$ and $q_{12}=r_2\mathrm{e}^{\mathrm{i}\theta_2}$. With $q_{11}q_{11}^*+q_{12}q_{12}^*=1$, which is equivalent to ${r_1}^2+{r_2}^2=1$, one can just write $r_1=\cos\frac{\theta}{2}$, so $r_2=\pm \sin\frac{\theta}{2}$. If choosing the minus sign, one can let $\theta\rightarrow4\pi-\theta$ to maintain the possibility of the expression: $r_1=\cos\frac{\theta}{2}$ and $r_2=\sin\frac{\theta}{2}$. From the expression, $\theta_1=-\frac{\varphi+\psi}{2}$ and $\theta_2=-\frac{\varphi-\psi+\pi}{2}$. For any real $\theta_1$, $\theta_2$, there is always a unique solution
\begin{equation}
\begin{cases}
\psi=\theta_2-\theta_1+\frac{\pi}{2}\\
\varphi=-\theta_1-\theta_2-\frac{\pi}{2}
\end{cases}
\end{equation}
And for the expression of $\mathbf{A}(-\theta,-\psi,-\varphi)$, it's simple to verify $q_{22}=q_{11}^*$, $q_{21}=-q_{12}^*$.
So the set of rotation matrices $\{\mathbf{A}(-\theta,-\psi,-\varphi)\}$ contains any SU(2) Group.
\end{proof}

Together with Theorem~\ref{trm:roteqsu2}, The equivalence between SU(2) Group and rotation is established.

Due to the equivalence, for any SU(2) element, as $\boldsymbol{X}$ takes all the traceless Hermitian matrices, $\boldsymbol{X'}$ will also take all of those, so any SU(2) element is an $\mathcal{H}_0\to\mathcal{H}_0$ linear mapping where $\mathcal{H}_0$ is the set of traceless Hermitian matrices.

\subsubsection{The connectivity of SU(2) Group and U(2) Group}
It is obvious to see that SU(2) Group is connective, because each SU(2) matrix can be generated by 3 successive continuous rotations correspondent to 3 Euler angles, which can be realized within SU(2) group. This can be concluded by the following theorem.
\begin{theorem}
$\mathrm{SU(2)}$ Group is a Lie Group.
\end{theorem}

For any U(2) matrix $\mathbf{Q}$ whose determinant is $\mathrm{e}^{\mathrm{i}\lambda}$, $\mathrm{e}^{-\mathrm{i}\frac{\lambda}{2}}\mathbf{Q}$ is a unitary matrix whose determinant is 1, i.e. an SU(2) matrix. Thus, a U(2) matrix can differ from some SU(2) matrix only in a global phase, which can be bridged through a continuous transformation within U(2) Group, because $\forall\alpha\in\mathbb{R}$,
$\begin{pmatrix}
\mathrm{e}^{\mathrm{i}\alpha}&0\\
0&\mathrm{e}^{\mathrm{i}\alpha}
\end{pmatrix}\in \mathrm{U(2)}$. That is to say, U(2) Group is also connective, which can be concluded by the following statement.

\begin{theorem}
$\mathrm{U(2)}$ Group is a Lie Group.
\end{theorem}

\subsubsection{The absence of reflection in U(2) Group}
Actually, for any matrix that belongs to U(2) Group but not in SU(2) Group written as $\mathbf{Q}$ whose determinant is $\mathrm{e}^{\mathrm{i}\lambda}$, denote $\mathrm{e}^{-\mathrm{i}\frac{\lambda}{2}}\mathbf{Q}$ as $\mathbf{\bar{Q}}$, and
\[
\mathbf{Q}\boldsymbol{X}\mathbf{Q}^\dagger=\mathrm{e}^{\mathrm{i}\frac{\lambda}{2}}\mathbf{\bar{Q}}\boldsymbol{X}\mathbf{\bar{Q}}^\dagger\mathrm{e}^{-\mathrm{i}\frac{\lambda}{2}}=\mathbf{\bar{Q}}\boldsymbol{X}\mathbf{\bar{Q}}^\dagger\]
That is to say, each U(2) matrix corresponds to a rotation not considering the gauge. Thus, reflections are absent in U(2) Group.

\subsubsection{Quantitative relationship between an SU(2) matrix and its SO(3) counterpart}
Because SO(3) Group and SU(2) Group both correspond to all rotations, any transformation can be seen as a coordinates transformation
\[
\mathbf{Q}(\boldsymbol{r}\cdot\boldsymbol{\vec{\sigma}})\mathbf{Q}^\dagger=\boldsymbol{r'}\cdot\boldsymbol{\vec{\sigma}}=(\mathbf{M}\boldsymbol{r})\cdot\boldsymbol{\vec{\sigma}}
\]
Write this expression in tensor index notation to avoid ambiguity when doing the derivation
\[\mathbf{Q}x_i\boldsymbol{\sigma}_i\mathbf{Q}^\dagger=M_{ji}x_i\boldsymbol{\sigma}_j\]
Because $\boldsymbol{r}$ is arbitrary,
\begin{equation}
\mathbf{Q}\boldsymbol{\sigma}_i\mathbf{Q}^\dagger=M_{ji}\boldsymbol{\sigma}_j
\end{equation}
Written back following the original notation style, the expression is
\begin{equation}
\mathbf{Q}\boldsymbol{\vec{\sigma}}\mathbf{Q}^\dagger=\mathbf{M}\boldsymbol{\vec{\sigma}}
\end{equation}

The derivations above in this subsubsection follow L\"{u} (2025)\cite{lu2025}. In fact, the vector $\boldsymbol{\vec{\sigma}}$ spanned by the three Pauli matrices is a third-order tensor, with one vector index and two spinor index. $\mathbf{Q}$ acts on one of its $2\times2$-dimension spinor indices, and $\mathbf{M}$ acts on its 3-dimension vector index. Since that the operation like LHS is understood as a rotation in the spinor space, while RHS as that in the vector space, the equation shows that the vector of Pauli matrices is covariant (a real vector).
\subsection{Rotation matrices in different representations}\label{sec:rmidr}
We have derived the formula $\mathbf{B}=\mathbf{M}\mathbf{F}\mathbf{M}^\mathrm{T}$ in Section~\ref{sec:repiner1} for rotations expressed in the inertial representation. But what about that representation transformation with regards to SU(2) elements?

The answer can't be easier. From Section~\ref{sec:anrottrlsrep}, there is the multiplication-preserving relationship between SO(3) Group and SU(2) Group. With $\mathbf{C}$ being the active rotation matrix expressed in the required representation corresponding to $\mathbf{B}$, the active rotation matrix $\mathbf{A}$ corresponding to $\mathbf{F}$, and $\mathbf{Q}$ being the coordinate transformation matrix for SU(2) representation transformation corresponding to $\mathbf{M}$ (thus $\mathbf{Q}^\dagger$ corresponds to $\mathbf{M}^\mathrm{T}$), $\mathbf{C}=\mathbf{Q}\mathbf{A}\mathbf{Q}^\dagger$ because $\mathbf{B}=\mathbf{M}\mathbf{F}\mathbf{M}^\mathrm{T}$.
\subsection{The exponential expression}\label{sec:expexp}
One can verify that
\begin{equation}
\begin{cases}
\boldsymbol{\sigma}_1\boldsymbol{\sigma}_2=\mathrm{i}\boldsymbol{\sigma}_3 & \boldsymbol{\sigma}_2\boldsymbol{\sigma}_1=-\mathrm{i}\boldsymbol{\sigma}_3 \\
\boldsymbol{\sigma}_2\boldsymbol{\sigma}_3=\mathrm{i}\boldsymbol{\sigma}_1 & \boldsymbol{\sigma}_3\boldsymbol{\sigma}_2=-\mathrm{i}\boldsymbol{\sigma}_1 \\
\boldsymbol{\sigma}_3\boldsymbol{\sigma}_1=\mathrm{i}\boldsymbol{\sigma}_2 & \boldsymbol{\sigma}_1\boldsymbol{\sigma}_3=-\mathrm{i}\boldsymbol{\sigma}_2 \\
{\boldsymbol{\sigma}_1}^2={\boldsymbol{\sigma}_2}^2={\boldsymbol{\sigma}_3}^2=\mathbf{I}
\end{cases}
\end{equation}

Introduce the axis of rotation like Section~\ref{sec:axangtomat}, $\boldsymbol{\hat{n}}=
\begin{pmatrix}
n_1&n_2&n_3
\end{pmatrix}^\mathrm{T}$, and notice ${n_1}^2+{n_2}^2+{n_3}^2=1$. Then let's calculate $(\boldsymbol{\hat{n}}\cdot\boldsymbol{\vec{\sigma}})^2$.
\begin{align*}
(\boldsymbol{\hat{n}}\cdot\boldsymbol{\vec{\sigma}})^2&=(n_1\boldsymbol{\sigma}_1+n_2\boldsymbol{\sigma}_2+n_3\boldsymbol{\sigma}_3)(n_1\boldsymbol{\sigma}_1+n_2\boldsymbol{\sigma}_2+n_3\boldsymbol{\sigma}_3)\\
&={n_1}^2{\boldsymbol{\sigma}_1}^2+{n_2}^2{\boldsymbol{\sigma}_2}^2+{n_3}^2{\boldsymbol{\sigma}_3}^2+n_1n_2(\boldsymbol{\sigma}_1\boldsymbol{\sigma}_2+\boldsymbol{\sigma}_2\boldsymbol{\sigma}_1)+n_2n_3(\boldsymbol{\sigma}_2\boldsymbol{\sigma}_3+\boldsymbol{\sigma}_3\boldsymbol{\sigma}_2)+n_3n_1(\boldsymbol{\sigma}_3\boldsymbol{\sigma}_1+\boldsymbol{\sigma}_1\boldsymbol{\sigma}_3)\\
&=({n_1}^2+{n_2}^2+{n_3}^2)\mathbf{I}=\mathbf{I}
\end{align*}
So,
\begin{equation}
\forall k=2m(m\in\mathbb{N}),\ (\boldsymbol{\hat{n}}\cdot\boldsymbol{\vec{\sigma}})^k=\mathbf{I}
\end{equation}
\begin{equation}
\forall k=2m+1(m\in\mathbb{N}),\ (\boldsymbol{\hat{n}}\cdot\boldsymbol{\vec{\sigma}})^k=\boldsymbol{\hat{n}}\cdot\boldsymbol{\vec{\sigma}}
\end{equation}

We will prove the result with these lemmas.
\begin{theorem}
\begin{equation}
\mathbf{A}(-\phi, \boldsymbol{\hat{n}})=
\begin{pmatrix}
\cos\frac{\phi}{2}-\mathrm{i}n_3\sin\frac{\phi}{2}&-\mathrm{i}(n_1-\mathrm{i}n_2)\sin\frac{\phi}{2}\\
-\mathrm{i}(n_1+\mathrm{i}n_2)\sin\frac{\phi}{2}&\cos\frac{\phi}{2}+\mathrm{i}n_3\sin\frac{\phi}{2}
\end{pmatrix}
=\mathrm{e}^{-\mathrm{i}\boldsymbol{\hat{n}}\cdot\boldsymbol{\vec{\sigma}}\frac{\phi}{2}}
\end{equation}
\end{theorem}
\begin{proof}
\begin{align*}
\mathrm{e}^{-\mathrm{i}\boldsymbol{\hat{n}}\cdot\boldsymbol{\vec{\sigma}}\frac{\phi}{2}}&=\sum_{k=0}^{\infty} \frac{1}{k!}(-\mathrm{i}\frac{\phi}{2})^k(\boldsymbol{\hat{n}}\cdot\boldsymbol{\vec{\sigma}})^k\\
&=\sum_{m=0}^{\infty}\frac{1}{(2m)!}(-\mathrm{i}\frac{\phi}{2})^{2m}\mathbf{I}+\sum_{m=0}^{\infty}\frac{1}{(2m+1)!}(-\mathrm{i}\frac{\phi}{2})^{2m+1}\boldsymbol{\hat{n}}\cdot\boldsymbol{\vec{\sigma}}\\
&=\cos\frac{\phi}{2}\mathbf{I}-\mathrm{i}\sin\frac{\phi}{2}(\boldsymbol{\hat{n}}\cdot\boldsymbol{\vec{\sigma}})\\
&=\begin{pmatrix}
\cos\frac{\phi}{2}-\mathrm{i}n_3\sin\frac{\phi}{2}&-\mathrm{i}(n_1-\mathrm{i}n_2)\sin\frac{\phi}{2}\\
-\mathrm{i}(n_1+\mathrm{i}n_2)\sin\frac{\phi}{2}&\cos\frac{\phi}{2}+\mathrm{i}n_3\sin\frac{\phi}{2}
\end{pmatrix}
\end{align*}
Next we will verify that the effects of $\mathrm{e}^{-\mathrm{i}\boldsymbol{\hat{n}}\cdot\boldsymbol{\sigma}\frac{\phi}{2}}$ on ($x$, $y$, $z$) is the same as $\mathbf{R}(-\phi, \boldsymbol{\hat{n}})$. Consider the following matrix, whose entry in the $i$-th row and $j$-th column is denoted as $h_{ij}$
\[
\begin{pmatrix}
\cos\frac{\phi}{2}-\mathrm{i}n_3\sin\frac{\phi}{2}&-\mathrm{i}(n_1-\mathrm{i}n_2)\sin\frac{\phi}{2}\\
-\mathrm{i}(n_1+\mathrm{i}n_2)\sin\frac{\phi}{2}&\cos\frac{\phi}{2}+\mathrm{i}n_3\sin\frac{\phi}{2}
\end{pmatrix}
\begin{pmatrix}
z&x-\mathrm{i}y\\
x+\mathrm{i}y&-z
\end{pmatrix}
\begin{pmatrix}
\cos\frac{\phi}{2}+\mathrm{i}n_3\sin\frac{\phi}{2}&\mathrm{i}(n_1-\mathrm{i}n_2)\sin\frac{\phi}{2}\\
\mathrm{i}(n_1+\mathrm{i}n_2)\sin\frac{\phi}{2}&\cos\frac{\phi}{2}-\mathrm{i}n_3\sin\frac{\phi}{2}
\end{pmatrix}
\]
\begin{adjustwidth}{-3em}{}
\begin{align*}
h_{11}&=(\cos\frac{\phi}{2}-\mathrm{i}n_3\sin\frac{\phi}{2})z(\cos\frac{\phi}{2}+\mathrm{i}n_3\sin\frac{\phi}{2})+(\cos\frac{\phi}{2}-\mathrm{i}n_3\sin\frac{\phi}{2})(x-\mathrm{i}y)\mathrm{i}(n_1+\mathrm{i}n_2)\sin\frac{\phi}{2}\\
&\ \ \ -\mathrm{i}(n_1-\mathrm{i}n_2)\sin\frac{\phi}{2}(x+\mathrm{i}y)(\cos\frac{\phi}{2}+\mathrm{i}n_3\sin\frac{\phi}{2})-\mathrm{i}(n_1-\mathrm{i}n_2)\sin\frac{\phi}{2}(-z)\mathrm{i}(n_1+\mathrm{i}n_2)\sin\frac{\phi}{2}\\
&=(2n_3n_1\sin^2\frac{\phi}{2}-2n_2\cos\frac{\phi}{2}\sin\frac{\phi}{2})x+(2n_2n_3\sin^2\frac{\phi}{2}+2n_1\cos\frac{\phi}{2}\sin\frac{\phi}{2})y+(\cos^2\frac{\phi}{2}-\sin^2\frac{\phi}{2}+2{n_3}^2\sin^2\frac{\phi}{2})z\\
&=[n_3n_1(1-\cos\phi)-n_2\sin\phi]x+[n_2n_3(1-\cos\phi)+n_1\sin\phi]y+[\cos\phi+{n_3}^2(1-\cos\phi)]z
\end{align*}
\begin{align*}
h_{12}&=(\cos\frac{\phi}{2}-\mathrm{i}n_3\sin\frac{\phi}{2})z\mathrm{i}(n_1-\mathrm{i}n_2)\sin\frac{\phi}{2}+(\cos\frac{\phi}{2}-\mathrm{i}n_3\sin\frac{\phi}{2})(x-\mathrm{i}y)(\cos\frac{\phi}{2}-\mathrm{i}n_3\sin\frac{\phi}{2})\\
&\ \ \ -\mathrm{i}(n_1-\mathrm{i}n_2)\sin\frac{\phi}{2}(x+\mathrm{i}y)\mathrm{i}(n_1-\mathrm{i}n_2)\sin\frac{\phi}{2}-\mathrm{i}(n_1-\mathrm{i}n_2)\sin\frac{\phi}{2}(-z)(\cos\frac{\phi}{2}-\mathrm{i}n_3\sin\frac{\phi}{2})\\
&=[(\cos^2\frac{\phi}{2}-\sin^2\frac{\phi}{2}+2{n_1}^2\sin^2\frac{\phi}{2})x+(2n_1n_2\sin^2\frac{\phi}{2}-2n_3\cos\frac{\phi}{2}\sin\frac{\phi}{2})y+(2n_3n_1\sin^2\frac{\phi}{2}+2n_2\cos\frac{\phi}{2}\sin\frac{\phi}{2})z]\\
&\ \ \ -\mathrm{i}[(2n_1n_2\sin^2\frac{\phi}{2}+2n_3\cos\frac{\phi}{2}\sin\frac{\phi}{2})x+(\cos^2\frac{\phi}{2}-\sin^2\frac{\phi}{2}+2{n_2}^2\sin^2\frac{\phi}{2})y+(2n_2n_3\sin^2\frac{\phi}{2}-2n_1\cos\frac{\phi}{2}\sin\frac{\phi}{2})z]\\
&=\{[\cos\phi+{n_1}^2(1-\cos\phi)]x+[n_1n_2(1-\cos\phi)-n_3\sin\phi]y+[n_3n_1(1-\cos\phi)+n_2\sin\phi]z\}\\
&\ \ \ -\mathrm{i}\{[n_1n_2(1-\cos\phi)+n_3\sin\phi]x+[\cos\phi+{n_2}^2(1-\cos\phi)]y+[n_2n_3(1-\cos\phi)-n_1\sin\phi]z\}
\end{align*}
\begin{align*}
h_{21}&=-\mathrm{i}(n_1+\mathrm{i}n_2)\sin\frac{\phi}{2}z(\cos\frac{\phi}{2}+\mathrm{i}n_3\sin\frac{\phi}{2})-\mathrm{i}(n_1+\mathrm{i}n_2)\sin\frac{\phi}{2}(x-\mathrm{i}y)\mathrm{i}(n_1+\mathrm{i}n_2)\sin\frac{\phi}{2}\\
&\ \ \ +(\cos\frac{\phi}{2}+\mathrm{i}n_3\sin\frac{\phi}{2})(x+\mathrm{i}y)(\cos\frac{\phi}{2}+\mathrm{i}n_3\sin\frac{\phi}{2})+(\cos\frac{\phi}{2}+\mathrm{i}n_3\sin\frac{\phi}{2})(-z)\mathrm{i}(n_1+\mathrm{i}n_2)\sin\frac{\phi}{2}\\
&=[(\cos^2\frac{\phi}{2}-\sin^2\frac{\phi}{2}+2{n_1}^2\sin^2\frac{\phi}{2})x+(2n_1n_2\sin^2\frac{\phi}{2}-2n_3\cos\frac{\phi}{2}\sin\frac{\phi}{2})y+(2n_3n_1\sin^2\frac{\phi}{2}+2n_2\cos\frac{\phi}{2}\sin\frac{\phi}{2})z]\\
&\ \ \ +\mathrm{i}[(2n_1n_2\sin^2\frac{\phi}{2}+2n_3\cos\frac{\phi}{2}\sin\frac{\phi}{2})x+(\cos^2\frac{\phi}{2}-\sin^2\frac{\phi}{2}+2{n_2}^2\sin^2\frac{\phi}{2})y+(2n_2n_3\sin^2\frac{\phi}{2}-2n_1\cos\frac{\phi}{2}\sin\frac{\phi}{2})z]\\
&=\{[\cos\phi+{n_1}^2(1-\cos\phi)]x+[n_1n_2(1-\cos\phi)-n_3\sin\phi]y+[n_3n_1(1-\cos\phi)+n_2\sin\phi]z\}\\
&\ \ \ +\mathrm{i}\{[n_1n_2(1-\cos\phi)+n_3\sin\phi]x+[\cos\phi+{n_2}^2(1-\cos\phi)]y+[n_2n_3(1-\cos\phi)-n_1\sin\phi]z\}
\end{align*}
\begin{align*}
h_{22}&=-\mathrm{i}(n_1+\mathrm{i}n_2)\sin\frac{\phi}{2}z\mathrm{i}(n_1-\mathrm{i}n_2)\sin\frac{\phi}{2}-\mathrm{i}(n_1+\mathrm{i}n_2)\sin\frac{\phi}{2}(x-\mathrm{i}y)(\cos\frac{\phi}{2}-\mathrm{i}n_3\sin\frac{\phi}{2})\\
&\ \ \ +(\cos\frac{\phi}{2}+\mathrm{i}n_3\sin\frac{\phi}{2})(x+\mathrm{i}y)\mathrm{i}(n_1-\mathrm{i}n_2)\sin\frac{\phi}{2}+(\cos\frac{\phi}{2}+\mathrm{i}n_3\sin\frac{\phi}{2})(-z)(\cos\frac{\phi}{2}-\mathrm{i}n_3\sin\frac{\phi}{2})\\
&=(-2n_3n_1\sin^2\frac{\phi}{2}+2n_2\cos\frac{\phi}{2}\sin\frac{\phi}{2})x+(-2n_2n_3\sin^2\frac{\phi}{2}-2n_1\cos\frac{\phi}{2}\sin\frac{\phi}{2})y+(-\cos^2\frac{\phi}{2}+\sin^2\frac{\phi}{2}-2{n_3}^2\sin^2\frac{\phi}{2})z\\
&=-[n_3n_1(1-\cos\phi)-n_2\sin\phi]x-[n_2n_3(1-\cos\phi)+n_1\sin\phi]y-[\cos\phi+{n_3}^2(1-\cos\phi)]z
\end{align*}
\end{adjustwidth}
With the relationship $h_{11}=z'$, $h_{12}=x'-\mathrm{i}y'$, $h_{21}=x'+\mathrm{i}y'$, $h_{22}=-z'$,
\[
\begin{cases}
x'=[\cos\phi+{n_1}^2(1-\cos\phi)]x+[n_1n_2(1-\cos\phi)-n_3\sin\phi]y+[n_3n_1(1-\cos\phi)+n_2\sin\phi]z\\
y'=[n_1n_2(1-\cos\phi)+n_3\sin\phi]x+[\cos\phi+{n_2}^2(1-\cos\phi)]y+[n_2n_3(1-\cos\phi)-n_1\sin\phi]z\\
z'=[n_3n_1(1-\cos\phi)-n_2\sin\phi]x+[n_2n_3(1-\cos\phi)+n_1\sin\phi]y+[\cos\phi+{n_3}^2(1-\cos\phi)]z
\end{cases}
\]
This is consistent with Rodrigues' rotation formula (See in~\ref{sec:axangtomat}). And because $\mathrm{e}^{-\mathrm{i}\boldsymbol{\hat{n}}\cdot\boldsymbol{\vec{\sigma}}\frac{\phi}{2}}\rightarrow\mathbf{I}$ when $\phi\rightarrow0$, $\mathbf{A}(-\phi, \boldsymbol{\hat{n}})=
\begin{pmatrix}
\cos\frac{\phi}{2}-\mathrm{i}n_3\sin\frac{\phi}{2}&-\mathrm{i}(n_1-\mathrm{i}n_2)\sin\frac{\phi}{2}\\
-\mathrm{i}(n_1+\mathrm{i}n_2)\sin\frac{\phi}{2}&\cos\frac{\phi}{2}+\mathrm{i}n_3\sin\frac{\phi}{2}
\end{pmatrix}
=\mathrm{e}^{-\mathrm{i}\boldsymbol{\hat{n}}\cdot\boldsymbol{\vec{\sigma}}\frac{\phi}{2}}
$.
\end{proof}
Correspondingly, under passive viewpoint, the matrix turns to
\[
\mathbf{A}(\phi, \boldsymbol{\hat{n}})=
\begin{pmatrix}
\cos\frac{\phi}{2}+\mathrm{i}n_3\sin\frac{\phi}{2}&\mathrm{i}(n_1-\mathrm{i}n_2)\sin\frac{\phi}{2}\\
\mathrm{i}(n_1+\mathrm{i}n_2)\sin\frac{\phi}{2}&\cos\frac{\phi}{2}-\mathrm{i}n_3\sin\frac{\phi}{2}
\end{pmatrix}
=\mathrm{e}^{\mathrm{i}\boldsymbol{\hat{n}}\cdot\boldsymbol{\vec{\sigma}}\frac{\phi}{2}}
\]

This kind of representation of rotation can simplify the calculation and reveal the essence for many problems, like proof to Wigner rotation theorem\cite{lu2025}.
\subsection{Some global properties of SU(2) Group}
Summarize our previous discussion and prove the others correspond to (\ref{eqt:period})\textendash(\ref{eqt:glb1}), (\ref{eqt:glb2})\textendash(\ref{eqt:lnindep}):
\begin{equation}
\begin{cases}
\mathbf{A}(-(\phi+2k\pi),\boldsymbol{\hat{n}})=-\mathbf{A}(-\phi,\boldsymbol{\hat{n}})\ \ \ (k\in\mathbb{Z})\\
\mathbf{A}(-(\phi+4k\pi),\boldsymbol{\hat{n}})=\mathbf{A}(-\phi,\boldsymbol{\hat{n}})\ \ \ (k\in\mathbb{Z})\\
\mathbf{A}(-\phi_1,\boldsymbol{\hat{n}_1})\neq\mathbf{A}(-\phi_2,\boldsymbol{\hat{n}_2})\ \ \ \ \ \mathrm{if\ }\boldsymbol{\hat{n}_1}\mathrm{\ and\ }\boldsymbol{\hat{n}_2}\mathrm{\ is\ linearly\ independent}\\
\mathbf{A}(\phi,-\boldsymbol{\hat{n}})=\mathbf{A}(-\phi,\boldsymbol{\hat{n}})\\
\mathbf{A}(-\phi_2,\boldsymbol{\hat{n}})\mathbf{A}(-\phi_1,\boldsymbol{\hat{n}})=\mathbf{A}(-(\phi_1+\phi_2),\boldsymbol{\hat{n}})=\mathbf{A}(-\phi_1,\boldsymbol{\hat{n}})\mathbf{A}(-\phi_2,\boldsymbol{\hat{n}})\\
\mathbf{A}^{-1}(-\phi,\boldsymbol{\hat{n}})=\mathbf{A}(\phi,\boldsymbol{\hat{n}})
\end{cases}
\end{equation}
and a correspondent theorem can be concluded.
\begin{theorem}
$\mathbf{A}(-\phi,\boldsymbol{\hat{n}})$ with $\boldsymbol{\hat{n}}$ fixed constitute an Abelian group.
\end{theorem}
Similarly, with variable substitution $\phi\to-\phi$, we will get corresponding formulae under passive viewpoints.

The composition of SU(2) rotations is like that of SO(3).
\subsection{Infinitesimal rotations of SU(2) Group}

\subsubsection{Lie algebras of SU(2) Group}

Considering an infinitesimal SU(2) transformation $\mathbf{Q}$ with $\mathbf{Q}\mathbf{Q}^\dagger=\mathbf{I}$ and $\mathrm{det}\ \mathbf{Q}=1$. Denote $\mathbf{Q}=\mathbf{I}+\mathrm{i}\varepsilon\mathbf{G}+\mathcal{O}(\varepsilon^2)$, then
\[
\mathbf{I}=\mathbf{Q}\mathbf{Q}^\dagger
=(\mathbf{I}+\mathrm{i}\varepsilon\mathbf{G}+\mathcal{O}(\varepsilon^2))(\mathbf{I}-\mathrm{i}\varepsilon\mathbf{G}^\dagger+\mathcal{O}(\varepsilon^2))
=\mathbf{I}+\mathrm{i}\varepsilon(\mathbf{G}-\mathbf{G}^\dagger)+\mathcal{O}(\varepsilon^2)
\]
So,
\begin{equation}
\mathbf{G}-\mathbf{G}^\dagger=\mathbf{O}\Leftrightarrow\mathbf{G}=\mathbf{G}^\dagger
\end{equation}

Denote the eigenvalues of $\mathbf{G}$ as $\lambda_1$, $\lambda_2$, then those of $\mathbf{I}+\mathrm{i}\varepsilon\mathbf{G}$ are $1+\mathrm{i}\varepsilon\lambda_1$, $1+\mathrm{i}\varepsilon\lambda_2$, so
\[
1=\mathrm{det}(\mathbf{I}+\mathrm{i}\varepsilon\mathbf{G})
=(1+\mathrm{i}\varepsilon\lambda_1)(1+\mathrm{i}\varepsilon\lambda_2)
=1+\mathrm{i}\varepsilon(\lambda_1+\lambda_2)-\varepsilon^2\lambda_1\lambda_2
=1+\mathrm{i}\varepsilon\ \mathrm{tr}\ \mathbf{G}+\mathcal{O}(\varepsilon^2)
\]
so $\mathrm{tr}\ \mathbf{G}=0$. In conclusion, all 2$\times$2 traceless Hermitian matrices are precisely the SU(2) generators, generally written as
\[
\mathbf{G}=\begin{pmatrix}
a&b-\mathrm{i}c\\
b+\mathrm{i}c&-a
\end{pmatrix}
=b\boldsymbol{\sigma}_1+c\boldsymbol{\sigma}_2+a\boldsymbol{\sigma}_3
\]
Then we can choose a set proportional to $\{\boldsymbol{\sigma}_1,\ \boldsymbol{\sigma}_2,\ \boldsymbol{\sigma}_3\}$ as the generating set.

Next we consider the normalization of the generating set. For infinitesimal rotations about the Cartesian axes,
\begin{equation}
\mathbf{A}(-\mathrm{d}\phi,\boldsymbol{\hat{x}})=
\begin{pmatrix}
1&-\frac{\mathrm{i}}{2}\mathrm{d}\phi\\
-\frac{\mathrm{i}}{2}\mathrm{d}\phi&1
\end{pmatrix}
,\
\mathbf{A}(-\mathrm{d}\phi,\boldsymbol{\hat{y}})=
\begin{pmatrix}
1&-\frac{1}{2}\mathrm{d}\phi\\
\frac{1}{2}\mathrm{d}\phi&1
\end{pmatrix}
,\
\mathbf{A}(-\mathrm{d}\phi,\boldsymbol{\hat{z}})=
\begin{pmatrix}
1-\frac{\mathrm{i}}{2}\mathrm{d}\phi&0\\
0&1+\frac{\mathrm{i}}{2}\mathrm{d}\phi
\end{pmatrix}
\end{equation}

Normalize $\varepsilon$ as $\mathrm{d}\phi$. Thus, the generators of the generating set are $-\frac{\mathrm{i}}{2}\boldsymbol{\sigma}_1$, $-\frac{\mathrm{i}}{2}\boldsymbol{\sigma}_2$, $-\frac{\mathrm{i}}{2}\boldsymbol{\sigma}_3$.

Then, we can calculate the matrix commutators of Pauli matrices, which is equivalent to calculating SU(2) Lie algebras.
\begin{align}
&[\boldsymbol{\sigma}_1,\boldsymbol{\sigma}_2]=
\begin{pmatrix}
0&1\\
1&0
\end{pmatrix}
\begin{pmatrix}
0&-\mathrm{i}\\
\mathrm{i}&0
\end{pmatrix}
-
\begin{pmatrix}
0&-\mathrm{i}\\
\mathrm{i}&0
\end{pmatrix}
\begin{pmatrix}
0&1\\
1&0
\end{pmatrix}
=
\begin{pmatrix}
2\mathrm{i}&0\\
0&-2\mathrm{i}
\end{pmatrix}
=2\mathrm{i}\boldsymbol{\sigma}_3
\label{eqt:numcase1}\\
&[\boldsymbol{\sigma}_2,\boldsymbol{\sigma}_3]=
\begin{pmatrix}
0&-\mathrm{i}\\
\mathrm{i}&0
\end{pmatrix}
\begin{pmatrix}
1&0\\
0&-1
\end{pmatrix}
-
\begin{pmatrix}
1&0\\
0&-1
\end{pmatrix}
\begin{pmatrix}
0&-\mathrm{i}\\
\mathrm{i}&0
\end{pmatrix}
=
\begin{pmatrix}
0&2\mathrm{i}\\
2\mathrm{i}&0
\end{pmatrix}
=2\mathrm{i}\boldsymbol{\sigma}_1\\
&[\boldsymbol{\sigma}_3,\boldsymbol{\sigma}_1]=
\begin{pmatrix}
1&0\\
0&-1
\end{pmatrix}
\begin{pmatrix}
0&1\\
1&0
\end{pmatrix}
-
\begin{pmatrix}
0&1\\
1&0
\end{pmatrix}
\begin{pmatrix}
1&0\\
0&-1
\end{pmatrix}
=
\begin{pmatrix}
0&2\\
-2&0
\end{pmatrix}
=2\mathrm{i}\boldsymbol{\sigma}_2
\label{eqt:numcase3}
\end{align}

Next, we can normalize any element in the generating set of SU(2), i.e. $\mathfrak{su}$(2). Apply the commuting property of rotation,
\begin{adjustwidth}{-1em}{}
\begin{align}
\mathbf{A}(-\mathrm{d}\phi,\boldsymbol{\hat{n}})&=\mathbf{A}(-\mathrm{d}\phi_x,\boldsymbol{\hat{x}})\mathbf{A}(-\mathrm{d}\phi_y,\boldsymbol{\hat{y}})\mathbf{A}(-\mathrm{d}\phi_z,\boldsymbol{\hat{z}})+\mathcal{O}(\mathrm{d}\phi^2)=\mathbf{I}-\frac{\mathrm{i}}{2}(\mathrm{d}\phi_x\boldsymbol{\sigma}_1+\mathrm{d}\phi_y\boldsymbol{\sigma}_2+\mathrm{d}\phi_z\boldsymbol{\sigma}_3)+\mathcal{O}(\mathrm{d}\phi^2)\nonumber\\
&=\mathbf{I}-\frac{\mathrm{i}}{2}\mathrm{d}\phi(n_1\boldsymbol{\sigma}_1+n_2\boldsymbol{\sigma}_2+n_3\boldsymbol{\sigma}_3)+\mathcal{O}(\mathrm{d}\phi^2)=\mathbf{I}-\frac{\mathrm{i}}{2}\mathrm{d}\phi\ \boldsymbol{\hat{n}}\cdot\boldsymbol{\vec{\sigma}}+\mathcal{O}(\mathrm{d}\phi^2)
\end{align}
\end{adjustwidth}
Thus, we determine the normalized $\mathfrak{su}$(2) elements about any axis. The effective matrix corresponding to Pauli matrices is
\begin{equation}
\boldsymbol{\sigma}=\boldsymbol{\hat{n}}\cdot\boldsymbol{\vec{\sigma}}
\end{equation}

Finally, calculate the exponential mapping.
\[
\mathbf{A}(-\phi,\boldsymbol{\hat{n}})=\lim\limits_{n\to\infty} \left[\mathbf{A}(-\frac{\phi}{n},\boldsymbol{\hat{n}})\right]^n=\lim\limits_{n\to\infty}(\mathbf{I}-\frac{\mathrm{i}\phi}{2n}\boldsymbol{\hat{n}}\cdot\boldsymbol{\vec{\sigma}})^n=\mathrm{e}^{{-\mathrm{i}\boldsymbol{\hat{n}}}\cdot\boldsymbol{\vec{\sigma}}\frac{\phi}{2}}
\]

The exponential mapping can express any SU(2) element. We can also see that the simple exponential expression in Section~\ref{sec:expexp} is not a coincidence. If there is any coincidence, that is the homomorphism between SO(3) Group and SU(2) Group.

\subsubsection{The angular velocity matrix of SU(2) Group}\label{sec:angvsu2g}
Since any rotation $\mathbf{Q}$ whose axis is $\boldsymbol{\hat{n}}$ and angle is $\phi$,
\begin{equation}\label{eqt:su2mat2}
\mathbf{Q}=\mathbf{A}(-\phi, \boldsymbol{\hat{n}})=\cos\frac{\phi}{2}\mathbf{I}-\mathrm{i}\sin\frac{\phi}{2}(\boldsymbol{\hat{n}}\cdot\boldsymbol{\vec{\sigma}})
\end{equation}
The Hermitian conjugate of $\mathbf{Q}$ is $\mathbf{Q}^\dagger=\cos\frac{\phi}{2}\mathbf{I}+\mathrm{i}\sin\frac{\phi}{2}(\boldsymbol{\hat{n}}\cdot\boldsymbol{\vec{\sigma}})$. Then we derive the expression of the derivative of rotation matrix $\mathbf{\dot{Q}}$. First of all, we should prove a theorem.
\begin{theorem}
For any two vectors $\boldsymbol{a}$ and $\boldsymbol{b}$,
\begin{equation}\label{eqt:quatop}
(\boldsymbol{a}\cdot\boldsymbol{\vec{\sigma}})(\boldsymbol{b}\cdot\boldsymbol{\vec{\sigma}})=(\boldsymbol{a}\cdot\boldsymbol{b})\mathbf{I}+\mathrm{i}(\boldsymbol{a}\times\boldsymbol{b})\cdot\boldsymbol{\vec{\sigma}}
\end{equation}
\end{theorem}
\begin{proof}
\begin{align*}
(\boldsymbol{a}\cdot\boldsymbol{\vec{\sigma}})(\boldsymbol{b}\cdot\boldsymbol{\vec{\sigma}})&=
\begin{pmatrix}
a_z&a_x-\mathrm{i}a_y\\a_x+\mathrm{i}a_y&-a_z
\end{pmatrix}
\begin{pmatrix}
b_z&b_x-\mathrm{i}b_y\\b_x+\mathrm{i}b_y&-b_z
\end{pmatrix}\\
&=\begin{pmatrix}
a_xb_x+a_yb_y+a_zb_z+\mathrm{i}(a_xb_y-a_yb_x)&a_zb_x-a_xb_z+\mathrm{i}(a_yb_z-a_zb_y)\\-(a_zb_x-a_xb_z)+\mathrm{i}(a_yb_z-a_zb_y)&a_xb_x+a_yb_y+a_zb_z-\mathrm{i}(a_xb_y-a_yb_x)
\end{pmatrix}\\
&=(\boldsymbol{a}\cdot\boldsymbol{b})\mathbf{I}+\mathrm{i}(\boldsymbol{a}\times\boldsymbol{b})\cdot\boldsymbol{\vec{\sigma}}
\end{align*}
\end{proof}

From (\ref{eqt:su2mat2}),
\begin{align*}
\mathbf{\dot{Q}}&=-\frac{\dot{\phi}}{2}\sin\frac{\phi}{2}\mathbf{I}-\mathrm{i}\left(\frac{\dot{\phi}}{2}\cos\frac{\phi}{2}\boldsymbol{\hat{n}}+\sin\frac{\phi}{2}\boldsymbol{\dot{\hat{n}}}\right)\cdot\boldsymbol{\vec{\sigma}}\nonumber\\
&=-\frac{1}{2}\left[\dot{\phi}\sin\frac{\phi}{2}\mathbf{I}+\mathrm{i}\left(\dot{\phi}\cos\frac{\phi}{2}\boldsymbol{\hat{n}}+2\sin\frac{\phi}{2}\boldsymbol{\dot{\hat{n}}}\right)\cdot\boldsymbol{\vec{\sigma}}\right]
\end{align*}
Define the SU(2) angular velocity matrix
\begin{equation}
\mathbf{W}=\mathbf{\dot{Q}}\mathbf{Q}^\dagger
\end{equation}
then
\begin{align*}
\mathbf{W}&=-\frac{1}{2}\left[\dot{\phi}\sin\frac{\phi}{2}\mathbf{I}+\mathrm{i}(\dot{\phi}\cos\frac{\phi}{2}\boldsymbol{\hat{n}}+2\sin\frac{\phi}{2}\boldsymbol{\dot{\hat{n}}})\cdot\boldsymbol{\vec{\sigma}}\right]\left[\cos\frac{\phi}{2}\mathbf{I}+\mathrm{i}\sin\frac{\phi}{2}(\boldsymbol{\hat{n}}\cdot\boldsymbol{\vec{\sigma}})\right]\\
&=-\frac{1}{2}\Big[\dot{\phi}\sin\frac{\phi}{2}\cos\frac{\phi}{2}\mathbf{I}+\mathrm{i}\dot{\phi}\sin^2\frac{\phi}{2}(\boldsymbol{\hat{n}}\cdot\boldsymbol{\vec{\sigma}})+\mathrm{i}\dot{\phi}\cos^2\frac{\phi}{2}(\boldsymbol{\hat{n}}\cdot\boldsymbol{\vec{\sigma}})+2\mathrm{i}\sin\frac{\phi}{2}\cos\frac{\phi}{2}(\boldsymbol{\dot{\hat{n}}}\cdot\boldsymbol{\vec{\sigma}})\\
&\ \ \ -\dot{\phi}\sin\frac{\phi}{2}\cos\frac{\phi}{2}(\boldsymbol{\hat{n}}\cdot\boldsymbol{\vec{\sigma}})^2+2\sin\frac{\phi}{2}(\boldsymbol{\dot{\hat{n}}}\cdot\boldsymbol{\vec{\sigma}})(\boldsymbol{\hat{n}}\cdot\boldsymbol{\vec{\sigma}})\Big]\\
&=-\frac{\mathrm{i}}{2}\left(\dot{\phi}\boldsymbol{\hat{n}}+2\sin\frac{\phi}{2}\cos\frac{\phi}{2}\boldsymbol{\dot{\hat{n}}}+2\sin^2\frac{\phi}{2}\boldsymbol{\hat{n}}\times\boldsymbol{\dot{\hat{n}}}\right)\cdot\boldsymbol{\vec{\sigma}}\\
&=-\frac{\mathrm{i}}{2}\left[\dot{\phi}\boldsymbol{\hat{n}}+\sin\phi\boldsymbol{\dot{\hat{n}}}+(1-\cos\phi)\boldsymbol{\hat{n}}\times\boldsymbol{\dot{\hat{n}}}\right]\cdot\boldsymbol{\vec{\sigma}}
\end{align*}
From (\ref{eqt:quaternion}), 
\begin{equation}\label{eqt:genw}
\mathbf{W}=-\frac{\mathrm{i}}{2}\boldsymbol{\omega}\cdot\boldsymbol{\vec{\sigma}}=-\frac{\mathrm{i}}{2}\boldsymbol{\sigma}
\end{equation}
i.e. 
\begin{equation}
\mathbf{\dot{Q}}=-\frac{\mathrm{i}}{2}(\boldsymbol{\omega}\cdot\boldsymbol{\vec{\sigma}})\mathbf{Q}
\end{equation}
That's why $\mathbf{W}$ is named SU(2) angular velocity matrix, because in previous discussions in Section~\ref{sec:angv}, the angular velocity matrix is defined to be correspondent to a generator, and $\mathbf{W}$ in (\ref{eqt:genw}) share the same property with $\mathbf{\Omega}$ in (\ref{eqt:ooj}) including the coefficient because the generating set is $-\frac{\mathrm{i}}{2}\boldsymbol{\vec{\sigma}}$ correspondent to $\boldsymbol{\vec{J}}$ in (\ref{eqt:ooj}).

Next, we analyze the physical meaning of $\mathbf{W}$. Consider a coordinate transformation $\boldsymbol{X}=\mathbf{Q}\boldsymbol{X_0}\mathbf{Q}^\dagger$. Since $\mathbf{Q}\mathbf{Q}^\dagger=\mathbf{I}$, the derivative $\mathbf{\dot{Q}}\mathbf{Q}^\dagger+\mathbf{Q}\mathbf{\dot{Q}}^\dagger=\mathbf{O}$, i.e. 
\begin{equation}
\mathbf{\dot{Q}}^\dagger=-\mathbf{Q}^\dagger\mathbf{\dot{Q}}\mathbf{Q}^\dagger
\end{equation}
So\cite{heard2008},
\begin{align}
\boldsymbol{\dot{X}}&=\mathbf{\dot{Q}}\boldsymbol{X_0}\mathbf{Q}^\dagger+\mathbf{Q}\boldsymbol{X_0}\mathbf{\dot{Q}}^\dagger\nonumber\\
&=\mathbf{\dot{Q}}\mathbf{Q}^\dagger\mathbf{Q}\boldsymbol{X_0}\mathbf{Q}^\dagger-\mathbf{Q}\boldsymbol{X_0}\mathbf{Q}^\dagger\mathbf{\dot{Q}}\mathbf{Q}^\dagger\nonumber\\
&=\mathbf{\dot{Q}}\mathbf{Q}^\dagger\boldsymbol{X}-\boldsymbol{X}\mathbf{\dot{Q}}\mathbf{Q}^\dagger\nonumber\\
&=[\mathbf{W},\boldsymbol{X}]
\end{align}
Consider a matrix transformation with an infinitesimal angle $\mathrm{d}\phi$
\[
\boldsymbol{X}\to\mathrm{e}^{{-\mathrm{i}\boldsymbol{\hat{n}}}\cdot\boldsymbol{\vec{\sigma}}\frac{\mathrm{d}\phi}{2}}\boldsymbol{X}\mathrm{e}^{{\mathrm{i}\boldsymbol{\hat{n}}}\cdot\boldsymbol{\vec{\sigma}}\frac{\mathrm{d}\phi}{2}}=\boldsymbol{X}-\mathrm{i}\frac{\mathrm{d}\phi}{2}[\boldsymbol{\sigma},\boldsymbol{X}]+\mathcal{O}(\mathrm{d}\phi^2)=\boldsymbol{X}+[\mathbf{W},\boldsymbol{X}]\mathrm{d}t+\mathcal{O}(\mathrm{d}t^2)
\]
Here, we can see the relationship (\ref{eqt:genw}) again.

Finally, we express the coordinate matrix by Caylay-Klein parameters, i.e. (\ref{eqt:caylayklein}), and the angular velocity matrix is given by\cite{heard2008}
\begin{equation}
\mathbf{W}=
\begin{pmatrix}
\dot{\alpha}\alpha^*+\dot{\beta}\beta^*&-\dot{\alpha}\beta+\alpha\dot{\beta}\\
\dot{\alpha}^*\beta^*-\alpha^*\dot{\beta}^*&\dot{\alpha}^*\alpha+\dot{\beta}^*\beta
\end{pmatrix}
\end{equation}
This expression will be extremely useful in Part II in discussing avoidance of gimbal lock problem.
\section{Conclusion}
In current textbooks, the foundation of rigid-body kinematics isn't often presented systematically, because literature often omits derivations and views rotations in an unchanged representation under an unchanged viewpoint. What's more, current literature tends to describe rotational phenomena with complex mathematical languages like quaternion. This paper can help overcome these problems. It applies the active viewpoint, replaces quaternion operations by equivalent but simpler ones, and illustrated Lie algebras in an intuitive way.

Furthermore, this paper covers some foundational contents that remain largely unexplored in the existing literature, especially those discussed in~\ref{sec:compo2}. So, this paper can serve as a supplement.

Part I includes rotational phenomena as much as it can. In the future, Part II would present the general form of rigid-body motion and derive the ways of quantity transformations. It would also reveal the essence of the difference between any two elements, degrees of freedom, and the gimbal lock problem, considering the topological structures and analyzing global properties systematically. Last, this article would talk about constrained rigid-body
motions extending the scope of discussion from groups to general manifolds embedded in the group of rigid-body transformations.
\section{Notes and Acknowledgements}
Most contents of Sections~\ref{sec:overv},~\ref{sec:o3so3},~\ref{sec:repro},~\ref{sec:compo}, and the last three subsections of Section~\ref{sec:propso3} are my independent derivations on April 2024 with the knowledge I learned, which laid the foundation of the articular frame. At that time I was learning a course about classical mechanics. The other parts are from my own derivations or my summaries of current literature since September 2025, the latter of which is noted in the reference part below. In these parts, Section~\ref{sec:view} was written combining relevant derivations in different parts after I corrected some fatal mistakes about them and derived the conclusions on my own starting from the beginning. I read much literature about the SU(2) description and found the derivations are not fully illustrated. I complete the missing derivations in Sections~\ref{sec:racaithr},~\ref{sec:arithr},~\ref{sec:rmidr} and~\ref{sec:expexp}, and combine my original thinking with the L\"{u} (2025)\cite{lu2025} in Section~\ref{sec:phymn}. Comparing different texts and figuring out some original insights and threads different with them, I write the first three subsections of Section~\ref{sec:propso3} and the last two subsections in Section~\ref{sec:descsu2}.

Thanks for all of my teachers and friends who taught me knowledge or inspired me. My teachers: Hong L\"{u} taught me the precise definition of vectors when I asked him, which is essential of Section~\ref{sec:commt}; Jun-Bao Wu helped me to do calculation of Section~\ref{sec:compo2} with Mathematica last year; Jiaju Zhang helped me to evaluate some threads; Daijun Tian's insight for an essential point enabled me to execute my idea for Section~\ref{sec:indep}. My friend Shaojie Wang inspired me on the writing of Sections~\ref{sec:racaithr} and~\ref{sec:arithr}. And thanks for some other friends and teachers who inspired me on some details for the topic indirectly.
\appendix
\refstepcounter{section}
\section*{Appendix A: 8 solutions of the rotation matrix for 8 cases}\label{sec:appendix}
\addcontentsline{toc}{section}{Appendix A: 8 solutions of the rotation matrix for 8 cases}
Denote $\Delta=(m_{11}+m_{22}+m_{33}+1)(-m_{11}+m_{22}+m_{33}-1)(m_{11}+m_{22}-m_{33}-1)(-m_{11}+m_{22}-m_{33}+1)$, and the roots for $\mathbf{M}$ are 
\subsection*{1$\ \ \ \ \boldsymbol{m_{11}m_{22}-m_{33}\geq0}$, $\boldsymbol{m_{22}m_{33}-m_{11}\geq0}$, $\boldsymbol{m_{33}m_{11}-m_{22}\geq0}$}
\begin{adjustwidth}{-3em}{}
\begin{small}
\[
\begin{pmatrix}
m_{11}&\sqrt{\frac{1}{2}\left[(-m_{11}^2-m_{22}^2+m_{33}^2+1)+\sqrt{\Delta}\right]}&\sqrt{\frac{1}{2}\left[(-m_{11}^2+m_{22}^2-m_{33}^2+1)-\sqrt{\Delta}\right]}\\[6pt]
\sqrt{\frac{1}{2}\left[(-m_{11}^2-m_{22}^2+m_{33}^2+1)-\sqrt{\Delta}\right]}&m_{22}&\sqrt{\frac{1}{2}\left[(m_{11}^2-m_{22}^2-m_{33}^2+1)+\sqrt{\Delta}\right]}\\[6pt]
\sqrt{\frac{1}{2}\left[(-m_{11}^2+m_{22}^2-m_{33}^2+1)+\sqrt{\Delta}\right]}&\sqrt{\frac{1}{2}\left[(m_{11}^2-m_{22}^2-m_{33}^2+1)-\sqrt{\Delta}\right]}&m_{33}
\end{pmatrix}
\]
\[
\begin{pmatrix}
m_{11}&-\sqrt{\frac{1}{2}\left[(-m_{11}^2-m_{22}^2+m_{33}^2+1)+\sqrt{\Delta}\right]}&-\sqrt{\frac{1}{2}\left[(-m_{11}^2+m_{22}^2-m_{33}^2+1)-\sqrt{\Delta}\right]}\\[6pt]
-\sqrt{\frac{1}{2}\left[(-m_{11}^2-m_{22}^2+m_{33}^2+1)-\sqrt{\Delta}\right]}&m_{22}&\sqrt{\frac{1}{2}\left[(m_{11}^2-m_{22}^2-m_{33}^2+1)+\sqrt{\Delta}\right]}\\[6pt]
-\sqrt{\frac{1}{2}\left[(-m_{11}^2+m_{22}^2-m_{33}^2+1)+\sqrt{\Delta}\right]}&\sqrt{\frac{1}{2}\left[(m_{11}^2-m_{22}^2-m_{33}^2+1)-\sqrt{\Delta}\right]}&m_{33}
\end{pmatrix}
\]
\[
\begin{pmatrix}
m_{11}&\sqrt{\frac{1}{2}\left[(-m_{11}^2-m_{22}^2+m_{33}^2+1)+\sqrt{\Delta}\right]}&-\sqrt{\frac{1}{2}\left[(-m_{11}^2+m_{22}^2-m_{33}^2+1)-\sqrt{\Delta}\right]}\\[6pt]
\sqrt{\frac{1}{2}\left[(-m_{11}^2-m_{22}^2+m_{33}^2+1)-\sqrt{\Delta}\right]}&m_{22}&-\sqrt{\frac{1}{2}\left[(m_{11}^2-m_{22}^2-m_{33}^2+1)+\sqrt{\Delta}\right]}\\[6pt]
-\sqrt{\frac{1}{2}\left[(-m_{11}^2+m_{22}^2-m_{33}^2+1)+\sqrt{\Delta}\right]}&-\sqrt{\frac{1}{2}\left[(m_{11}^2-m_{22}^2-m_{33}^2+1)-\sqrt{\Delta}\right]}&m_{33}
\end{pmatrix}
\]
\[
\begin{pmatrix}
m_{11}&-\sqrt{\frac{1}{2}\left[(-m_{11}^2-m_{22}^2+m_{33}^2+1)+\sqrt{\Delta}\right]}&\sqrt{\frac{1}{2}\left[(-m_{11}^2+m_{22}^2-m_{33}^2+1)-\sqrt{\Delta}\right]}\\[6pt]
-\sqrt{\frac{1}{2}\left[(-m_{11}^2-m_{22}^2+m_{33}^2+1)-\sqrt{\Delta}\right]}&m_{22}&-\sqrt{\frac{1}{2}\left[(m_{11}^2-m_{22}^2-m_{33}^2+1)+\sqrt{\Delta}\right]}\\[6pt]
\sqrt{\frac{1}{2}\left[(-m_{11}^2+m_{22}^2-m_{33}^2+1)+\sqrt{\Delta}\right]}&-\sqrt{\frac{1}{2}\left[(m_{11}^2-m_{22}^2-m_{33}^2+1)-\sqrt{\Delta}\right]}&m_{33}
\end{pmatrix}
\]
\[
\begin{pmatrix}
m_{11}&\sqrt{\frac{1}{2}\left[(-m_{11}^2-m_{22}^2+m_{33}^2+1)-\sqrt{\Delta}\right]}&\sqrt{\frac{1}{2}\left[(-m_{11}^2+m_{22}^2-m_{33}^2+1)+\sqrt{\Delta}\right]}\\[6pt]
\sqrt{\frac{1}{2}\left[(-m_{11}^2-m_{22}^2+m_{33}^2+1)+\sqrt{\Delta}\right]}&m_{22}&\sqrt{\frac{1}{2}\left[(m_{11}^2-m_{22}^2-m_{33}^2+1)-\sqrt{\Delta}\right]}\\[6pt]
\sqrt{\frac{1}{2}\left[(-m_{11}^2+m_{22}^2-m_{33}^2+1)-\sqrt{\Delta}\right]}&\sqrt{\frac{1}{2}\left[(m_{11}^2-m_{22}^2-m_{33}^2+1)+\sqrt{\Delta}\right]}&m_{33}
\end{pmatrix}
\]
\[
\begin{pmatrix}
m_{11}&-\sqrt{\frac{1}{2}\left[(-m_{11}^2-m_{22}^2+m_{33}^2+1)-\sqrt{\Delta}\right]}&-\sqrt{\frac{1}{2}\left[(-m_{11}^2+m_{22}^2-m_{33}^2+1)+\sqrt{\Delta}\right]}\\[6pt]
-\sqrt{\frac{1}{2}\left[(-m_{11}^2-m_{22}^2+m_{33}^2+1)+\sqrt{\Delta}\right]}&m_{22}&\sqrt{\frac{1}{2}\left[(m_{11}^2-m_{22}^2-m_{33}^2+1)-\sqrt{\Delta}\right]}\\[6pt]
-\sqrt{\frac{1}{2}\left[(-m_{11}^2+m_{22}^2-m_{33}^2+1)-\sqrt{\Delta}\right]}&\sqrt{\frac{1}{2}\left[(m_{11}^2-m_{22}^2-m_{33}^2+1)+\sqrt{\Delta}\right]}&m_{33}
\end{pmatrix}
\]
\[
\begin{pmatrix}
m_{11}&\sqrt{\frac{1}{2}\left[(-m_{11}^2-m_{22}^2+m_{33}^2+1)-\sqrt{\Delta}\right]}&-\sqrt{\frac{1}{2}\left[(-m_{11}^2+m_{22}^2-m_{33}^2+1)+\sqrt{\Delta}\right]}\\[6pt]
\sqrt{\frac{1}{2}\left[(-m_{11}^2-m_{22}^2+m_{33}^2+1)+\sqrt{\Delta}\right]}&m_{22}&-\sqrt{\frac{1}{2}\left[(m_{11}^2-m_{22}^2-m_{33}^2+1)-\sqrt{\Delta}\right]}\\[6pt]
-\sqrt{\frac{1}{2}\left[(-m_{11}^2+m_{22}^2-m_{33}^2+1)-\sqrt{\Delta}\right]}&-\sqrt{\frac{1}{2}\left[(m_{11}^2-m_{22}^2-m_{33}^2+1)+\sqrt{\Delta}\right]}&m_{33}
\end{pmatrix}
\]
\[
\begin{pmatrix}
m_{11}&-\sqrt{\frac{1}{2}\left[(-m_{11}^2-m_{22}^2+m_{33}^2+1)-\sqrt{\Delta}\right]}&\sqrt{\frac{1}{2}\left[(-m_{11}^2+m_{22}^2-m_{33}^2+1)+\sqrt{\Delta}\right]}\\[6pt]
-\sqrt{\frac{1}{2}\left[(-m_{11}^2-m_{22}^2+m_{33}^2+1)+\sqrt{\Delta}\right]}&m_{22}&-\sqrt{\frac{1}{2}\left[(m_{11}^2-m_{22}^2-m_{33}^2+1)-\sqrt{\Delta}\right]}\\[6pt]
\sqrt{\frac{1}{2}\left[(-m_{11}^2+m_{22}^2-m_{33}^2+1)-\sqrt{\Delta}\right]}&-\sqrt{\frac{1}{2}\left[(m_{11}^2-m_{22}^2-m_{33}^2+1)+\sqrt{\Delta}\right]}&m_{33}
\end{pmatrix}
\]
\end{small}
\end{adjustwidth}
\subsection*{2$\ \ \ \ \boldsymbol{m_{11}m_{22}-m_{33}<0}$, $\boldsymbol{m_{22}m_{33}-m_{11}\geq0}$, $\boldsymbol{m_{33}m_{11}-m_{22}\geq0}$}
\begin{adjustwidth}{-3em}{}
\begin{small}
\[
\begin{pmatrix}
m_{11}&-\sqrt{\frac{1}{2}\left[(-m_{11}^2-m_{22}^2+m_{33}^2+1)+\sqrt{\Delta}\right]}&\sqrt{\frac{1}{2}\left[(-m_{11}^2+m_{22}^2-m_{33}^2+1)-\sqrt{\Delta}\right]}\\[6pt]
\sqrt{\frac{1}{2}\left[(-m_{11}^2-m_{22}^2+m_{33}^2+1)-\sqrt{\Delta}\right]}&m_{22}&-\sqrt{\frac{1}{2}\left[(m_{11}^2-m_{22}^2-m_{33}^2+1)+\sqrt{\Delta}\right]}\\[6pt]
\sqrt{\frac{1}{2}\left[(-m_{11}^2+m_{22}^2-m_{33}^2+1)+\sqrt{\Delta}\right]}&-\sqrt{\frac{1}{2}\left[(m_{11}^2-m_{22}^2-m_{33}^2+1)-\sqrt{\Delta}\right]}&m_{33}
\end{pmatrix}
\]
\[
\begin{pmatrix}
m_{11}&\sqrt{\frac{1}{2}\left[(-m_{11}^2-m_{22}^2+m_{33}^2+1)+\sqrt{\Delta}\right]}&-\sqrt{\frac{1}{2}\left[(-m_{11}^2+m_{22}^2-m_{33}^2+1)-\sqrt{\Delta}\right]}\\[6pt]
-\sqrt{\frac{1}{2}\left[(-m_{11}^2-m_{22}^2+m_{33}^2+1)-\sqrt{\Delta}\right]}&m_{22}&-\sqrt{\frac{1}{2}\left[(m_{11}^2-m_{22}^2-m_{33}^2+1)+\sqrt{\Delta}\right]}\\[6pt]
-\sqrt{\frac{1}{2}\left[(-m_{11}^2+m_{22}^2-m_{33}^2+1)+\sqrt{\Delta}\right]}&-\sqrt{\frac{1}{2}\left[(m_{11}^2-m_{22}^2-m_{33}^2+1)-\sqrt{\Delta}\right]}&m_{33}
\end{pmatrix}
\]
\[
\begin{pmatrix}
m_{11}&-\sqrt{\frac{1}{2}\left[(-m_{11}^2-m_{22}^2+m_{33}^2+1)+\sqrt{\Delta}\right]}&-\sqrt{\frac{1}{2}\left[(-m_{11}^2+m_{22}^2-m_{33}^2+1)-\sqrt{\Delta}\right]}\\[6pt]
\sqrt{\frac{1}{2}\left[(-m_{11}^2-m_{22}^2+m_{33}^2+1)-\sqrt{\Delta}\right]}&m_{22}&\sqrt{\frac{1}{2}\left[(m_{11}^2-m_{22}^2-m_{33}^2+1)+\sqrt{\Delta}\right]}\\[6pt]
-\sqrt{\frac{1}{2}\left[(-m_{11}^2+m_{22}^2-m_{33}^2+1)+\sqrt{\Delta}\right]}&\sqrt{\frac{1}{2}\left[(m_{11}^2-m_{22}^2-m_{33}^2+1)-\sqrt{\Delta}\right]}&m_{33}
\end{pmatrix}
\]
\[
\begin{pmatrix}
m_{11}&\sqrt{\frac{1}{2}\left[(-m_{11}^2-m_{22}^2+m_{33}^2+1)+\sqrt{\Delta}\right]}&\sqrt{\frac{1}{2}\left[(-m_{11}^2+m_{22}^2-m_{33}^2+1)-\sqrt{\Delta}\right]}\\[6pt]
-\sqrt{\frac{1}{2}\left[(-m_{11}^2-m_{22}^2+m_{33}^2+1)-\sqrt{\Delta}\right]}&m_{22}&\sqrt{\frac{1}{2}\left[(m_{11}^2-m_{22}^2-m_{33}^2+1)+\sqrt{\Delta}\right]}\\[6pt]
\sqrt{\frac{1}{2}\left[(-m_{11}^2+m_{22}^2-m_{33}^2+1)+\sqrt{\Delta}\right]}&\sqrt{\frac{1}{2}\left[(m_{11}^2-m_{22}^2-m_{33}^2+1)-\sqrt{\Delta}\right]}&m_{33}
\end{pmatrix}
\]
\[
\begin{pmatrix}
m_{11}&-\sqrt{\frac{1}{2}\left[(-m_{11}^2-m_{22}^2+m_{33}^2+1)-\sqrt{\Delta}\right]}&\sqrt{\frac{1}{2}\left[(-m_{11}^2+m_{22}^2-m_{33}^2+1)+\sqrt{\Delta}\right]}\\[6pt]
\sqrt{\frac{1}{2}\left[(-m_{11}^2-m_{22}^2+m_{33}^2+1)+\sqrt{\Delta}\right]}&m_{22}&-\sqrt{\frac{1}{2}\left[(m_{11}^2-m_{22}^2-m_{33}^2+1)-\sqrt{\Delta}\right]}\\[6pt]
\sqrt{\frac{1}{2}\left[(-m_{11}^2+m_{22}^2-m_{33}^2+1)-\sqrt{\Delta}\right]}&-\sqrt{\frac{1}{2}\left[(m_{11}^2-m_{22}^2-m_{33}^2+1)+\sqrt{\Delta}\right]}&m_{33}
\end{pmatrix}
\]
\[
\begin{pmatrix}
m_{11}&\sqrt{\frac{1}{2}\left[(-m_{11}^2-m_{22}^2+m_{33}^2+1)-\sqrt{\Delta}\right]}&-\sqrt{\frac{1}{2}\left[(-m_{11}^2+m_{22}^2-m_{33}^2+1)+\sqrt{\Delta}\right]}\\[6pt]
-\sqrt{\frac{1}{2}\left[(-m_{11}^2-m_{22}^2+m_{33}^2+1)+\sqrt{\Delta}\right]}&m_{22}&-\sqrt{\frac{1}{2}\left[(m_{11}^2-m_{22}^2-m_{33}^2+1)-\sqrt{\Delta}\right]}\\[6pt]
-\sqrt{\frac{1}{2}\left[(-m_{11}^2+m_{22}^2-m_{33}^2+1)-\sqrt{\Delta}\right]}&-\sqrt{\frac{1}{2}\left[(m_{11}^2-m_{22}^2-m_{33}^2+1)+\sqrt{\Delta}\right]}&m_{33}
\end{pmatrix}
\]
\[
\begin{pmatrix}
m_{11}&-\sqrt{\frac{1}{2}\left[(-m_{11}^2-m_{22}^2+m_{33}^2+1)-\sqrt{\Delta}\right]}&-\sqrt{\frac{1}{2}\left[(-m_{11}^2+m_{22}^2-m_{33}^2+1)+\sqrt{\Delta}\right]}\\[6pt]
\sqrt{\frac{1}{2}\left[(-m_{11}^2-m_{22}^2+m_{33}^2+1)+\sqrt{\Delta}\right]}&m_{22}&\sqrt{\frac{1}{2}\left[(m_{11}^2-m_{22}^2-m_{33}^2+1)-\sqrt{\Delta}\right]}\\[6pt]
-\sqrt{\frac{1}{2}\left[(-m_{11}^2+m_{22}^2-m_{33}^2+1)-\sqrt{\Delta}\right]}&\sqrt{\frac{1}{2}\left[(m_{11}^2-m_{22}^2-m_{33}^2+1)+\sqrt{\Delta}\right]}&m_{33}
\end{pmatrix}
\]
\[
\begin{pmatrix}
m_{11}&\sqrt{\frac{1}{2}\left[(-m_{11}^2-m_{22}^2+m_{33}^2+1)-\sqrt{\Delta}\right]}&\sqrt{\frac{1}{2}\left[(-m_{11}^2+m_{22}^2-m_{33}^2+1)+\sqrt{\Delta}\right]}\\[6pt]
-\sqrt{\frac{1}{2}\left[(-m_{11}^2-m_{22}^2+m_{33}^2+1)+\sqrt{\Delta}\right]}&m_{22}&\sqrt{\frac{1}{2}\left[(m_{11}^2-m_{22}^2-m_{33}^2+1)-\sqrt{\Delta}\right]}\\[6pt]
\sqrt{\frac{1}{2}\left[(-m_{11}^2+m_{22}^2-m_{33}^2+1)-\sqrt{\Delta}\right]}&\sqrt{\frac{1}{2}\left[(m_{11}^2-m_{22}^2-m_{33}^2+1)+\sqrt{\Delta}\right]}&m_{33}
\end{pmatrix}
\]
\end{small}
\end{adjustwidth}
\subsection*{3$\ \ \ \ \boldsymbol{m_{11}m_{22}-m_{33}\geq0}$, $\boldsymbol{m_{22}m_{33}-m_{11}<0}$, $\boldsymbol{m_{33}m_{11}-m_{22}\geq0}$}
\begin{adjustwidth}{-3em}{}
\begin{small}
\[
\begin{pmatrix}
m_{11}&\sqrt{\frac{1}{2}\left[(-m_{11}^2-m_{22}^2+m_{33}^2+1)+\sqrt{\Delta}\right]}&-\sqrt{\frac{1}{2}\left[(-m_{11}^2+m_{22}^2-m_{33}^2+1)-\sqrt{\Delta}\right]}\\[6pt]
\sqrt{\frac{1}{2}\left[(-m_{11}^2-m_{22}^2+m_{33}^2+1)-\sqrt{\Delta}\right]}&m_{22}&-\sqrt{\frac{1}{2}\left[(m_{11}^2-m_{22}^2-m_{33}^2+1)+\sqrt{\Delta}\right]}\\[6pt]
-\sqrt{\frac{1}{2}\left[(-m_{11}^2+m_{22}^2-m_{33}^2+1)+\sqrt{\Delta}\right]}&\sqrt{\frac{1}{2}\left[(m_{11}^2-m_{22}^2-m_{33}^2+1)-\sqrt{\Delta}\right]}&m_{33}
\end{pmatrix}
\]
\[
\begin{pmatrix}
m_{11}&-\sqrt{\frac{1}{2}\left[(-m_{11}^2-m_{22}^2+m_{33}^2+1)+\sqrt{\Delta}\right]}&\sqrt{\frac{1}{2}\left[(-m_{11}^2+m_{22}^2-m_{33}^2+1)-\sqrt{\Delta}\right]}\\[6pt]
-\sqrt{\frac{1}{2}\left[(-m_{11}^2-m_{22}^2+m_{33}^2+1)-\sqrt{\Delta}\right]}&m_{22}&-\sqrt{\frac{1}{2}\left[(m_{11}^2-m_{22}^2-m_{33}^2+1)+\sqrt{\Delta}\right]}\\[6pt]
\sqrt{\frac{1}{2}\left[(-m_{11}^2+m_{22}^2-m_{33}^2+1)+\sqrt{\Delta}\right]}&\sqrt{\frac{1}{2}\left[(m_{11}^2-m_{22}^2-m_{33}^2+1)-\sqrt{\Delta}\right]}&m_{33}
\end{pmatrix}
\]
\[
\begin{pmatrix}
m_{11}&\sqrt{\frac{1}{2}\left[(-m_{11}^2-m_{22}^2+m_{33}^2+1)+\sqrt{\Delta}\right]}&\sqrt{\frac{1}{2}\left[(-m_{11}^2+m_{22}^2-m_{33}^2+1)-\sqrt{\Delta}\right]}\\[6pt]
\sqrt{\frac{1}{2}\left[(-m_{11}^2-m_{22}^2+m_{33}^2+1)-\sqrt{\Delta}\right]}&m_{22}&\sqrt{\frac{1}{2}\left[(m_{11}^2-m_{22}^2-m_{33}^2+1)+\sqrt{\Delta}\right]}\\[6pt]
\sqrt{\frac{1}{2}\left[(-m_{11}^2+m_{22}^2-m_{33}^2+1)+\sqrt{\Delta}\right]}&-\sqrt{\frac{1}{2}\left[(m_{11}^2-m_{22}^2-m_{33}^2+1)-\sqrt{\Delta}\right]}&m_{33}
\end{pmatrix}
\]
\[
\begin{pmatrix}
m_{11}&-\sqrt{\frac{1}{2}\left[(-m_{11}^2-m_{22}^2+m_{33}^2+1)+\sqrt{\Delta}\right]}&-\sqrt{\frac{1}{2}\left[(-m_{11}^2+m_{22}^2-m_{33}^2+1)-\sqrt{\Delta}\right]}\\[6pt]
-\sqrt{\frac{1}{2}\left[(-m_{11}^2-m_{22}^2+m_{33}^2+1)-\sqrt{\Delta}\right]}&m_{22}&\sqrt{\frac{1}{2}\left[(m_{11}^2-m_{22}^2-m_{33}^2+1)+\sqrt{\Delta}\right]}\\[6pt]
-\sqrt{\frac{1}{2}\left[(-m_{11}^2+m_{22}^2-m_{33}^2+1)+\sqrt{\Delta}\right]}&-\sqrt{\frac{1}{2}\left[(m_{11}^2-m_{22}^2-m_{33}^2+1)-\sqrt{\Delta}\right]}&m_{33}
\end{pmatrix}
\]
\[
\begin{pmatrix}
m_{11}&\sqrt{\frac{1}{2}\left[(-m_{11}^2-m_{22}^2+m_{33}^2+1)-\sqrt{\Delta}\right]}&-\sqrt{\frac{1}{2}\left[(-m_{11}^2+m_{22}^2-m_{33}^2+1)+\sqrt{\Delta}\right]}\\[6pt]
\sqrt{\frac{1}{2}\left[(-m_{11}^2-m_{22}^2+m_{33}^2+1)+\sqrt{\Delta}\right]}&m_{22}&-\sqrt{\frac{1}{2}\left[(m_{11}^2-m_{22}^2-m_{33}^2+1)-\sqrt{\Delta}\right]}\\[6pt]
-\sqrt{\frac{1}{2}\left[(-m_{11}^2+m_{22}^2-m_{33}^2+1)-\sqrt{\Delta}\right]}&\sqrt{\frac{1}{2}\left[(m_{11}^2-m_{22}^2-m_{33}^2+1)+\sqrt{\Delta}\right]}&m_{33}
\end{pmatrix}
\]
\[
\begin{pmatrix}
m_{11}&-\sqrt{\frac{1}{2}\left[(-m_{11}^2-m_{22}^2+m_{33}^2+1)-\sqrt{\Delta}\right]}&\sqrt{\frac{1}{2}\left[(-m_{11}^2+m_{22}^2-m_{33}^2+1)+\sqrt{\Delta}\right]}\\[6pt]
-\sqrt{\frac{1}{2}\left[(-m_{11}^2-m_{22}^2+m_{33}^2+1)+\sqrt{\Delta}\right]}&m_{22}&-\sqrt{\frac{1}{2}\left[(m_{11}^2-m_{22}^2-m_{33}^2+1)-\sqrt{\Delta}\right]}\\[6pt]
\sqrt{\frac{1}{2}\left[(-m_{11}^2+m_{22}^2-m_{33}^2+1)-\sqrt{\Delta}\right]}&\sqrt{\frac{1}{2}\left[(m_{11}^2-m_{22}^2-m_{33}^2+1)+\sqrt{\Delta}\right]}&m_{33}
\end{pmatrix}
\]
\[
\begin{pmatrix}
m_{11}&\sqrt{\frac{1}{2}\left[(-m_{11}^2-m_{22}^2+m_{33}^2+1)-\sqrt{\Delta}\right]}&\sqrt{\frac{1}{2}\left[(-m_{11}^2+m_{22}^2-m_{33}^2+1)+\sqrt{\Delta}\right]}\\[6pt]
\sqrt{\frac{1}{2}\left[(-m_{11}^2-m_{22}^2+m_{33}^2+1)+\sqrt{\Delta}\right]}&m_{22}&\sqrt{\frac{1}{2}\left[(m_{11}^2-m_{22}^2-m_{33}^2+1)-\sqrt{\Delta}\right]}\\[6pt]
\sqrt{\frac{1}{2}\left[(-m_{11}^2+m_{22}^2-m_{33}^2+1)-\sqrt{\Delta}\right]}&-\sqrt{\frac{1}{2}\left[(m_{11}^2-m_{22}^2-m_{33}^2+1)+\sqrt{\Delta}\right]}&m_{33}
\end{pmatrix}
\]
\[
\begin{pmatrix}
m_{11}&-\sqrt{\frac{1}{2}\left[(-m_{11}^2-m_{22}^2+m_{33}^2+1)-\sqrt{\Delta}\right]}&-\sqrt{\frac{1}{2}\left[(-m_{11}^2+m_{22}^2-m_{33}^2+1)+\sqrt{\Delta}\right]}\\[6pt]
-\sqrt{\frac{1}{2}\left[(-m_{11}^2-m_{22}^2+m_{33}^2+1)+\sqrt{\Delta}\right]}&m_{22}&\sqrt{\frac{1}{2}\left[(m_{11}^2-m_{22}^2-m_{33}^2+1)-\sqrt{\Delta}\right]}\\[6pt]
-\sqrt{\frac{1}{2}\left[(-m_{11}^2+m_{22}^2-m_{33}^2+1)-\sqrt{\Delta}\right]}&-\sqrt{\frac{1}{2}\left[(m_{11}^2-m_{22}^2-m_{33}^2+1)+\sqrt{\Delta}\right]}&m_{33}
\end{pmatrix}
\]
\end{small}
\end{adjustwidth}
\subsection*{4$\ \ \ \ \boldsymbol{m_{11}m_{22}-m_{33}\geq0}$, $\boldsymbol{m_{22}m_{33}-m_{11}\geq0}$, $\boldsymbol{m_{33}m_{11}-m_{22}<0}$}
\begin{adjustwidth}{-3em}{}
\begin{small}
\[
\begin{pmatrix}
m_{11}&-\sqrt{\frac{1}{2}\left[(-m_{11}^2-m_{22}^2+m_{33}^2+1)+\sqrt{\Delta}\right]}&\sqrt{\frac{1}{2}\left[(-m_{11}^2+m_{22}^2-m_{33}^2+1)-\sqrt{\Delta}\right]}\\[6pt]
-\sqrt{\frac{1}{2}\left[(-m_{11}^2-m_{22}^2+m_{33}^2+1)-\sqrt{\Delta}\right]}&m_{22}&\sqrt{\frac{1}{2}\left[(m_{11}^2-m_{22}^2-m_{33}^2+1)+\sqrt{\Delta}\right]}\\[6pt]
-\sqrt{\frac{1}{2}\left[(-m_{11}^2+m_{22}^2-m_{33}^2+1)+\sqrt{\Delta}\right]}&\sqrt{\frac{1}{2}\left[(m_{11}^2-m_{22}^2-m_{33}^2+1)-\sqrt{\Delta}\right]}&m_{33}
\end{pmatrix}
\]
\[
\begin{pmatrix}
m_{11}&\sqrt{\frac{1}{2}\left[(-m_{11}^2-m_{22}^2+m_{33}^2+1)+\sqrt{\Delta}\right]}&-\sqrt{\frac{1}{2}\left[(-m_{11}^2+m_{22}^2-m_{33}^2+1)-\sqrt{\Delta}\right]}\\[6pt]
\sqrt{\frac{1}{2}\left[(-m_{11}^2-m_{22}^2+m_{33}^2+1)-\sqrt{\Delta}\right]}&m_{22}&\sqrt{\frac{1}{2}\left[(m_{11}^2-m_{22}^2-m_{33}^2+1)+\sqrt{\Delta}\right]}\\[6pt]
\sqrt{\frac{1}{2}\left[(-m_{11}^2+m_{22}^2-m_{33}^2+1)+\sqrt{\Delta}\right]}&\sqrt{\frac{1}{2}\left[(m_{11}^2-m_{22}^2-m_{33}^2+1)-\sqrt{\Delta}\right]}&m_{33}
\end{pmatrix}
\]
\[
\begin{pmatrix}
m_{11}&-\sqrt{\frac{1}{2}\left[(-m_{11}^2-m_{22}^2+m_{33}^2+1)+\sqrt{\Delta}\right]}&-\sqrt{\frac{1}{2}\left[(-m_{11}^2+m_{22}^2-m_{33}^2+1)-\sqrt{\Delta}\right]}\\[6pt]
-\sqrt{\frac{1}{2}\left[(-m_{11}^2-m_{22}^2+m_{33}^2+1)-\sqrt{\Delta}\right]}&m_{22}&-\sqrt{\frac{1}{2}\left[(m_{11}^2-m_{22}^2-m_{33}^2+1)+\sqrt{\Delta}\right]}\\[6pt]
\sqrt{\frac{1}{2}\left[(-m_{11}^2+m_{22}^2-m_{33}^2+1)+\sqrt{\Delta}\right]}&-\sqrt{\frac{1}{2}\left[(m_{11}^2-m_{22}^2-m_{33}^2+1)-\sqrt{\Delta}\right]}&m_{33}
\end{pmatrix}
\]
\[
\begin{pmatrix}
m_{11}&\sqrt{\frac{1}{2}\left[(-m_{11}^2-m_{22}^2+m_{33}^2+1)+\sqrt{\Delta}\right]}&\sqrt{\frac{1}{2}\left[(-m_{11}^2+m_{22}^2-m_{33}^2+1)-\sqrt{\Delta}\right]}\\[6pt]
\sqrt{\frac{1}{2}\left[(-m_{11}^2-m_{22}^2+m_{33}^2+1)-\sqrt{\Delta}\right]}&m_{22}&-\sqrt{\frac{1}{2}\left[(m_{11}^2-m_{22}^2-m_{33}^2+1)+\sqrt{\Delta}\right]}\\[6pt]
-\sqrt{\frac{1}{2}\left[(-m_{11}^2+m_{22}^2-m_{33}^2+1)+\sqrt{\Delta}\right]}&-\sqrt{\frac{1}{2}\left[(m_{11}^2-m_{22}^2-m_{33}^2+1)-\sqrt{\Delta}\right]}&m_{33}
\end{pmatrix}
\]
\[
\begin{pmatrix}
m_{11}&-\sqrt{\frac{1}{2}\left[(-m_{11}^2-m_{22}^2+m_{33}^2+1)-\sqrt{\Delta}\right]}&\sqrt{\frac{1}{2}\left[(-m_{11}^2+m_{22}^2-m_{33}^2+1)+\sqrt{\Delta}\right]}\\[6pt]
-\sqrt{\frac{1}{2}\left[(-m_{11}^2-m_{22}^2+m_{33}^2+1)+\sqrt{\Delta}\right]}&m_{22}&\sqrt{\frac{1}{2}\left[(m_{11}^2-m_{22}^2-m_{33}^2+1)-\sqrt{\Delta}\right]}\\[6pt]
-\sqrt{\frac{1}{2}\left[(-m_{11}^2+m_{22}^2-m_{33}^2+1)-\sqrt{\Delta}\right]}&\sqrt{\frac{1}{2}\left[(m_{11}^2-m_{22}^2-m_{33}^2+1)+\sqrt{\Delta}\right]}&m_{33}
\end{pmatrix}
\]
\[
\begin{pmatrix}
m_{11}&\sqrt{\frac{1}{2}\left[(-m_{11}^2-m_{22}^2+m_{33}^2+1)-\sqrt{\Delta}\right]}&-\sqrt{\frac{1}{2}\left[(-m_{11}^2+m_{22}^2-m_{33}^2+1)+\sqrt{\Delta}\right]}\\[6pt]
\sqrt{\frac{1}{2}\left[(-m_{11}^2-m_{22}^2+m_{33}^2+1)+\sqrt{\Delta}\right]}&m_{22}&\sqrt{\frac{1}{2}\left[(m_{11}^2-m_{22}^2-m_{33}^2+1)-\sqrt{\Delta}\right]}\\[6pt]
\sqrt{\frac{1}{2}\left[(-m_{11}^2+m_{22}^2-m_{33}^2+1)-\sqrt{\Delta}\right]}&\sqrt{\frac{1}{2}\left[(m_{11}^2-m_{22}^2-m_{33}^2+1)+\sqrt{\Delta}\right]}&m_{33}
\end{pmatrix}
\]
\[
\begin{pmatrix}
m_{11}&-\sqrt{\frac{1}{2}\left[(-m_{11}^2-m_{22}^2+m_{33}^2+1)-\sqrt{\Delta}\right]}&-\sqrt{\frac{1}{2}\left[(-m_{11}^2+m_{22}^2-m_{33}^2+1)+\sqrt{\Delta}\right]}\\[6pt]
-\sqrt{\frac{1}{2}\left[(-m_{11}^2-m_{22}^2+m_{33}^2+1)+\sqrt{\Delta}\right]}&m_{22}&-\sqrt{\frac{1}{2}\left[(m_{11}^2-m_{22}^2-m_{33}^2+1)-\sqrt{\Delta}\right]}\\[6pt]
\sqrt{\frac{1}{2}\left[(-m_{11}^2+m_{22}^2-m_{33}^2+1)-\sqrt{\Delta}\right]}&-\sqrt{\frac{1}{2}\left[(m_{11}^2-m_{22}^2-m_{33}^2+1)+\sqrt{\Delta}\right]}&m_{33}
\end{pmatrix}
\]
\[
\begin{pmatrix}
m_{11}&\sqrt{\frac{1}{2}\left[(-m_{11}^2-m_{22}^2+m_{33}^2+1)-\sqrt{\Delta}\right]}&\sqrt{\frac{1}{2}\left[(-m_{11}^2+m_{22}^2-m_{33}^2+1)+\sqrt{\Delta}\right]}\\[6pt]
\sqrt{\frac{1}{2}\left[(-m_{11}^2-m_{22}^2+m_{33}^2+1)+\sqrt{\Delta}\right]}&m_{22}&-\sqrt{\frac{1}{2}\left[(m_{11}^2-m_{22}^2-m_{33}^2+1)-\sqrt{\Delta}\right]}\\[6pt]
-\sqrt{\frac{1}{2}\left[(-m_{11}^2+m_{22}^2-m_{33}^2+1)-\sqrt{\Delta}\right]}&-\sqrt{\frac{1}{2}\left[(m_{11}^2-m_{22}^2-m_{33}^2+1)+\sqrt{\Delta}\right]}&m_{33}
\end{pmatrix}
\]
\end{small}
\end{adjustwidth}
\subsection*{5$\ \ \ \ \boldsymbol{m_{11}m_{22}-m_{33}<0}$, $\boldsymbol{m_{22}m_{33}-m_{11}<0}$, $\boldsymbol{m_{33}m_{11}-m_{22}\geq0}$}
\begin{adjustwidth}{-3em}{}
\begin{small}
\[
\begin{pmatrix}
m_{11}&-\sqrt{\frac{1}{2}\left[(-m_{11}^2-m_{22}^2+m_{33}^2+1)+\sqrt{\Delta}\right]}&-\sqrt{\frac{1}{2}\left[(-m_{11}^2+m_{22}^2-m_{33}^2+1)-\sqrt{\Delta}\right]}\\[6pt]
\sqrt{\frac{1}{2}\left[(-m_{11}^2-m_{22}^2+m_{33}^2+1)-\sqrt{\Delta}\right]}&m_{22}&\sqrt{\frac{1}{2}\left[(m_{11}^2-m_{22}^2-m_{33}^2+1)+\sqrt{\Delta}\right]}\\[6pt]
-\sqrt{\frac{1}{2}\left[(-m_{11}^2+m_{22}^2-m_{33}^2+1)+\sqrt{\Delta}\right]}&-\sqrt{\frac{1}{2}\left[(m_{11}^2-m_{22}^2-m_{33}^2+1)-\sqrt{\Delta}\right]}&m_{33}
\end{pmatrix}
\]
\[
\begin{pmatrix}
m_{11}&\sqrt{\frac{1}{2}\left[(-m_{11}^2-m_{22}^2+m_{33}^2+1)+\sqrt{\Delta}\right]}&\sqrt{\frac{1}{2}\left[(-m_{11}^2+m_{22}^2-m_{33}^2+1)-\sqrt{\Delta}\right]}\\[6pt]
-\sqrt{\frac{1}{2}\left[(-m_{11}^2-m_{22}^2+m_{33}^2+1)-\sqrt{\Delta}\right]}&m_{22}&\sqrt{\frac{1}{2}\left[(m_{11}^2-m_{22}^2-m_{33}^2+1)+\sqrt{\Delta}\right]}\\[6pt]
\sqrt{\frac{1}{2}\left[(-m_{11}^2+m_{22}^2-m_{33}^2+1)+\sqrt{\Delta}\right]}&-\sqrt{\frac{1}{2}\left[(m_{11}^2-m_{22}^2-m_{33}^2+1)-\sqrt{\Delta}\right]}&m_{33}
\end{pmatrix}
\]
\[
\begin{pmatrix}
m_{11}&-\sqrt{\frac{1}{2}\left[(-m_{11}^2-m_{22}^2+m_{33}^2+1)+\sqrt{\Delta}\right]}&\sqrt{\frac{1}{2}\left[(-m_{11}^2+m_{22}^2-m_{33}^2+1)-\sqrt{\Delta}\right]}\\[6pt]
\sqrt{\frac{1}{2}\left[(-m_{11}^2-m_{22}^2+m_{33}^2+1)-\sqrt{\Delta}\right]}&m_{22}&-\sqrt{\frac{1}{2}\left[(m_{11}^2-m_{22}^2-m_{33}^2+1)+\sqrt{\Delta}\right]}\\[6pt]
\sqrt{\frac{1}{2}\left[(-m_{11}^2+m_{22}^2-m_{33}^2+1)+\sqrt{\Delta}\right]}&\sqrt{\frac{1}{2}\left[(m_{11}^2-m_{22}^2-m_{33}^2+1)-\sqrt{\Delta}\right]}&m_{33}
\end{pmatrix}
\]
\[
\begin{pmatrix}
m_{11}&\sqrt{\frac{1}{2}\left[(-m_{11}^2-m_{22}^2+m_{33}^2+1)+\sqrt{\Delta}\right]}&-\sqrt{\frac{1}{2}\left[(-m_{11}^2+m_{22}^2-m_{33}^2+1)-\sqrt{\Delta}\right]}\\[6pt]
-\sqrt{\frac{1}{2}\left[(-m_{11}^2-m_{22}^2+m_{33}^2+1)-\sqrt{\Delta}\right]}&m_{22}&-\sqrt{\frac{1}{2}\left[(m_{11}^2-m_{22}^2-m_{33}^2+1)+\sqrt{\Delta}\right]}\\[6pt]
-\sqrt{\frac{1}{2}\left[(-m_{11}^2+m_{22}^2-m_{33}^2+1)+\sqrt{\Delta}\right]}&\sqrt{\frac{1}{2}\left[(m_{11}^2-m_{22}^2-m_{33}^2+1)-\sqrt{\Delta}\right]}&m_{33}
\end{pmatrix}
\]
\[
\begin{pmatrix}
m_{11}&-\sqrt{\frac{1}{2}\left[(-m_{11}^2-m_{22}^2+m_{33}^2+1)-\sqrt{\Delta}\right]}&-\sqrt{\frac{1}{2}\left[(-m_{11}^2+m_{22}^2-m_{33}^2+1)+\sqrt{\Delta}\right]}\\[6pt]
\sqrt{\frac{1}{2}\left[(-m_{11}^2-m_{22}^2+m_{33}^2+1)+\sqrt{\Delta}\right]}&m_{22}&\sqrt{\frac{1}{2}\left[(m_{11}^2-m_{22}^2-m_{33}^2+1)-\sqrt{\Delta}\right]}\\[6pt]
-\sqrt{\frac{1}{2}\left[(-m_{11}^2+m_{22}^2-m_{33}^2+1)-\sqrt{\Delta}\right]}&-\sqrt{\frac{1}{2}\left[(m_{11}^2-m_{22}^2-m_{33}^2+1)+\sqrt{\Delta}\right]}&m_{33}
\end{pmatrix}
\]
\[
\begin{pmatrix}
m_{11}&\sqrt{\frac{1}{2}\left[(-m_{11}^2-m_{22}^2+m_{33}^2+1)-\sqrt{\Delta}\right]}&\sqrt{\frac{1}{2}\left[(-m_{11}^2+m_{22}^2-m_{33}^2+1)+\sqrt{\Delta}\right]}\\[6pt]
-\sqrt{\frac{1}{2}\left[(-m_{11}^2-m_{22}^2+m_{33}^2+1)+\sqrt{\Delta}\right]}&m_{22}&\sqrt{\frac{1}{2}\left[(m_{11}^2-m_{22}^2-m_{33}^2+1)-\sqrt{\Delta}\right]}\\[6pt]
\sqrt{\frac{1}{2}\left[(-m_{11}^2+m_{22}^2-m_{33}^2+1)-\sqrt{\Delta}\right]}&-\sqrt{\frac{1}{2}\left[(m_{11}^2-m_{22}^2-m_{33}^2+1)+\sqrt{\Delta}\right]}&m_{33}
\end{pmatrix}
\]
\[
\begin{pmatrix}
m_{11}&-\sqrt{\frac{1}{2}\left[(-m_{11}^2-m_{22}^2+m_{33}^2+1)-\sqrt{\Delta}\right]}&\sqrt{\frac{1}{2}\left[(-m_{11}^2+m_{22}^2-m_{33}^2+1)+\sqrt{\Delta}\right]}\\[6pt]
\sqrt{\frac{1}{2}\left[(-m_{11}^2-m_{22}^2+m_{33}^2+1)+\sqrt{\Delta}\right]}&m_{22}&-\sqrt{\frac{1}{2}\left[(m_{11}^2-m_{22}^2-m_{33}^2+1)-\sqrt{\Delta}\right]}\\[6pt]
\sqrt{\frac{1}{2}\left[(-m_{11}^2+m_{22}^2-m_{33}^2+1)-\sqrt{\Delta}\right]}&\sqrt{\frac{1}{2}\left[(m_{11}^2-m_{22}^2-m_{33}^2+1)+\sqrt{\Delta}\right]}&m_{33}
\end{pmatrix}
\]
\[
\begin{pmatrix}
m_{11}&\sqrt{\frac{1}{2}\left[(-m_{11}^2-m_{22}^2+m_{33}^2+1)-\sqrt{\Delta}\right]}&-\sqrt{\frac{1}{2}\left[(-m_{11}^2+m_{22}^2-m_{33}^2+1)+\sqrt{\Delta}\right]}\\[6pt]
-\sqrt{\frac{1}{2}\left[(-m_{11}^2-m_{22}^2+m_{33}^2+1)+\sqrt{\Delta}\right]}&m_{22}&-\sqrt{\frac{1}{2}\left[(m_{11}^2-m_{22}^2-m_{33}^2+1)-\sqrt{\Delta}\right]}\\[6pt]
-\sqrt{\frac{1}{2}\left[(-m_{11}^2+m_{22}^2-m_{33}^2+1)-\sqrt{\Delta}\right]}&\sqrt{\frac{1}{2}\left[(m_{11}^2-m_{22}^2-m_{33}^2+1)+\sqrt{\Delta}\right]}&m_{33}
\end{pmatrix}
\]
\end{small}
\end{adjustwidth}
\subsection*{6$\ \ \ \ \boldsymbol{m_{11}m_{22}-m_{33}\geq0}$, $\boldsymbol{m_{22}m_{33}-m_{11}<0}$, $\boldsymbol{m_{33}m_{11}-m_{22}<0}$}
\begin{adjustwidth}{-3em}{}
\begin{small}
\[
\begin{pmatrix}
m_{11}&-\sqrt{\frac{1}{2}\left[(-m_{11}^2-m_{22}^2+m_{33}^2+1)+\sqrt{\Delta}\right]}&-\sqrt{\frac{1}{2}\left[(-m_{11}^2+m_{22}^2-m_{33}^2+1)-\sqrt{\Delta}\right]}\\[6pt]
-\sqrt{\frac{1}{2}\left[(-m_{11}^2-m_{22}^2+m_{33}^2+1)-\sqrt{\Delta}\right]}&m_{22}&-\sqrt{\frac{1}{2}\left[(m_{11}^2-m_{22}^2-m_{33}^2+1)+\sqrt{\Delta}\right]}\\[6pt]
\sqrt{\frac{1}{2}\left[(-m_{11}^2+m_{22}^2-m_{33}^2+1)+\sqrt{\Delta}\right]}&\sqrt{\frac{1}{2}\left[(m_{11}^2-m_{22}^2-m_{33}^2+1)-\sqrt{\Delta}\right]}&m_{33}
\end{pmatrix}
\]
\[
\begin{pmatrix}
m_{11}&\sqrt{\frac{1}{2}\left[(-m_{11}^2-m_{22}^2+m_{33}^2+1)+\sqrt{\Delta}\right]}&\sqrt{\frac{1}{2}\left[(-m_{11}^2+m_{22}^2-m_{33}^2+1)-\sqrt{\Delta}\right]}\\[6pt]
\sqrt{\frac{1}{2}\left[(-m_{11}^2-m_{22}^2+m_{33}^2+1)-\sqrt{\Delta}\right]}&m_{22}&-\sqrt{\frac{1}{2}\left[(m_{11}^2-m_{22}^2-m_{33}^2+1)+\sqrt{\Delta}\right]}\\[6pt]
-\sqrt{\frac{1}{2}\left[(-m_{11}^2+m_{22}^2-m_{33}^2+1)+\sqrt{\Delta}\right]}&\sqrt{\frac{1}{2}\left[(m_{11}^2-m_{22}^2-m_{33}^2+1)-\sqrt{\Delta}\right]}&m_{33}
\end{pmatrix}
\]
\[
\begin{pmatrix}
m_{11}&-\sqrt{\frac{1}{2}\left[(-m_{11}^2-m_{22}^2+m_{33}^2+1)+\sqrt{\Delta}\right]}&\sqrt{\frac{1}{2}\left[(-m_{11}^2+m_{22}^2-m_{33}^2+1)-\sqrt{\Delta}\right]}\\[6pt]
-\sqrt{\frac{1}{2}\left[(-m_{11}^2-m_{22}^2+m_{33}^2+1)-\sqrt{\Delta}\right]}&m_{22}&\sqrt{\frac{1}{2}\left[(m_{11}^2-m_{22}^2-m_{33}^2+1)+\sqrt{\Delta}\right]}\\[6pt]
-\sqrt{\frac{1}{2}\left[(-m_{11}^2+m_{22}^2-m_{33}^2+1)+\sqrt{\Delta}\right]}&-\sqrt{\frac{1}{2}\left[(m_{11}^2-m_{22}^2-m_{33}^2+1)-\sqrt{\Delta}\right]}&m_{33}
\end{pmatrix}
\]
\[
\begin{pmatrix}
m_{11}&\sqrt{\frac{1}{2}\left[(-m_{11}^2-m_{22}^2+m_{33}^2+1)+\sqrt{\Delta}\right]}&-\sqrt{\frac{1}{2}\left[(-m_{11}^2+m_{22}^2-m_{33}^2+1)-\sqrt{\Delta}\right]}\\[6pt]
\sqrt{\frac{1}{2}\left[(-m_{11}^2-m_{22}^2+m_{33}^2+1)-\sqrt{\Delta}\right]}&m_{22}&\sqrt{\frac{1}{2}\left[(m_{11}^2-m_{22}^2-m_{33}^2+1)+\sqrt{\Delta}\right]}\\[6pt]
\sqrt{\frac{1}{2}\left[(-m_{11}^2+m_{22}^2-m_{33}^2+1)+\sqrt{\Delta}\right]}&-\sqrt{\frac{1}{2}\left[(m_{11}^2-m_{22}^2-m_{33}^2+1)-\sqrt{\Delta}\right]}&m_{33}
\end{pmatrix}
\]
\[
\begin{pmatrix}
m_{11}&-\sqrt{\frac{1}{2}\left[(-m_{11}^2-m_{22}^2+m_{33}^2+1)-\sqrt{\Delta}\right]}&-\sqrt{\frac{1}{2}\left[(-m_{11}^2+m_{22}^2-m_{33}^2+1)+\sqrt{\Delta}\right]}\\[6pt]
-\sqrt{\frac{1}{2}\left[(-m_{11}^2-m_{22}^2+m_{33}^2+1)+\sqrt{\Delta}\right]}&m_{22}&-\sqrt{\frac{1}{2}\left[(m_{11}^2-m_{22}^2-m_{33}^2+1)-\sqrt{\Delta}\right]}\\[6pt]
\sqrt{\frac{1}{2}\left[(-m_{11}^2+m_{22}^2-m_{33}^2+1)-\sqrt{\Delta}\right]}&\sqrt{\frac{1}{2}\left[(m_{11}^2-m_{22}^2-m_{33}^2+1)+\sqrt{\Delta}\right]}&m_{33}
\end{pmatrix}
\]
\[
\begin{pmatrix}
m_{11}&\sqrt{\frac{1}{2}\left[(-m_{11}^2-m_{22}^2+m_{33}^2+1)-\sqrt{\Delta}\right]}&\sqrt{\frac{1}{2}\left[(-m_{11}^2+m_{22}^2-m_{33}^2+1)+\sqrt{\Delta}\right]}\\[6pt]
\sqrt{\frac{1}{2}\left[(-m_{11}^2-m_{22}^2+m_{33}^2+1)+\sqrt{\Delta}\right]}&m_{22}&-\sqrt{\frac{1}{2}\left[(m_{11}^2-m_{22}^2-m_{33}^2+1)-\sqrt{\Delta}\right]}\\[6pt]
-\sqrt{\frac{1}{2}\left[(-m_{11}^2+m_{22}^2-m_{33}^2+1)-\sqrt{\Delta}\right]}&\sqrt{\frac{1}{2}\left[(m_{11}^2-m_{22}^2-m_{33}^2+1)+\sqrt{\Delta}\right]}&m_{33}
\end{pmatrix}
\]
\[
\begin{pmatrix}
m_{11}&-\sqrt{\frac{1}{2}\left[(-m_{11}^2-m_{22}^2+m_{33}^2+1)-\sqrt{\Delta}\right]}&\sqrt{\frac{1}{2}\left[(-m_{11}^2+m_{22}^2-m_{33}^2+1)+\sqrt{\Delta}\right]}\\[6pt]
-\sqrt{\frac{1}{2}\left[(-m_{11}^2-m_{22}^2+m_{33}^2+1)+\sqrt{\Delta}\right]}&m_{22}&\sqrt{\frac{1}{2}\left[(m_{11}^2-m_{22}^2-m_{33}^2+1)-\sqrt{\Delta}\right]}\\[6pt]
-\sqrt{\frac{1}{2}\left[(-m_{11}^2+m_{22}^2-m_{33}^2+1)-\sqrt{\Delta}\right]}&-\sqrt{\frac{1}{2}\left[(m_{11}^2-m_{22}^2-m_{33}^2+1)+\sqrt{\Delta}\right]}&m_{33}
\end{pmatrix}
\]
\[
\begin{pmatrix}
m_{11}&\sqrt{\frac{1}{2}\left[(-m_{11}^2-m_{22}^2+m_{33}^2+1)-\sqrt{\Delta}\right]}&-\sqrt{\frac{1}{2}\left[(-m_{11}^2+m_{22}^2-m_{33}^2+1)+\sqrt{\Delta}\right]}\\[6pt]
\sqrt{\frac{1}{2}\left[(-m_{11}^2-m_{22}^2+m_{33}^2+1)+\sqrt{\Delta}\right]}&m_{22}&\sqrt{\frac{1}{2}\left[(m_{11}^2-m_{22}^2-m_{33}^2+1)-\sqrt{\Delta}\right]}\\[6pt]
\sqrt{\frac{1}{2}\left[(-m_{11}^2+m_{22}^2-m_{33}^2+1)-\sqrt{\Delta}\right]}&-\sqrt{\frac{1}{2}\left[(m_{11}^2-m_{22}^2-m_{33}^2+1)+\sqrt{\Delta}\right]}&m_{33}
\end{pmatrix}
\]
\end{small}
\end{adjustwidth}
\subsection*{7$\ \ \ \ \boldsymbol{m_{11}m_{22}-m_{33}<0}$, $\boldsymbol{m_{22}m_{33}-m_{11}\geq0}$, $\boldsymbol{m_{33}m_{11}-m_{22}<0}$}
\begin{adjustwidth}{-3em}{}
\begin{small}
\[
\begin{pmatrix}
m_{11}&\sqrt{\frac{1}{2}\left[(-m_{11}^2-m_{22}^2+m_{33}^2+1)+\sqrt{\Delta}\right]}&\sqrt{\frac{1}{2}\left[(-m_{11}^2+m_{22}^2-m_{33}^2+1)-\sqrt{\Delta}\right]}\\[6pt]
-\sqrt{\frac{1}{2}\left[(-m_{11}^2-m_{22}^2+m_{33}^2+1)-\sqrt{\Delta}\right]}&m_{22}&-\sqrt{\frac{1}{2}\left[(m_{11}^2-m_{22}^2-m_{33}^2+1)+\sqrt{\Delta}\right]}\\[6pt]
-\sqrt{\frac{1}{2}\left[(-m_{11}^2+m_{22}^2-m_{33}^2+1)+\sqrt{\Delta}\right]}&-\sqrt{\frac{1}{2}\left[(m_{11}^2-m_{22}^2-m_{33}^2+1)-\sqrt{\Delta}\right]}&m_{33}
\end{pmatrix}
\]
\[
\begin{pmatrix}
m_{11}&-\sqrt{\frac{1}{2}\left[(-m_{11}^2-m_{22}^2+m_{33}^2+1)+\sqrt{\Delta}\right]}&-\sqrt{\frac{1}{2}\left[(-m_{11}^2+m_{22}^2-m_{33}^2+1)-\sqrt{\Delta}\right]}\\[6pt]
\sqrt{\frac{1}{2}\left[(-m_{11}^2-m_{22}^2+m_{33}^2+1)-\sqrt{\Delta}\right]}&m_{22}&-\sqrt{\frac{1}{2}\left[(m_{11}^2-m_{22}^2-m_{33}^2+1)+\sqrt{\Delta}\right]}\\[6pt]
\sqrt{\frac{1}{2}\left[(-m_{11}^2+m_{22}^2-m_{33}^2+1)+\sqrt{\Delta}\right]}&-\sqrt{\frac{1}{2}\left[(m_{11}^2-m_{22}^2-m_{33}^2+1)-\sqrt{\Delta}\right]}&m_{33}
\end{pmatrix}
\]
\[
\begin{pmatrix}
m_{11}&\sqrt{\frac{1}{2}\left[(-m_{11}^2-m_{22}^2+m_{33}^2+1)+\sqrt{\Delta}\right]}&-\sqrt{\frac{1}{2}\left[(-m_{11}^2+m_{22}^2-m_{33}^2+1)-\sqrt{\Delta}\right]}\\[6pt]
-\sqrt{\frac{1}{2}\left[(-m_{11}^2-m_{22}^2+m_{33}^2+1)-\sqrt{\Delta}\right]}&m_{22}&\sqrt{\frac{1}{2}\left[(m_{11}^2-m_{22}^2-m_{33}^2+1)+\sqrt{\Delta}\right]}\\[6pt]
\sqrt{\frac{1}{2}\left[(-m_{11}^2+m_{22}^2-m_{33}^2+1)+\sqrt{\Delta}\right]}&\sqrt{\frac{1}{2}\left[(m_{11}^2-m_{22}^2-m_{33}^2+1)-\sqrt{\Delta}\right]}&m_{33}
\end{pmatrix}
\]
\[
\begin{pmatrix}
m_{11}&-\sqrt{\frac{1}{2}\left[(-m_{11}^2-m_{22}^2+m_{33}^2+1)+\sqrt{\Delta}\right]}&\sqrt{\frac{1}{2}\left[(-m_{11}^2+m_{22}^2-m_{33}^2+1)-\sqrt{\Delta}\right]}\\[6pt]
\sqrt{\frac{1}{2}\left[(-m_{11}^2-m_{22}^2+m_{33}^2+1)-\sqrt{\Delta}\right]}&m_{22}&\sqrt{\frac{1}{2}\left[(m_{11}^2-m_{22}^2-m_{33}^2+1)+\sqrt{\Delta}\right]}\\[6pt]
-\sqrt{\frac{1}{2}\left[(-m_{11}^2+m_{22}^2-m_{33}^2+1)+\sqrt{\Delta}\right]}&\sqrt{\frac{1}{2}\left[(m_{11}^2-m_{22}^2-m_{33}^2+1)-\sqrt{\Delta}\right]}&m_{33}
\end{pmatrix}
\]
\[
\begin{pmatrix}
m_{11}&\sqrt{\frac{1}{2}\left[(-m_{11}^2-m_{22}^2+m_{33}^2+1)-\sqrt{\Delta}\right]}&\sqrt{\frac{1}{2}\left[(-m_{11}^2+m_{22}^2-m_{33}^2+1)+\sqrt{\Delta}\right]}\\[6pt]
-\sqrt{\frac{1}{2}\left[(-m_{11}^2-m_{22}^2+m_{33}^2+1)+\sqrt{\Delta}\right]}&m_{22}&-\sqrt{\frac{1}{2}\left[(m_{11}^2-m_{22}^2-m_{33}^2+1)-\sqrt{\Delta}\right]}\\[6pt]
-\sqrt{\frac{1}{2}\left[(-m_{11}^2+m_{22}^2-m_{33}^2+1)-\sqrt{\Delta}\right]}&-\sqrt{\frac{1}{2}\left[(m_{11}^2-m_{22}^2-m_{33}^2+1)+\sqrt{\Delta}\right]}&m_{33}
\end{pmatrix}
\]
\[
\begin{pmatrix}
m_{11}&-\sqrt{\frac{1}{2}\left[(-m_{11}^2-m_{22}^2+m_{33}^2+1)-\sqrt{\Delta}\right]}&-\sqrt{\frac{1}{2}\left[(-m_{11}^2+m_{22}^2-m_{33}^2+1)+\sqrt{\Delta}\right]}\\[6pt]
\sqrt{\frac{1}{2}\left[(-m_{11}^2-m_{22}^2+m_{33}^2+1)+\sqrt{\Delta}\right]}&m_{22}&-\sqrt{\frac{1}{2}\left[(m_{11}^2-m_{22}^2-m_{33}^2+1)-\sqrt{\Delta}\right]}\\[6pt]
\sqrt{\frac{1}{2}\left[(-m_{11}^2+m_{22}^2-m_{33}^2+1)-\sqrt{\Delta}\right]}&-\sqrt{\frac{1}{2}\left[(m_{11}^2-m_{22}^2-m_{33}^2+1)+\sqrt{\Delta}\right]}&m_{33}
\end{pmatrix}
\]
\[
\begin{pmatrix}
m_{11}&\sqrt{\frac{1}{2}\left[(-m_{11}^2-m_{22}^2+m_{33}^2+1)-\sqrt{\Delta}\right]}&-\sqrt{\frac{1}{2}\left[(-m_{11}^2+m_{22}^2-m_{33}^2+1)+\sqrt{\Delta}\right]}\\[6pt]
-\sqrt{\frac{1}{2}\left[(-m_{11}^2-m_{22}^2+m_{33}^2+1)+\sqrt{\Delta}\right]}&m_{22}&\sqrt{\frac{1}{2}\left[(m_{11}^2-m_{22}^2-m_{33}^2+1)-\sqrt{\Delta}\right]}\\[6pt]
\sqrt{\frac{1}{2}\left[(-m_{11}^2+m_{22}^2-m_{33}^2+1)-\sqrt{\Delta}\right]}&\sqrt{\frac{1}{2}\left[(m_{11}^2-m_{22}^2-m_{33}^2+1)+\sqrt{\Delta}\right]}&m_{33}
\end{pmatrix}
\]
\[
\begin{pmatrix}
m_{11}&-\sqrt{\frac{1}{2}\left[(-m_{11}^2-m_{22}^2+m_{33}^2+1)-\sqrt{\Delta}\right]}&\sqrt{\frac{1}{2}\left[(-m_{11}^2+m_{22}^2-m_{33}^2+1)+\sqrt{\Delta}\right]}\\[6pt]
\sqrt{\frac{1}{2}\left[(-m_{11}^2-m_{22}^2+m_{33}^2+1)+\sqrt{\Delta}\right]}&m_{22}&\sqrt{\frac{1}{2}\left[(m_{11}^2-m_{22}^2-m_{33}^2+1)-\sqrt{\Delta}\right]}\\[6pt]
-\sqrt{\frac{1}{2}\left[(-m_{11}^2+m_{22}^2-m_{33}^2+1)-\sqrt{\Delta}\right]}&\sqrt{\frac{1}{2}\left[(m_{11}^2-m_{22}^2-m_{33}^2+1)+\sqrt{\Delta}\right]}&m_{33}
\end{pmatrix}
\]
\end{small}
\end{adjustwidth}
\subsection*{8$\ \ \ \ \boldsymbol{m_{11}m_{22}-m_{33}<0}$, $\boldsymbol{m_{22}m_{33}-m_{11}<0}$, $\boldsymbol{m_{33}m_{11}-m_{22}<0}$}
\begin{adjustwidth}{-3em}{}
\begin{small}
\[
\begin{pmatrix}
m_{11}&-\sqrt{\frac{1}{2}\left[(-m_{11}^2-m_{22}^2+m_{33}^2+1)+\sqrt{\Delta}\right]}&-\sqrt{\frac{1}{2}\left[(-m_{11}^2+m_{22}^2-m_{33}^2+1)-\sqrt{\Delta}\right]}\\[6pt]
\sqrt{\frac{1}{2}\left[(-m_{11}^2-m_{22}^2+m_{33}^2+1)-\sqrt{\Delta}\right]}&m_{22}&-\sqrt{\frac{1}{2}\left[(m_{11}^2-m_{22}^2-m_{33}^2+1)+\sqrt{\Delta}\right]}\\[6pt]
\sqrt{\frac{1}{2}\left[(-m_{11}^2+m_{22}^2-m_{33}^2+1)+\sqrt{\Delta}\right]}&\sqrt{\frac{1}{2}\left[(m_{11}^2-m_{22}^2-m_{33}^2+1)-\sqrt{\Delta}\right]}&m_{33}
\end{pmatrix}
\]
\[
\begin{pmatrix}
m_{11}&\sqrt{\frac{1}{2}\left[(-m_{11}^2-m_{22}^2+m_{33}^2+1)+\sqrt{\Delta}\right]}&\sqrt{\frac{1}{2}\left[(-m_{11}^2+m_{22}^2-m_{33}^2+1)-\sqrt{\Delta}\right]}\\[6pt]
-\sqrt{\frac{1}{2}\left[(-m_{11}^2-m_{22}^2+m_{33}^2+1)-\sqrt{\Delta}\right]}&m_{22}&-\sqrt{\frac{1}{2}\left[(m_{11}^2-m_{22}^2-m_{33}^2+1)+\sqrt{\Delta}\right]}\\[6pt]
-\sqrt{\frac{1}{2}\left[(-m_{11}^2+m_{22}^2-m_{33}^2+1)+\sqrt{\Delta}\right]}&\sqrt{\frac{1}{2}\left[(m_{11}^2-m_{22}^2-m_{33}^2+1)-\sqrt{\Delta}\right]}&m_{33}
\end{pmatrix}
\]
\[
\begin{pmatrix}
m_{11}&-\sqrt{\frac{1}{2}\left[(-m_{11}^2-m_{22}^2+m_{33}^2+1)+\sqrt{\Delta}\right]}&\sqrt{\frac{1}{2}\left[(-m_{11}^2+m_{22}^2-m_{33}^2+1)-\sqrt{\Delta}\right]}\\[6pt]
\sqrt{\frac{1}{2}\left[(-m_{11}^2-m_{22}^2+m_{33}^2+1)-\sqrt{\Delta}\right]}&m_{22}&\sqrt{\frac{1}{2}\left[(m_{11}^2-m_{22}^2-m_{33}^2+1)+\sqrt{\Delta}\right]}\\[6pt]
-\sqrt{\frac{1}{2}\left[(-m_{11}^2+m_{22}^2-m_{33}^2+1)+\sqrt{\Delta}\right]}&-\sqrt{\frac{1}{2}\left[(m_{11}^2-m_{22}^2-m_{33}^2+1)-\sqrt{\Delta}\right]}&m_{33}
\end{pmatrix}
\]
\[
\begin{pmatrix}
m_{11}&\sqrt{\frac{1}{2}\left[(-m_{11}^2-m_{22}^2+m_{33}^2+1)+\sqrt{\Delta}\right]}&-\sqrt{\frac{1}{2}\left[(-m_{11}^2+m_{22}^2-m_{33}^2+1)-\sqrt{\Delta}\right]}\\[6pt]
-\sqrt{\frac{1}{2}\left[(-m_{11}^2-m_{22}^2+m_{33}^2+1)-\sqrt{\Delta}\right]}&m_{22}&\sqrt{\frac{1}{2}\left[(m_{11}^2-m_{22}^2-m_{33}^2+1)+\sqrt{\Delta}\right]}\\[6pt]
\sqrt{\frac{1}{2}\left[(-m_{11}^2+m_{22}^2-m_{33}^2+1)+\sqrt{\Delta}\right]}&-\sqrt{\frac{1}{2}\left[(m_{11}^2-m_{22}^2-m_{33}^2+1)-\sqrt{\Delta}\right]}&m_{33}
\end{pmatrix}
\]
\[
\begin{pmatrix}
m_{11}&-\sqrt{\frac{1}{2}\left[(-m_{11}^2-m_{22}^2+m_{33}^2+1)-\sqrt{\Delta}\right]}&-\sqrt{\frac{1}{2}\left[(-m_{11}^2+m_{22}^2-m_{33}^2+1)+\sqrt{\Delta}\right]}\\[6pt]
\sqrt{\frac{1}{2}\left[(-m_{11}^2-m_{22}^2+m_{33}^2+1)+\sqrt{\Delta}\right]}&m_{22}&-\sqrt{\frac{1}{2}\left[(m_{11}^2-m_{22}^2-m_{33}^2+1)-\sqrt{\Delta}\right]}\\[6pt]
\sqrt{\frac{1}{2}\left[(-m_{11}^2+m_{22}^2-m_{33}^2+1)-\sqrt{\Delta}\right]}&\sqrt{\frac{1}{2}\left[(m_{11}^2-m_{22}^2-m_{33}^2+1)+\sqrt{\Delta}\right]}&m_{33}
\end{pmatrix}
\]
\[
\begin{pmatrix}
m_{11}&\sqrt{\frac{1}{2}\left[(-m_{11}^2-m_{22}^2+m_{33}^2+1)-\sqrt{\Delta}\right]}&\sqrt{\frac{1}{2}\left[(-m_{11}^2+m_{22}^2-m_{33}^2+1)+\sqrt{\Delta}\right]}\\[6pt]
-\sqrt{\frac{1}{2}\left[(-m_{11}^2-m_{22}^2+m_{33}^2+1)+\sqrt{\Delta}\right]}&m_{22}&-\sqrt{\frac{1}{2}\left[(m_{11}^2-m_{22}^2-m_{33}^2+1)-\sqrt{\Delta}\right]}\\[6pt]
-\sqrt{\frac{1}{2}\left[(-m_{11}^2+m_{22}^2-m_{33}^2+1)-\sqrt{\Delta}\right]}&\sqrt{\frac{1}{2}\left[(m_{11}^2-m_{22}^2-m_{33}^2+1)+\sqrt{\Delta}\right]}&m_{33}
\end{pmatrix}
\]
\[
\begin{pmatrix}
m_{11}&-\sqrt{\frac{1}{2}\left[(-m_{11}^2-m_{22}^2+m_{33}^2+1)-\sqrt{\Delta}\right]}&\sqrt{\frac{1}{2}\left[(-m_{11}^2+m_{22}^2-m_{33}^2+1)+\sqrt{\Delta}\right]}\\[6pt]
\sqrt{\frac{1}{2}\left[(-m_{11}^2-m_{22}^2+m_{33}^2+1)+\sqrt{\Delta}\right]}&m_{22}&\sqrt{\frac{1}{2}\left[(m_{11}^2-m_{22}^2-m_{33}^2+1)-\sqrt{\Delta}\right]}\\[6pt]
-\sqrt{\frac{1}{2}\left[(-m_{11}^2+m_{22}^2-m_{33}^2+1)-\sqrt{\Delta}\right]}&-\sqrt{\frac{1}{2}\left[(m_{11}^2-m_{22}^2-m_{33}^2+1)+\sqrt{\Delta}\right]}&m_{33}
\end{pmatrix}
\]
\[
\begin{pmatrix}
m_{11}&\sqrt{\frac{1}{2}\left[(-m_{11}^2-m_{22}^2+m_{33}^2+1)-\sqrt{\Delta}\right]}&-\sqrt{\frac{1}{2}\left[(-m_{11}^2+m_{22}^2-m_{33}^2+1)+\sqrt{\Delta}\right]}\\[6pt]
-\sqrt{\frac{1}{2}\left[(-m_{11}^2-m_{22}^2+m_{33}^2+1)+\sqrt{\Delta}\right]}&m_{22}&\sqrt{\frac{1}{2}\left[(m_{11}^2-m_{22}^2-m_{33}^2+1)-\sqrt{\Delta}\right]}\\[6pt]
\sqrt{\frac{1}{2}\left[(-m_{11}^2+m_{22}^2-m_{33}^2+1)-\sqrt{\Delta}\right]}&-\sqrt{\frac{1}{2}\left[(m_{11}^2-m_{22}^2-m_{33}^2+1)+\sqrt{\Delta}\right]}&m_{33}
\end{pmatrix}
\]
\end{small}
\end{adjustwidth}
\bibliographystyle{unsrt}
\bibliography{refs}
\end{document}